\documentclass[twocolumn,hyperpdf,amsmath,amssymb,aps,prd,10pt,superscriptaddress,nofootinbib,noeprint,preprintnumbers, floatfix]{revtex4-1}

\usepackage{graphicx}
\usepackage{dcolumn}
\usepackage{bm}
\usepackage{amsmath}
\usepackage{amsfonts}
\usepackage{slashed}
\usepackage{hyperref}
\usepackage{multirow}
\usepackage{soul}
\usepackage{mathtools}
\usepackage{float}
\usepackage{afterpage}
\usepackage{color}
\usepackage[utf8]{inputenc} 

\definecolor{jlab_red}{RGB}{192,39,45}
\definecolor{jlab_orange}{RGB}{249,102,0}
\definecolor{jlab_blue}{RGB}{47,122,121}
\definecolor{jlab_green}{RGB}{65,125,10}

\hypersetup{%
pdftitle = {Hadron Spectrum Collaboration},
pdfsubject = {QCD},
pdfkeywords = {QCD, Hadron, Physics, JLab, Lattice, Meson, Scattering},
pdfauthor = {Hadron Spectrum Collaboration},
colorlinks = {true},
filecolor = {black},
linkcolor = {jlab_blue},
menucolor = {black},
citecolor = {jlab_green},
urlcolor = {jlab_green},
}{}

\newcommand{\cm}{\ensuremath{\mathsf{cm}}}
\newcommand{\sm}{\ensuremath{\text{-}}}
\newcommand{\kk}{\ensuremath{{K\overline{K}}}}
\newcommand{\kkb}{\ensuremath{{K\!\bar{K}}}}
\newcommand{\kb}{\ensuremath{{\bar{K}}}}
\newcommand{\pe}{\ensuremath{{\pi\eta}}}
\newcommand{\pep}{\ensuremath{{\pi\eta^\prime}}}
\newcommand{\ep}{\ensuremath{{\eta^\prime}}}

\newcommand{\minitab}[2][l]{\begin{tabular}{#1}#2\end{tabular}}

\begin{document}

\preprint{JLAB-THY-16-2216}
\preprint{DAMTP-2016-18}

\title{An $a_0$ resonance in strongly coupled $\pi\eta,K\overline{K}$ scattering from lattice QCD}


\author{Jozef~J.~Dudek}
\email{dudek@jlab.org}
\affiliation{Theory Center, Jefferson Lab, 12000 Jefferson Avenue, Newport News, VA 23606, USA}
\affiliation{Department of Physics, Old Dominion University, Norfolk, VA 23529, USA}

\author{Robert~G.~Edwards}
\email{edwards@jlab.org}
\affiliation{Theory Center, Jefferson Lab, 12000 Jefferson Avenue, Newport News, VA 23606, USA}

\author{David~J.~Wilson}
\email{d.j.wilson@damtp.cam.ac.uk}
\affiliation{Department of Physics, Old Dominion University, Norfolk, VA 23529, USA}
\affiliation{Department of Applied Mathematics and Theoretical Physics, Centre for Mathematical Sciences, University of Cambridge, Wilberforce Road, Cambridge, CB3 0WA, UK}


\collaboration{for the Hadron Spectrum Collaboration}
\date{\today}

\begin{abstract}

We present the first calculation of coupled-channel meson-meson scattering in the isospin $=1$, \mbox{$G$-parity} negative sector, with channels \pe, \kk~and \pep, in a first-principles approach to QCD. From the discrete spectrum of eigenstates in three volumes extracted from lattice QCD correlation functions we determine the energy dependence of the $S$-matrix, and find that the \mbox{$S$-wave} features a prominent cusp-like structure in $\pe \to \pe$ close to $K\overline{K}$ threshold coupled with a rapid turn on of amplitudes leading to the \kk~final-state. This behavior is traced to an $a_0(980)$-like resonance, strongly coupled to both \pe~and \kk, which is identified with a pole in the complex energy plane, appearing on only a single unphysical Riemann sheet. Consideration of $D$-wave scattering suggests a narrow tensor resonance at higher energy.
\end{abstract}

\maketitle

\section{Introduction} \label{intro}

The spectrum of excited mesons is one of \emph{resonances} appearing in the scattering of lighter pseudoscalar mesons, like $\pi$, $K$ and $\eta$, which are stable against decay by the strong interaction, and in most cases these resonances decay to more than one final state, necessitating study of \emph{coupled-channel} scattering.
One of the lightest experimentally observed resonances is the $a_0(980)$, an isovector ($I=1$), scalar ($J^P=0^+$) meson, seen primarily as an enhancement in \pe ~final states very close to the \kk ~threshold in, amongst other processes, $p\overline{p}$ annihilation~\cite{Conforto:1967zz,Astier:1967zz}, $\pi p$ scattering~\cite{Teige:1996fi}, $pp$ collisions~\cite{Barberis:2000cx}, radiative decays of the $\phi$~\cite{Ambrosino:2009py}, $\gamma\gamma$ fusion~\cite{Uehara:2009cf}, and in decays of heavy-quark mesons~\cite{Ablikim:2014dnh, Aaij:2015lsa}. Because the observed enhancement straddles the \kk ~threshold, any attempt to describe this state must consider coupled \pe,~\kk ~scattering, and parameterizations which describe the available data~\cite{Flatte:1976xv,Flatte:1976xu,Janssen:1994wn,Chung:1995dx,Oller:2000ma,Achasov:2002ir,Bargiotti:2003ev,Baru:2004xg,Bugg:2008ig,Albaladejo:2015aca} typically find a large coupling of the resonance to the \kk ~channel. This finding, combined with proximity of the state to the \kk~threshold, and the claimed presence of a second isovector scalar resonance near 1450 MeV~\cite{Agashe:2014kda, Amsler:1995bf}, has led to suggestions that the $a_0(980)$ should not be identified with the $q\bar{q}$ state expected within $q\bar{q}$ quark models, but rather that it might be dominated by a \kk ~molecular configuration~\cite{Weinstein:1990gu}. Model studies have considered the effects of meson loops on simple $q\overline{q}$-like states~\cite{Tornqvist:1995kr,Boglione:2002vv,Wolkanowski:2015lsa}, typically resulting in significant effects due to virtual $\kk$ configurations.

Ultimately, questions pertaining to the structure of hadron resonances must be addressed within Quantum Chromodynamics (QCD), the theory that describes the interactions of the quarks and gluons that make up all hadrons, but doing so is in general not a simple matter. Quarks and gluons interact non-perturbatively at the energy scales relevant to hadrons, and confinement ensures that the asymptotic states of the theory are not free quarks and gluons, but rather combinations of these bound into hadrons stable against strong decay. 

A powerful calculational tool comes in the form of \mbox{\emph{lattice QCD}}, a rigorous approximation to QCD in which the fields are considered on a discrete space-time grid of finite extent. By averaging over a large number of Monte-Carlo sampled field configurations, correlation functions with hadronic quantum numbers can be computed, and from these we may extract a discrete spectrum of eigenstates of QCD in the finite volume of the lattice. The precise relationship between the spectrum in a finite periodic volume and hadron-hadron scattering amplitudes is known~\cite{Luscher:1990ux,Luscher:1991cf, Rummukainen:1995vs, Feng:2004ua, He:2005ey, Christ:2005gi, Kim:2005gf, Bernard:2008ax, Bernard:2010fp, Leskovec:2012gb, Gockeler:2012yj, Guo:2012hv, Hansen:2012tf, Briceno:2012yi, Briceno:2014oea}, and it follows that from sufficiently accurate determinations of the spectrum in one or more volumes, we can extract details of hadron scattering and any resonances which may contribute.

The simple case of elastic $\pi\pi$ scattering in $P$-wave, where the narrow $\rho$ resonance appears, has been considered extensively in lattice QCD calculations~\cite{Aoki:2007rd,Feng:2010es,Lang:2011mn,Aoki:2011yj,Pelissier:2012pi,Dudek:2012xn,Wilson:2015dqa,Bali:2015gji}, and recently the first extraction of a coupled-channel \mbox{$S$-matrix} from QCD was reported~\cite{Dudek:2014qha,Wilson:2014cna}, for the case of $\pi K,\, \eta K$ scattering in $S$, $P$ and $D$-waves. This calculation, performed at a larger than physical light-quark mass, demonstrated that resonance properties can be extracted from such calculations, albeit in a case where the two channels prove to be relatively weakly coupled.

In this paper we will report on the determination of $S$-wave scattering amplitudes for the \pe,~\kk ~system, where we find these two channels to be strongly coupled and to feature a resonance coupled to both channels whose properties we will discuss. We will also explore the three-channel \pe, \kk, \pep ~system, and the behavior of $P$ and $D$ waves in the low-energy region. The remainder of the paper is structured as follows:

In Section \ref{levels} we review the techniques used to determine the finite-volume spectrum in our lattice QCD calculation. After discussion of the composite operators used to interpolate hadronic states from the vacuum, and the construction and variational analysis of matrices of correlation functions, we present the spectra obtained in a range of moving frames in three lattice volumes.

In Section \ref{amps} we present the scattering amplitudes for the coupled-channel \pe,~\kk ~system determined from the finite-volume spectrum using a variety of unitary \mbox{$K$-matrix} parameterizations. The $S$-wave amplitudes show a prominent cusp-like behavior in \pe ~at \kk ~threshold, and a rapid turn on of amplitudes leading to \kk ~final states. We also consider the three-channel system including the \pep ~channel and present results for higher partial waves.

In Section \ref{poles} we examine the singularity structure of our scattering amplitudes, and find that the strong cusp at \kk ~threshold is due to a narrow resonance, strongly coupled to both \pe ~and \kk ~channels. We find that this resonance is dominated by a single nearby pole, lying close to the real axis, slightly above the \kk ~threshold. We also consider Jost-style parameterizations of the scattering $S$-matrix~\cite{Jost:1947,LeCouteur:1960,Newton:1961,Kato:1965}, in which the pole structure can be specified.

In Section \ref{interpret} we consider physical interpretations of the extracted amplitudes and the corresponding distribution of pole singularities before we summarize our findings and lay out future prospects for studying excited hadron resonances using lattice QCD techniques.

Appendix \ref{app_phases} considers the behavior of a successful amplitude parameterization under variation of two key parameter values, exploring the migration of a resonance pole between Riemann sheets and a discontinuous change in the character of the phase-shift curves. Appendix~\ref{app_op_tables} presents a list of the operators used to compute the finite-volume spectra in each lattice irrep.

\section{Computing the spectrum in a finite volume using lattice QCD} \label{levels}

In lattice QCD, we obtain Euclidean correlation functions on a finite cubic grid by evaluating averages over an ensemble of gluon field configurations obtained through Monte-Carlo sampling. Since we extract the energy spectrum of states from the time-dependence of correlation functions, there is an advantage in having a fine temporal lattice spacing. Through use of an anisotropic lattice having a finer temporal than spatial spacing, we obtain improved energy resolution for only a moderate computational cost increase. The configurations used here feature three flavors of dynamical quarks: two degenerate light quarks, leading to exact isospin symmetry, and a heavier strange quark. The strange quark is approximately tuned to the physical strange quark mass, while the light quarks are somewhat heavier than their physical counterparts, leading to a pion mass of 391 MeV. Details of the discretized anisotropic action  and the corresponding parameter tuning may be found in refs.~\cite{Lin:2008pr,Edwards:2008ja}. In this study we make use of three ensembles of configurations with $a_s \sim 0.12\,\mathrm{fm}$ corresponding to three volumes with spatial extents between 2 and 3 fm. The temporal lattice spacing, $a_t$, expressed in physical units, is determined by computing the $\Omega$ baryon mass on these lattices, $a_tm_\Omega=0.2951$, and matching this to the experimentally determined value, $m_\Omega^{\mathrm{phys.}}=1672$ MeV using $a_t=\frac{a_t m_\Omega}{m_\Omega^{\mathrm{phys.}}}$.

These configurations have been used in several other studies to obtain a picture of the highly excited meson and baryon spectrum~\cite{Dudek:2009qf,Dudek:2010wm, Dudek:2011tt,Edwards:2011jj,Edwards:2012fx, Dudek:2012ag, Liu:2012ze, Moir:2013ub, Dudek:2013yja, Padmanath:2013zfa,Padmanath:2015jea}, and to determine elastic~\cite{Dudek:2010ew, Dudek:2012gj,Dudek:2012xn} and coupled-channel scattering amplitudes~\cite{Dudek:2014qha,Wilson:2014cna,Wilson:2015dqa}.



Our approach to extract the discrete spectrum of eigenstates follows closely that presented in the references above -- we form large matrices of two-point correlation functions, $C_{ij}(t) = \langle 0 | \mathcal{O}_i(t) \, \mathcal{O}^\dag_j(0) | 0 \rangle$, using a large basis of operators featuring some of ``single-meson-like'' construction, and others of ``meson-meson-like'' construction. It was shown in the papers referenced above that inclusion of both types is required in order to reliably extract the complete spectrum of states in the region below three-meson thresholds. We analyze the resulting correlation matrices using a variational method~\cite{Michael:1985ne,Luscher:1990ck,Dudek:2007wv,Blossier:2009kd} which amounts to solving a generalized eigenvalue problem of the form $C(t)v^\mathfrak{n}= C(t_0)v^\mathfrak{n}\lambda_\mathfrak{n}(t)$, where the spectrum of eigenstates, $\left\{E_\mathfrak{n}\right\}$, is obtained from the large time behavior of the ``principal correlators'', $\lambda_\mathfrak{n}(t)\sim e^{-E_\mathfrak{n}(t-t_0)}$.


In evaluating the correlation matrices we must account for all the quark-field Wick contractions specified by QCD, and many of these involve quark annihilation. In order to efficiently include all such contributions, we make use of the \emph{distillation} framework~\cite{Peardon:2009gh}. All the propagation objects required for the present study had already been computed for use in previous studies, and they are reused here. The lattice volumes, ensemble size and the rank of the distillation vector space are provided in Table~\ref{tab_lattices}.

\begin{table}
  \begin{tabular}{c|c|ccc}
  $(L/a_s)^3 \times (T/a_t)$	& $L$ (fm) & $N_\mathrm{cfgs}$ 	& $N_{t_\mathrm{srcs}}$	& $N_\mathrm{vecs}$ \\
  \hline
  $16^3 \times 128$		& 1.9 & 479				& 8				& 64 \\
  $20^3 \times 128$		& 2.4 & 603				& 2--4				& 128 \\
  $24^3 \times 128$		& 2.9 & 553				& 2--4				& 162
  \end{tabular}
  \caption{The three volumes used, the number of configurations on each, the number of independent time-sources averaged over (which varies somewhat according to irrep), and the number of vectors in distillation space.}
  \label{tab_lattices}
\end{table}



The cubic symmetry of the lattice grid and the spatial boundary of the lattice break the full rotational symmetry of QCD down to a smaller group, and as such states are classified not by spin and parity, but rather by irreducible representations (irreps) of the cubic symmetry group at rest, and the relevant ``little-group'' in the case of moving frames. In Table~\ref{tab_pwa_irrep} we list the contributions of each infinite-volume partial-wave into each lattice irrep. This table may be constructed from the unequal mass pseudoscalar-pseudoscalar scattering case given in~\cite{Wilson:2014cna}, and the equal mass case given in Refs.~\cite{Dudek:2012gj} -- the derivation is described in those works.

\begin{table}
  \begin{center}
    \begin{tabular}{cc|c|l|l}
      \hline
      \hline
           &&&&\\[-1.7ex]
      \multirow{2}{*}{$\vec{P}$} & \multirow{2}{*}{LG$(\vec{P})  $ }  & \;  \multirow{2}{*}{$\Lambda$} \;& $\,\,J^P (\vec{P}=\vec{0})$    & \; $\pi \eta,\, \pep$ $\ell^N$ \\
      &               &       & $\left|\lambda\right|^{({\tilde{\eta}})}(\vec{P}\neq\vec{0})$ & \; $\pi \eta, \kk,\, \pep$ $\bm{\ell^N}$ \\[0.5ex]
      \hline \hline
           &&&&\\[-1.5ex]
      \multirow{7}{*}{$\left[0,0,0\right]$}&     \multirow{7}{*}{$\textrm{O}_h^\textrm{D}$ ($\textrm{O}_h$)} & $A_1^+$
      & $0^+,\, 4^+$          &\; $\bm{0^1},\, \bm{4^1}$\\
      && $T_1^-$    & $1^-,\, 3^-,\, \mathit{(4^-)}$ &\; $1^1,\, 3^1$\\
      && $E^+$      & $2^+,\, 4^+$          &\; $\bm{2^1},\, \bm{4^1}$\\
      && $T_2^+$    & $2^+,\, 4^+,\,  \mathit{(3^+)}\,$  &\; $\bm{2^1},\, \bm{4^1}$\\
      && $T_1^+$    & $4^+, \, \mathit{(1^+,3^+)}$ &\; $\bm{4^1}$ \\
      && $T_2^-$    & $3^-,\, \mathit{(2^-,4^-)}$ &\; $3^1$ \\
      && $A_2^-$    & $3^-$               &\; $3^1$\\[0.5ex]
      \hline
      \hline
      &&&&\\[-1.5ex]

      \multirow{5}{*}{$\left[0,0,n\right]$} & \multirow{5}{*}{Dic$_4$ ($\textrm{C}_{4\textrm{v}}$)}
      & $A_1$       & $0^+,\, 4$        &\; $\bm 0^1,\, 1^1,\, \bm 2^1,\, 3^1,\, \bm 4^2$ \\
      && $E_2$      & $1,\, 3$          &\; $1^1,\, \bm 2^1,\, 3^2,\, \bm 4^2$\\
      && $B_1$      & $2$               &\; $\bm 2^1,\, 3^1,\, \bm 4^1$ \\
      && $B_2$      & $2$               &\; $\bm 2^1,\, 3^1,\, \bm 4^1$ \\
      && $A_2$      & $4,\, \mathit{(0^-)}$      &\; $\bm 4^1$ \\[0.5ex]
      \hline
      &&&&\\[-1.5ex]

      \multirow{4}{*}{$\left[0,n,n\right]$} & \multirow{4}{*}{Dic$_2$ ($\textrm{C}_{2\textrm{v}}$)}
      & $A_1$     & $0^+,\, 2,\, 4$   &\; $\bm 0^1,\, 1^1,\, \bm 2^2,\, 3^2,\, \bm 4^3$ \\
      && $B_1$     & $1,\, 3$          &\; $1^1,\, \bm 2^1,\, 3^2,\, \bm 4^2$ \\
      && $B_2$     & $1,\, 3$          &\; $1^1,\, \bm 2^1,\, 3^2,\, \bm 4^2$ \\
      && $A_2$     & $2,\, 4, \, \mathit{(0^-)}$ &\; $\bm 2^1,\, 3^1,\, \bm 4^2$ \\[0.5ex]
      \hline
      &&&&\\[-1.5ex]

      \multirow{3}{*}{$\left[n,n,n\right]$} &       \multirow{3}{*}{Dic$_3$ ($\textrm{C}_{3\textrm{v}}$)}
      &  $A_1$     & $0^+,\, 3$        &\;  $\bm 0^1,\, 1^1,\, \bm 2^1,\, 3^2,\,  \bm 4^2$\\
      && $E_2$     & $1,\, 2,\, 4$     &\;  $1^1,\, \bm 2^2,\, 3^2,\,  \bm 4^3$ \\
      && $A_2$     & $3, \, \mathit{(0^-)}$      &\;  $3^1,\, \bm 4^1$\\[0.5ex]
      \hline
      \hline
    \end{tabular}
  \end{center}
  \caption{The subductions of $\pi\eta$, $\kk$ and $\pi \eta^\prime$ partial-waves with $\ell \leq 4$ into lattice irreps, $\Lambda$. $N$ is the number of embeddings of each $\ell$ in the irrep. This table is derived from Table II of Ref.~\cite{Dudek:2012gj} where more detailed discussion is presented. The LG$(\vec{P})$ column shows the double-cover little group (the corresponding single-cover little group relevant for only integer spin is given in parentheses).  Also shown are the various spins, $J \leq 4$, or helicites, $|\lambda| \leq 4$, that appear in each of the relevant irreps.  The $J^P$ values and $|\lambda|^{\tilde{\eta}}=0^-$ in italics are in the ``unnatural parity'' [$P = (-1)^{J+1}$] series and do not contribute to pseudoscalar-pseudoscalar scattering. The $\kk$ operators we use are constructed with good $G$-parity and these $I^G=1^-$ combinations do not subduce into odd partial waves.}
  \label{tab_pwa_irrep}
\end{table}



The mass and dispersion relations for the pion and the kaon on the current lattices were presented in previous papers~\cite{Dudek:2012xn,Wilson:2014cna}. For the $\eta$ and $\eta^\prime$, in the current study, we more precisely determined the mass and dispersion relation than in our previous work. Using a large basis constructed from ``single-meson-like'' operators, considering momenta up to $[2,0,0]$, we find $a_t \, m_\eta=0.10364(19)$ with anisotropy $\xi_\eta=3.436(6)$ and $a_t \, m_{\eta^\prime}=0.1641(10)$ with anisotropy $\xi_{\eta^\prime}=3.36(3)$. These anisotropies are statistically compatible with those found for the pion and kaon~\cite{Dudek:2012gj}. The $\eta^\prime$, which is stable on these lattices, is reliably extracted, but not as precisely determined as the $\pi$, $K$ and $\eta$, and its mass displays somewhat larger volume dependence than the other light states, as identified in Ref.~\cite{Dudek:2013yja}, and in this first study we will only make limited use of the region above $\pi\eta^\prime$ threshold. The masses of low-lying stable hadrons and corresponding thresholds are presented in Table \ref{tab_masses}.


\begin{table}
\begin{tabular}{r|c}
meson ($J^P$)          & $a_t\, m$ \\
\hline
$\pi(0^-)$         & 0.06906(13) \\
$K(0^-)$           & 0.09698(9)  \\
$\eta(0^-)$        & 0.10364(19) \\
$\omega(1^-) $     & 0.15678(41) \\
$\eta^\prime(0^-)$ & 0.1641(10)  \\
\end{tabular}
\begin{tabular}{c|c}
threshold         & $a_t\, E_\mathrm{thr.}$ \\
\hline
$\pi\eta$         & 0.17270(23) \\
$K\overline{K}$        & 0.19396(13) \\
$\pi\pi\pi$       & {\it 0.20718(23)} \\
$\pi\eta^\prime$  & 0.2332(11)  \\
$\pi K \overline{K}$   & {\it 0.26302(18)} \\
$\pi\eta\eta$     & {\it 0.27634(30)} \\
$\omega\pi\pi$    & 0.29490(45) \\
$\eta K \overline{K}$  & {\it 0.29760(23)} \\
$\pi\pi\pi\eta$   & 0.31082(30) \\
\end{tabular}
\caption{Left: Stable meson masses. Right: Multi-meson kinematic thresholds. Channels whose threshold is shown in italics do not contribute to scattering with $J^P=0^+$.}
\label{tab_masses}
\end{table}



The ``meson-meson-like'' constructions that we utilize take the form $\sum_{\vec{p}_1,\vec{p}_2}\mathcal{C}(\vec{p}_1,\vec{p}_2)\,\Omega^\dag(\vec{p}_1)\,\Omega^\dag(\vec{p}_2)$, where $\Omega^\dag(\vec{p})$ is a variationally-optimal momentum-projected operator~\cite{Thomas:2011rh} that is obtained as a linear superposition of the ``single-particle-like'' basis for each of the $\pi$, $K$, $\eta$ and $\eta^\prime$. We combine these with the Clebsch-Gordon coefficients for the desired lattice irrep to produce \mbox{``$\pe$-like''}, \mbox{``$\kk$-like''} and \mbox{``$\pep$-like''} operators in the manner described in Ref.~\cite{Dudek:2012gj}. The $\kk$ operators are constructed to have definite \mbox{$G$-parity}, ${G=-}$, as is described in Ref.~\cite{Wilson:2015dqa} -- there is no $\kk$ $P$-wave scattering with $G=-$.

\begin{figure*}
\includegraphics[width=0.99\textwidth]{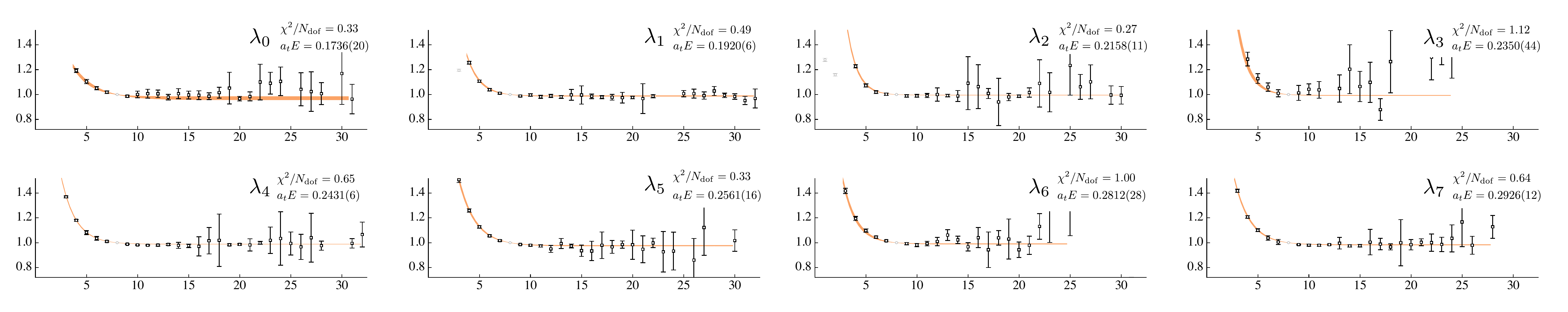}
\caption{The principal correlators of the lowest 8 states obtained from the variational analysis of 21 operators in the $[000]A_1^+$ irrep on the $24^3$ volume plotted as $e^{E_\mathfrak{n}(t-t_0)}\lambda_\mathfrak{n}(t)$, with $t_0=8a_t$. State $\mathfrak{n}=3$ is dominated by a ``$\pep$-like'' operator and has the largest uncertainty.}
\label{fig_prin_corrs}
\end{figure*}

The operator sets used to form the matrix of correlation functions in each irrep are listed in Appendix~\ref{app_op_tables}. Before solving the generalized eigenvalue problem, we remove the largest of the unwanted effects arising from the finite temporal extent of the lattice using the weighting-shifting procedure described in Ref.~\cite{Dudek:2012gj}. An example of the statistical quality of signals is presented in Figure \ref{fig_prin_corrs} which shows the principal correlators for the eight lowest-lying states in the $[000]$ $A_1^+$ irrep on the $24^3$ lattice -- one notable feature is that the $\mathfrak{n}=3$ level, which is dominantly produced by a ``$\pep$-like'' operator, is significantly less well-determined than the other levels, in line with our discussion of the $\eta'$ on these lattices above.

\begin{figure}
\includegraphics[width=\columnwidth]{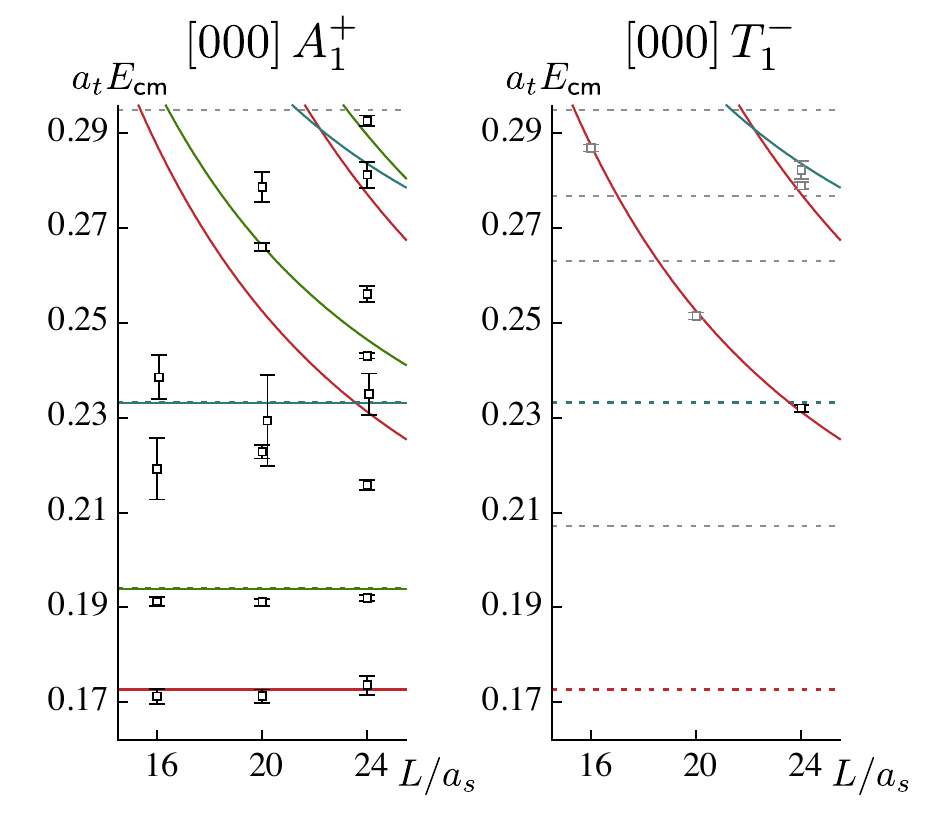}

\caption{
	Lattice QCD spectra in three volumes -- left panel: rest-frame $A_1^+$ irrep, dominated by $J^{P(C)}=0^{+(+)}$, right panel: rest-frame $T_1^-$ panel, dominated by $J^{P(C)} = 1^{-(+)}$. Red, green, blue curves represent non-interacting \pe, \kk, \pep ~levels respectively. Horizontal dashed lines show kinematic thresholds (see Table~\ref{tab_masses}).
  \label{fig_spectra_A1_T1}
}
\end{figure}

\begin{figure*}
\includegraphics[width=0.60\textwidth]{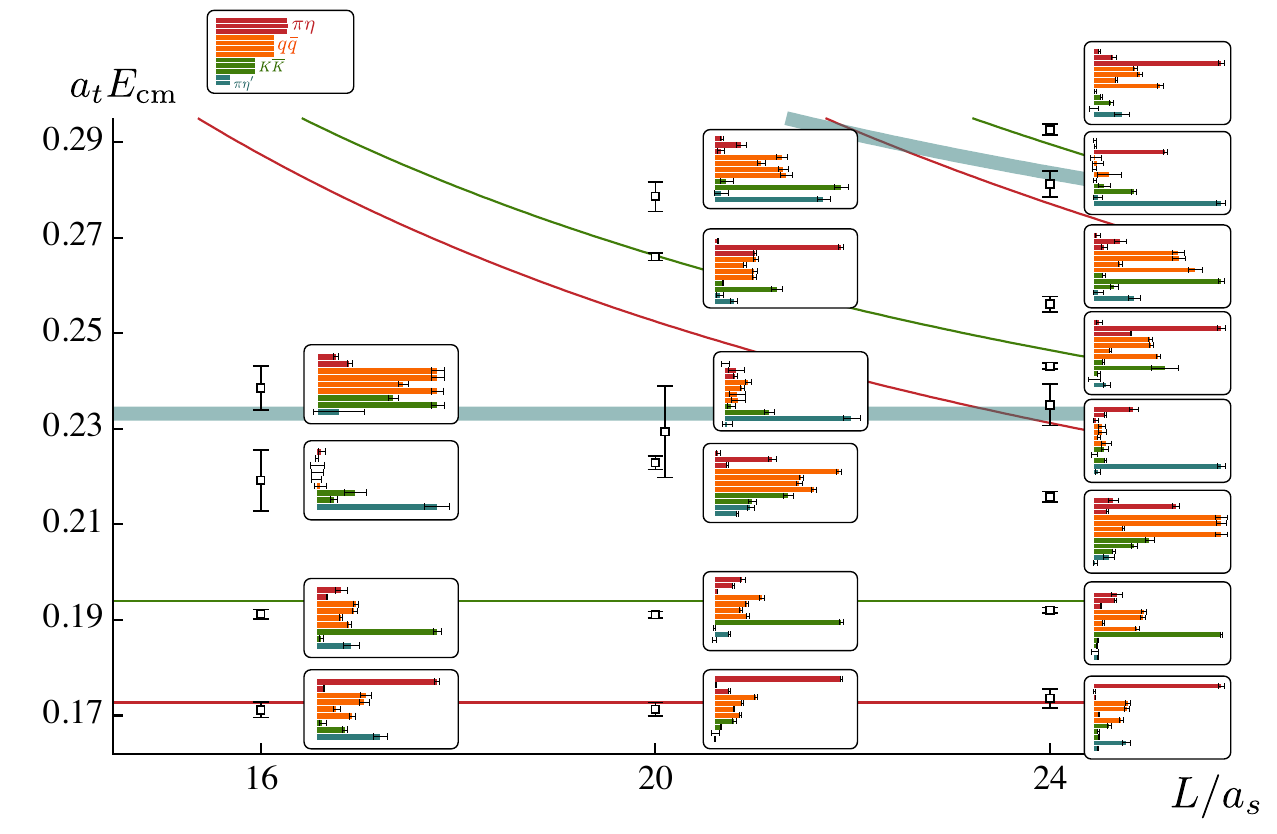}
\caption{
  The low-lying $[000] A_1^{+}$ spectra for each volume. The relative operator overlaps $\big| \left\langle\mathfrak{n}\right|\mathcal{O}_i\left| 0\right\rangle \big|$ are also shown as histograms.
  \label{fig_spectrum_with_histograms}
}
\end{figure*}

In Fig.~\ref{fig_spectra_A1_T1} we present the spectra in the rest-frame irreps $A_1^+$ and $T_1^-$ -- these are dominated by $S$-wave and $P$-wave meson-meson scattering respectively. The $T_1^-$ spectrum is observed to feature only levels lying very close to the curves which show the position of non-interacting \pe ~and \pep ~pairs. This suggests that the $P$-wave interactions at low energy are weak, as we might expect in the exotic $J^{PC} = 1^{-+}$ channel. The $A_1^+$ spectrum shows more structure, and the utility of the large basis of operators in determining this spectrum can be inferred from Figure~\ref{fig_spectrum_with_histograms} which shows the relative overlap of operators onto each extracted level. The lowest two levels on each volume have largest overlap with the $\pi_{[000]}\,  \eta_{[000]}$, $K_{[000]} \overline{K}_{[000]}$ operators respectively. The third level (fourth level on $16^3$) is dominated by the ``single-meson-like'' operators, but with significant, and volume-dependent, admixture of \pe ~and \kk ~operators. We observe that the fourth level (third level on $16^3$) lying near \pep ~threshold, which has larger statistical uncertainty than the others, is dominantly produced by the $\pi_{[000]}\,  \eta'_{[000]}$ operator. The levels above these are seen to be shifted significantly from the non-interacting meson-pair energy curves, suggesting strong scattering.

\begin{figure*}
\includegraphics[width=0.82\textwidth]{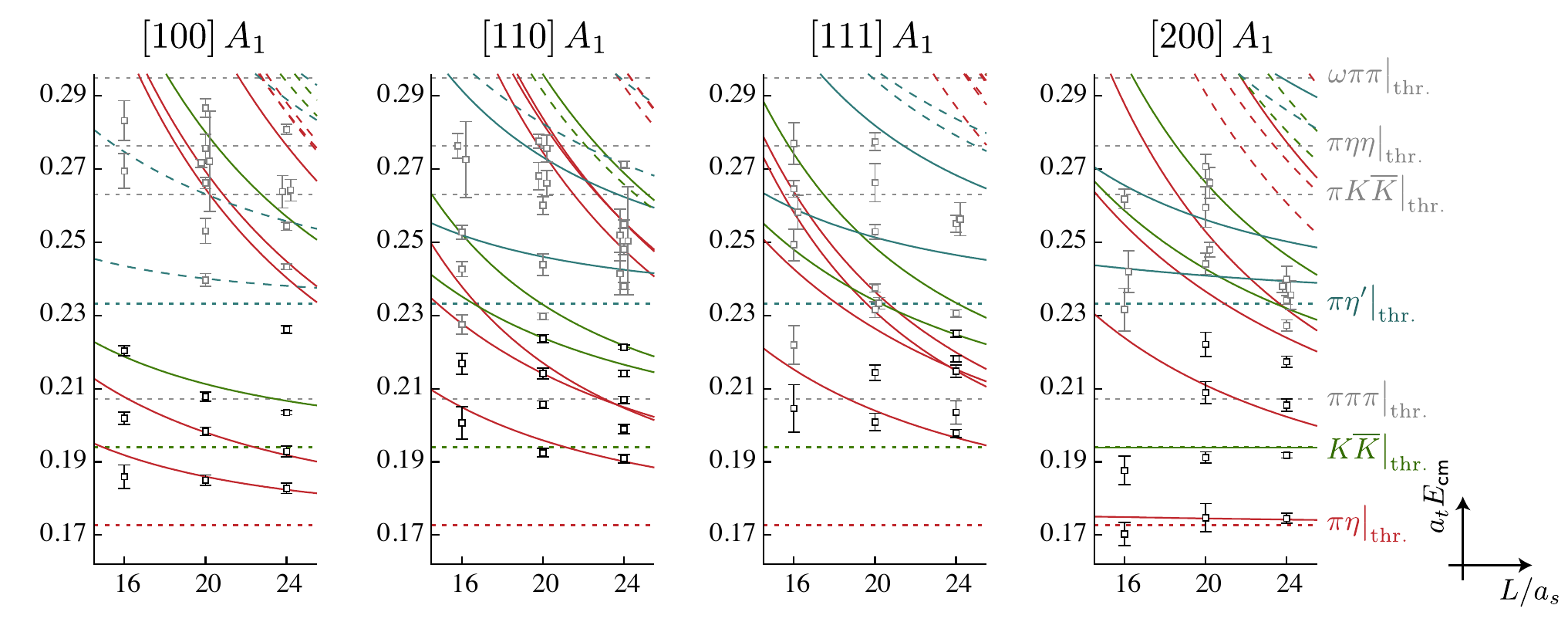}
\caption{
Lattice QCD spectra in three volumes for $A_1$ irreps in moving frames. Curves represent non-interacting meson-meson energy levels with the coloring defined in the caption to Figure~\ref{fig_spectra_A1_T1} -- a solid curve indicates that the corresponding ``meson-meson-like'' operator was included in the basis (see Table~\ref{tab_opsused}), and a long-dashed curve indicates that it was not included in the basis. Levels in gray are not used in the determination of scattering amplitudes.
  \label{fig_spectra_A1_flight}
}
\end{figure*}

\begin{figure*}
\includegraphics[width=0.82\textwidth]{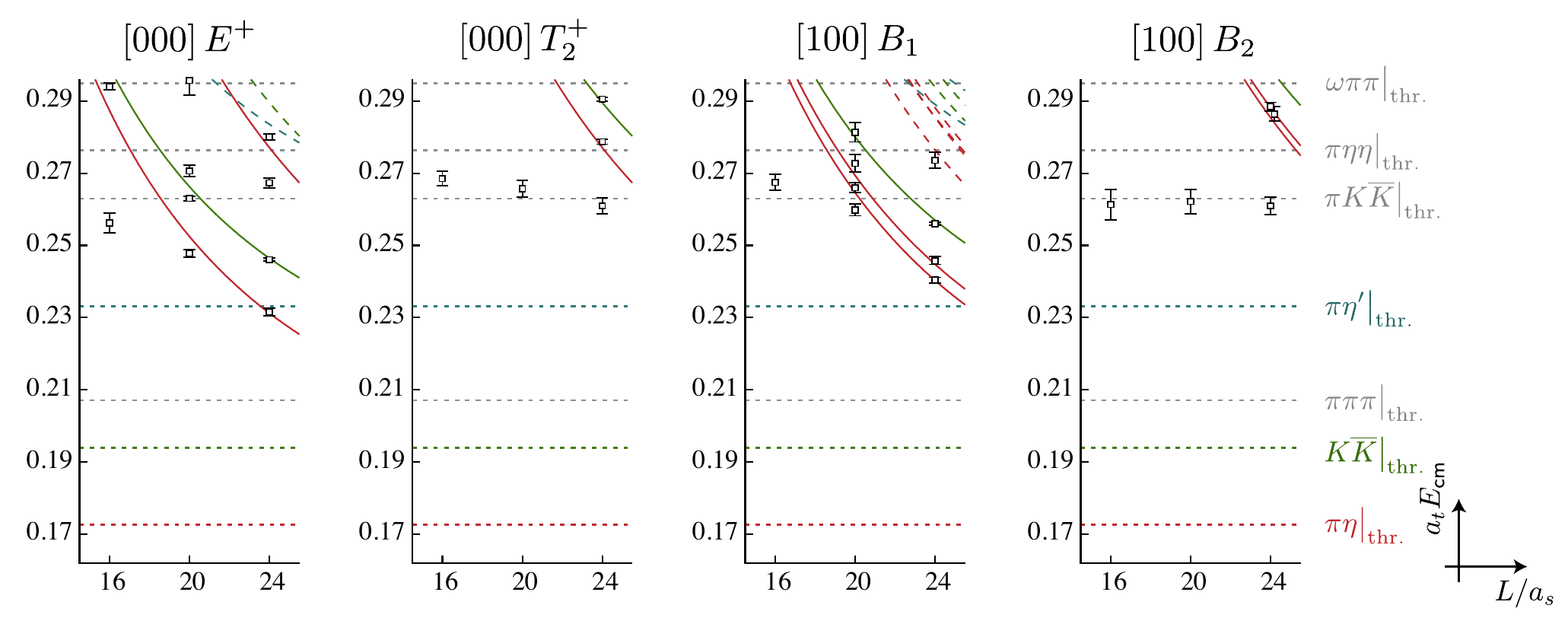}
\caption{
As Figure~\ref{fig_spectra_A1_flight} for moving-frame irreps having $J^P=2^+$ as the lowest partial-wave.
  \label{fig_spectra_Dwave}
}
\end{figure*}

Fig.~\ref{fig_spectra_A1_flight} shows the spectra in those moving-frame irreps which feature $S$-wave scattering, and Fig.~\ref{fig_spectra_Dwave} spectra in irreps which have $D$-wave scattering as the lowest partial-wave. We do not present spectra in moving-frame irreps having $P$-wave scattering as the leading partial-wave. The $T_1^-$ spectrum indicates that $P$-wave scattering is likely to be weak at low energies, and owing to contributions from opposite parity in these moving-frame irreps, the spectrum will actually be dominated by the $J^P=1^+$ scattering amplitudes. These do not appear in pseudoscalar-pseudoscalar scattering, rather being a feature of, for example, $\pi \pi \pi$ scattering. Since we have not included operators resembling three pions we do not expect to obtain a reliable determination of the spectrum here, and as such we do not make use of these irreps.

With the finite-volume spectra presented in Figures \ref{fig_spectra_A1_T1}, \ref{fig_spectra_A1_flight}, \ref{fig_spectra_Dwave} in hand, we move to the problem of determining coupled-channel partial-wave scattering amplitudes.


\section{Determining coupled-channel scattering amplitudes}\label{amps}

In order to rigorously understand these finite volume spectra in terms of infinite volume scattering amplitudes we use the formalism first derived for elastic scattering of equal mass particles in field theories with periodic cubic boundaries by L\"uscher~\cite{Luscher:1990ux,Luscher:1991cf}. Later extensions ~\cite{Rummukainen:1995vs, Feng:2004ua, Christ:2005gi, Kim:2005gf, Bernard:2008ax, Leskovec:2012gb, Gockeler:2012yj} to the elastic formalism deal with moving frames, non-zero spin, and scattering of particles of unequal mass. 
The relationship to the finite-volume spectrum for \emph{coupled-channel} scattering of hadron pairs \cite{He:2005ey, Guo:2012hv, Hansen:2012tf, Briceno:2012yi, Briceno:2014oea}, described by the scattering $t$-matrix, $\mathbf{t}(E_\cm)$, can be written,
\begin{align}
\det[\mathbf{1}+ i\bm{\rho}\cdot \mathbf{t}\cdot(\mathbf{1}+i\bm{\mathcal{M}})]=0,
\label{eq_det}
\end{align}
where $E_\cm$ is the $\mathsf{cm}$-frame energy, $\bm{\rho}(E_\mathsf{cm})$ is the diagonal matrix of phase space factors $\rho_{ij}(E_\mathsf{cm})=\delta_{ij}\rho_j(E_\mathsf{cm})=\delta_{ij}\frac{2k_j}{E_\cm}$, and where $\bm{\mathcal{M}}(E_\mathsf{cm}, L)$ is a matrix of known functions of essentially kinematic origin. The finite-volume spectrum in a cubic $L\times L \times L$ volume, corresponds to the set of solutions,
$\{ E_\mathsf{cm}(\mathfrak{n}) \}$, of Eq.~\ref{eq_det}, for a unitarity-preserving $\mathbf{t}(E_\mathsf{cm})$.

The matrix space under the determinant is over scattering channels and all partial waves subduced into the relevant irrep, $\Lambda$. 
For a partial wave of angular momentum $\ell$, the threshold barrier, $k_j^{2\ell}$, suppresses the contribution of amplitudes for all but the lowest partial waves at low energy.  The subduction of $\bm{\mathcal{M}}$ into the appropriate irreps is described in Refs.~\cite{Dudek:2012gj}, where further discussion of Eq.~\ref{eq_det} can be found.

In coupled-channel scattering with two or more channels and one or more partial waves, Eq.~\ref{eq_det} depends on several unknowns at any given value of energy, and it is clear that from knowledge of a single energy level value, not all these unknowns can be determined. An efficient way of extracting information from \emph{all} of the energy levels in a given region is to use a parameterization of $\bm{t}(E_\cm)$ with a limited number of parameters. Provided sufficiently many energy levels are present then it is possible to constrain those parameters by performing a $\chi^2$ minimization comparing the lattice-determined spectrum and the spectrum provided by the solutions of Eq.~\ref{eq_det} for our parameterized $\bm{t}(E_\cm)$. This procedure was followed in Ref.~\cite{Wilson:2014cna}, and the correlated $\chi^2$ we minimize is defined in Eq.~8 of that reference.

Parameterizations of $\bm{t}$ typically feature the $s$-channel Mandelstam variable, $s=E_\mathsf{cm}^2$, and the $\cm$-frame momentum,
\begin{align}
k_i^2(s)=\frac{1}{4s}\left(s-(m_{i,1}+m_{i,2})^2\right)\left(s-(m_{i,1}-m_{i,2})^2\right),
\label{eq_kcm}
\end{align}
where $m_{i,1}$ and $m_{i,2}$ are the scattering particle masses in channel $i$. Our parameterizations must satisfy unitarity if they are to solve Eq.~\ref{eq_det}, and since we intend eventually to explore the singularities of the $t$-matrix at complex values of $s$, they should also respect certain analyticity properties. The $K$-matrix approach provides a convenient parameterization of coupled-channel scattering that manifestly ensures a unitarity-preserving $t$-matrix. In general, for $\ell$-wave scattering, we may write the elements (with $i$ and $j$ labelling scattering channels) of the inverse of $t$ as,
\begin{equation}
	t^{-1}_{ij}(s) = \frac{1}{(2k_i)^\ell} K^{-1}_{ij}(s) \frac{1}{(2k_j)^\ell} + I_{ij}(s)\,,
\label{eq_t_matrix_k}
\end{equation}
where the factors $(2k_i)^{-\ell}$ provide the required kinematic behavior at thresholds~\cite{Guo:2010gx}. The elements $K_{ij}(s)$ form a symmetric\footnote{thus ensuring time-reversal invariance} matrix that is real\footnote{away from any ``left-hand'' cut} for real $s$. The elements $I_{ij}(s)$ form a diagonal matrix whose imaginary part is fixed by unitarity to be $-\rho_i(s)$ above threshold in channel $i$ and zero below threshold. The real part of $I_{ij}(s)$ is not fixed by unitarity, however the analyticity of the amplitude motivates a logarithmic form~\cite{Chew:1960iv}, which follows from a dispersion relation relating the real part to the known imaginary part. The resulting $I_{ij}(s)$ function behaves reasonably below threshold and off the real energy axis -- our implementation of this \mbox{``Chew-Mandelstam''} phase space is described in the appendices of Ref.~\cite{Wilson:2014cna}. We will explore a range of parameterizations for $\mathbf{K}(s)$ when we consider coupled-channel scattering to ensure that the results are not dependent on any particular choice.

Before considering the case of coupled-channel scattering, we begin by examining the limited energy region below $\kk$~threshold where $\pe$~scattering is elastic.

\subsection{Elastic $S$-wave \pe ~scattering}

\begin{figure}[b]
\includegraphics[width=0.95\columnwidth]{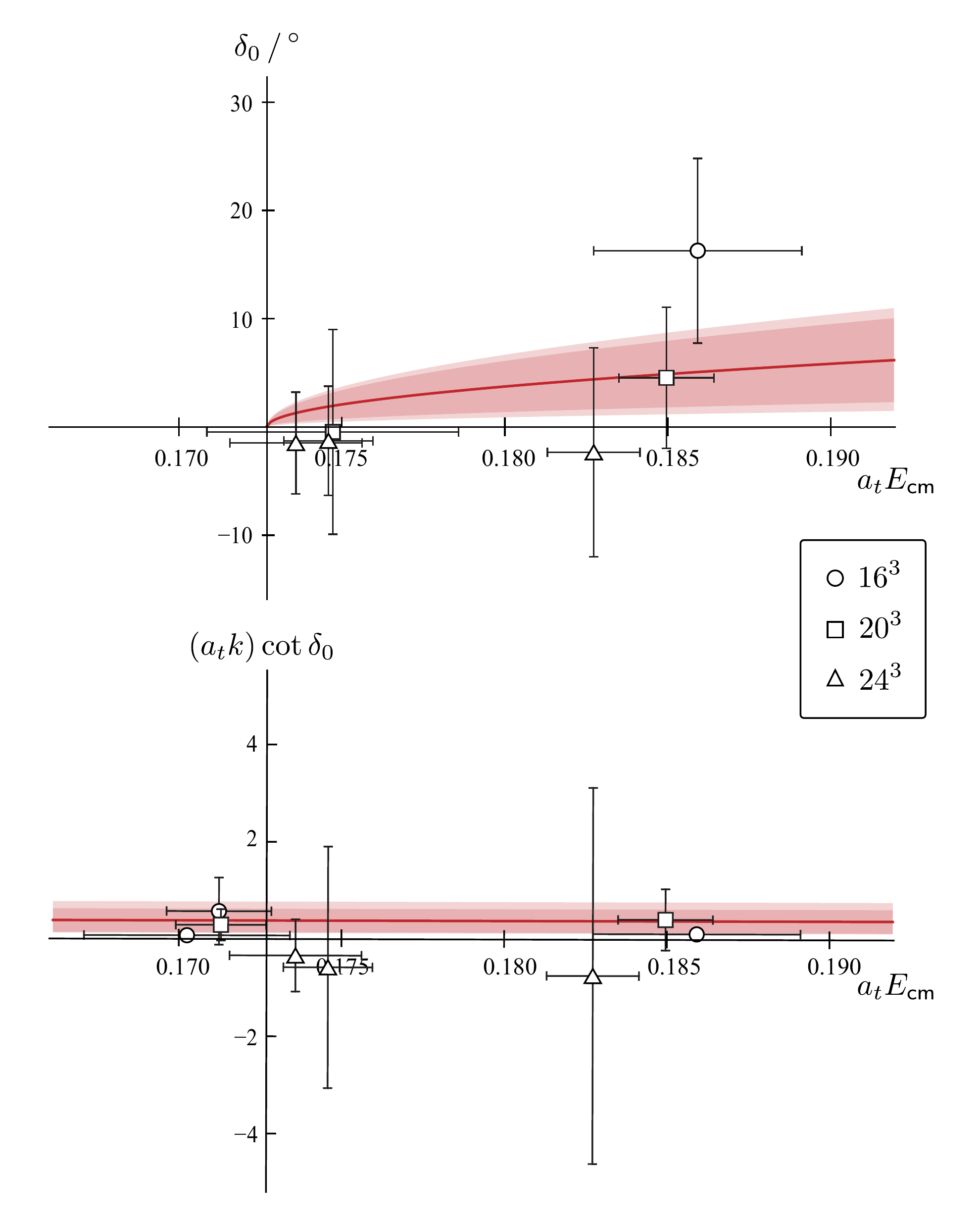}
\caption{$S$-wave \pe ~elastic scattering amplitude determined using the $[000]A_1^+$, $[100]A_1$ and $[200]A_1$ spectra in three volumes. Upper panel: phase-shift $\delta_0$. Lower panel: $k\cot\delta_0$. Discrete points from application of the L\"uscher condition assuming no contribution from partial waves above $\ell = 0$, curves from describing the spectrum using a scattering length.}
\label{fig_elastic_combined}
\end{figure}

Below $\kk$ threshold, where only elastic $\pe$ scattering occurs, the $S$-wave scattering $t$-matrix reduces to a single scattering amplitude, $t = \frac{1}{\rho} e^{i\delta} \sin \delta$, that can be described by a single real parameter, the phase-shift, $\delta_0(E_\mathsf{cm})$. The determinant condition Eq.~\ref{eq_det} reduces to a single equation\footnote{Up to higher partial-wave contributions which we later determine to be negligible in this energy region. Amplitudes featuring the \kk ~channel can influence the spectrum in a limited energy region below \kk ~threshold, so we are careful to exclude energy levels which lie too close to the threshold.} for $\delta_0(E_\mathsf{cm})$. The points shown in Figure \ref{fig_elastic_combined} correspond to solving this equation for nine levels well below \kk ~threshold in the $[000]A_1^+$, $[100]A_1$ and $[200]A_1$ irreps (we include three levels which appear below \pe ~threshold). The small values of $\delta_0$ appear to indicate rather weak \pe ~scattering at low energy.

An alternative approach is to parameterize $t(E_\mathsf{cm})$ and describe the finite-volume spectra by varying the parameters. At low energies we may expand the scattering amplitude in a power series in $k^2$ known as the effective-range expansion. Retaining only the first term in this expansion, the scattering length, we have
\begin{align}
k_\pe^{2\ell+1}\cot\delta_\ell = \frac{1}{a_\ell}  +\mathcal{O}\!\left(k_\pe^2\right).
\label{eq_ere}
\end{align}
For $S$-wave scattering the spectra in the elastic region are best described (with $\chi^2/{N_\mathrm{dof}}=0.52$) by ${a_{\ell=0}= ( 2.6 \pm 1.6 \pm 0.4) \cdot a_t}$, where the first uncertainty reflects the statistical uncertainty in the discrete energies and the second the uncertainty on the pion mass, $\eta$ mass and the anisotropy. Adding an effective range term does not improve the description. A closely related form uses a one-channel implementation of Eq.~\ref{eq_t_matrix_k} with a constant $K=\gamma_\pe$, and the Chew-Mandelstam phase space. Using this form to describe the energy spectrum we find $\gamma_\pe= 0.22\pm 0.14\pm 0.04$ with a similar $\chi^2/{N_\mathrm{dof}}=0.54$. The energy-dependence of the phase-shift curves for the two forms are consistent and we plot the scattering length form along with the discrete points in Fig.~\ref{fig_elastic_combined}.

We note in passing that this scattering length is significantly smaller than that obtained in the $I=1/2$ $\pi K$ channel on the same lattices~\cite{Wilson:2014cna,Dudek:2014qha}, and we discuss this further in Section~\ref{interpret}.

\subsection{Coupled-channel $S$-wave \pe, \kk ~scattering}\label{2chan}

We now investigate the region below $\pep$ threshold, where two channels, $\pe$ and $\kk$, are kinematically open. The left panel of Figure~\ref{fig_spectra_A1_T1} and all panels of Figure~\ref{fig_spectra_A1_flight} show the finite-volume spectra we will use to constrain the scattering $t$-matrix. We use the lowest 3 levels from each of the $[000]A_1^+$ irreps (2 on $16^3$) and all of the points shown in black in the moving frames. These 47 energy levels are sufficiently far below \pep~threshold that we believe we can neglect the effect of that kinematically closed channel\footnote{and see the next section for a study of the influence of \pep}.

We will initially ignore partial wave contributions with $\ell > 0$, and we will later show them to be negligible in this energy region. Below the \pep~threshold, in the energy region of interest, lies the $\pi\pi\pi$ threshold, and this channel does subduce into the in-flight $A_1$ irreps, not however in $J^P=0^+$ scattering, but only through higher partial waves. These higher partial waves necessarily involve some $\ell$ and the threshold suppression that comes from that. We did not include any operators resembling three pions in our basis, and we will proceed assuming that $\pi\pi\pi$ amplitudes do not contribute significantly to determining the in-flight $A_1$ spectra at low energy.

We will attempt to describe the finite-volume spectra using parameterizations of the energy dependence of the $t$-matrix that are as simple as possible and which feature relatively few variable parameters. An example of a form we find to be successful expresses the $K$-matrix as,
\begin{equation}
 \mathbf{K} = \frac{1}{m^2 -s} \begin{bmatrix} g_\pe^2 & g_\pe g_\kkb \\ g_\pe g_\kkb & g_\kkb^2 \end{bmatrix} + \begin{bmatrix} \gamma_{\pe, \pe} & \gamma_{\pe, \kkb}\\ \gamma_{\pe, \kkb} & \gamma_{\kkb, \kkb} \end{bmatrix}, \label{K_ppc}
\end{equation}
i.e. as the sum of a pole in $s$ and a matrix of constants. The six real parameters, $m$, $g_\pe$, $g_\kkb$, $\gamma_{\pe, \pe}$, $\gamma_{\pe, \kkb}$, $\gamma_{\kkb, \kkb}$, can be varied to describe the spectra. As described in Refs.~\cite{Guo:2012hv,Wilson:2014cna}, we choose to subtract the dispersive integral in the Chew-Mandelstam phase-space at $s=m^2$. 

\begin{figure*}
\includegraphics[width=0.92\textwidth]{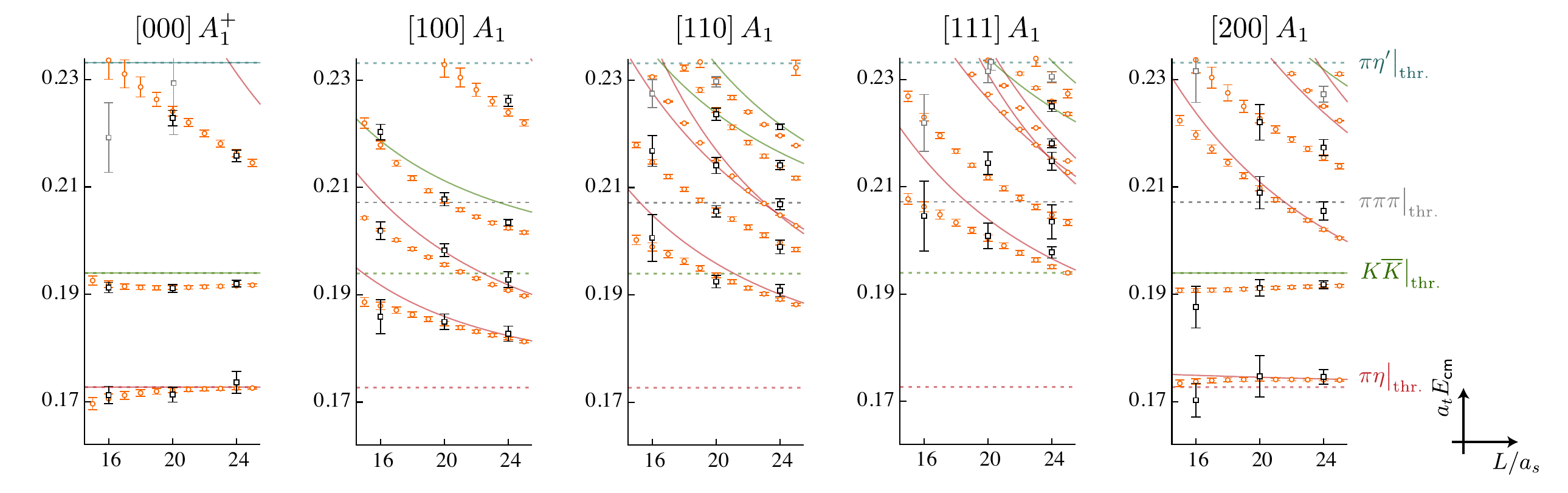}
\caption{The $A_1$ irrep spectra: black points show the spectra of Figures~\ref{fig_spectra_A1_T1} and \ref{fig_spectra_A1_flight} and the orange points show the volume-dependent spectra corresponding to the $S$-wave amplitude presented in Eq.~\ref{fit_Kmatrix_const_2x2}. Energy levels not used in the minimization are shown in grey.}
\label{fig_A1_spec_comparison}
\end{figure*}

The best description of the 47 energy levels is given by,
\begin{widetext}
\begin{center}
\begin{tabular}{rll}
$m =$                         & $(0.2214 \pm 0.0029 \pm 0.0004) \cdot a_t^{-1}$   &
\multirow{5}{*}{ $\begin{bmatrix*}[r] 1 &  0.58 & -0.06 & -0.51 &  0.39 &  0.02 \\
                                    	&  1    & -0.63 & -0.87 &  0.84 & -0.49 \\
                                    	&       & 1     &  0.52 & -0.68 &  0.83 \\
                                    	&       &       & 1     & -0.90 &  0.53 \\
                                    	&       &       &       & 1     & -0.78 \\
                                    	&       &       &       &       &  1    \end{bmatrix*}$ } \\
$g_{\pi \eta} =$                  & $(\;\;\; 0.091 \pm 0.016 \pm 0.009) \cdot a_t^{-1}$   & \\
$g_{\kkb}     =$                  & $(-0.129 \pm 0.015 \pm 0.002) \cdot a_t^{-1}$   & \\
$\gamma_{\pi \eta,\,\pi \eta} = $ & $-0.16 \pm 0.24 \pm 0.03 $   & \\
$\gamma_{\pi \eta,\,\kkb}     = $ & $-0.56 \pm 0.29 \pm 0.04 $   & \\
$\gamma_{\kkb,    \,\kkb}     = $ & $\;\;\, 0.12 \pm 0.38 \pm 0.08 $   & \\[1.3ex]
&\multicolumn{2}{l}{ $\chi^2/ N_\mathrm{dof} = \frac{58.0}{47-6} = 1.41 $\,,}
\end{tabular}
\end{center}
\vspace{-1cm}
\begin{equation} \label{fit_Kmatrix_const_2x2}\end{equation}
\end{widetext}
where the uncertainties are first statistical and second due to variation of $m_\pi, m_K, m_\eta$ and $\xi$ within their uncertainties. The matrix shows the parameter correlations. In Figure~\ref{fig_A1_spec_comparison} we show the finite volume spectra obtained from the lattice QCD computation (shown in black) alongside the spectrum corresponding to the minimization in Eq.~\ref{fit_Kmatrix_const_2x2} (shown in orange), where we observe the good agreement suggested by the small $\chi^2$.

\begin{figure}
\includegraphics[width=0.95\columnwidth]{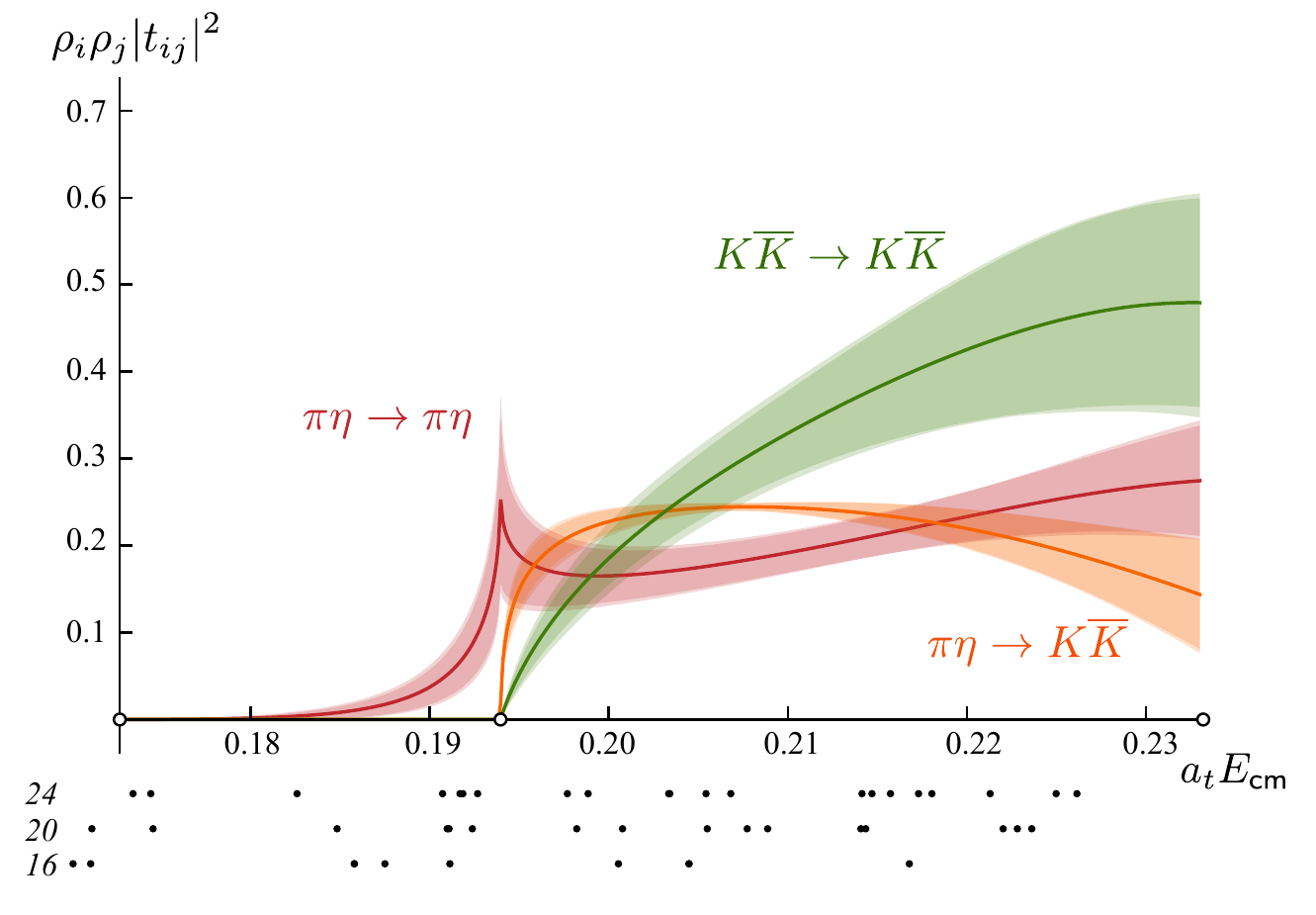}
\caption{The $S$-wave scattering amplitude expressed as $\rho_{\pi\eta}^2 |t_{\pi\eta, \pi\eta}|^2$ (red), $\rho_{\pi\eta} \rho_{K\bar{K}} |t_{\pi\eta, K\bar{K}}|^2$ (orange) and $\rho_{K\bar{K}}^2 |t_{K\bar{K}, K\bar{K}}|^2$ (green). The inner bands are determined by statistically sampling the space of correlated errors in Eq.~\ref{fit_Kmatrix_const_2x2}, and the outer bands show the variation with hadron masses and anisotropy as explained in the text. The small black dots indicate the positions of the energy levels on $16^3, 20^3$ and $24^3$ lattices (plotted in Figures~\ref{fig_spectra_A1_T1} and \ref{fig_spectra_A1_flight}) used to constrain the amplitude.}
\label{fig_kmat2x2_rho_t}
\end{figure}

The cross-sections for $\pe \to \pe$, $\pe \to \kk$ and $\kk \to \kk$ scattering are proportional to $\rho_i \rho_j | t_{ij}|^2$ -- we plot these for this amplitude in Figure \ref{fig_kmat2x2_rho_t}. A clear cusp structure is observed in $\pe \to \pe$ at the opening of the \kk ~threshold, and the amplitudes to produce \kk ~are seen to turn on rapidly at threshold.

\begin{figure}
\includegraphics[width=0.95\columnwidth]{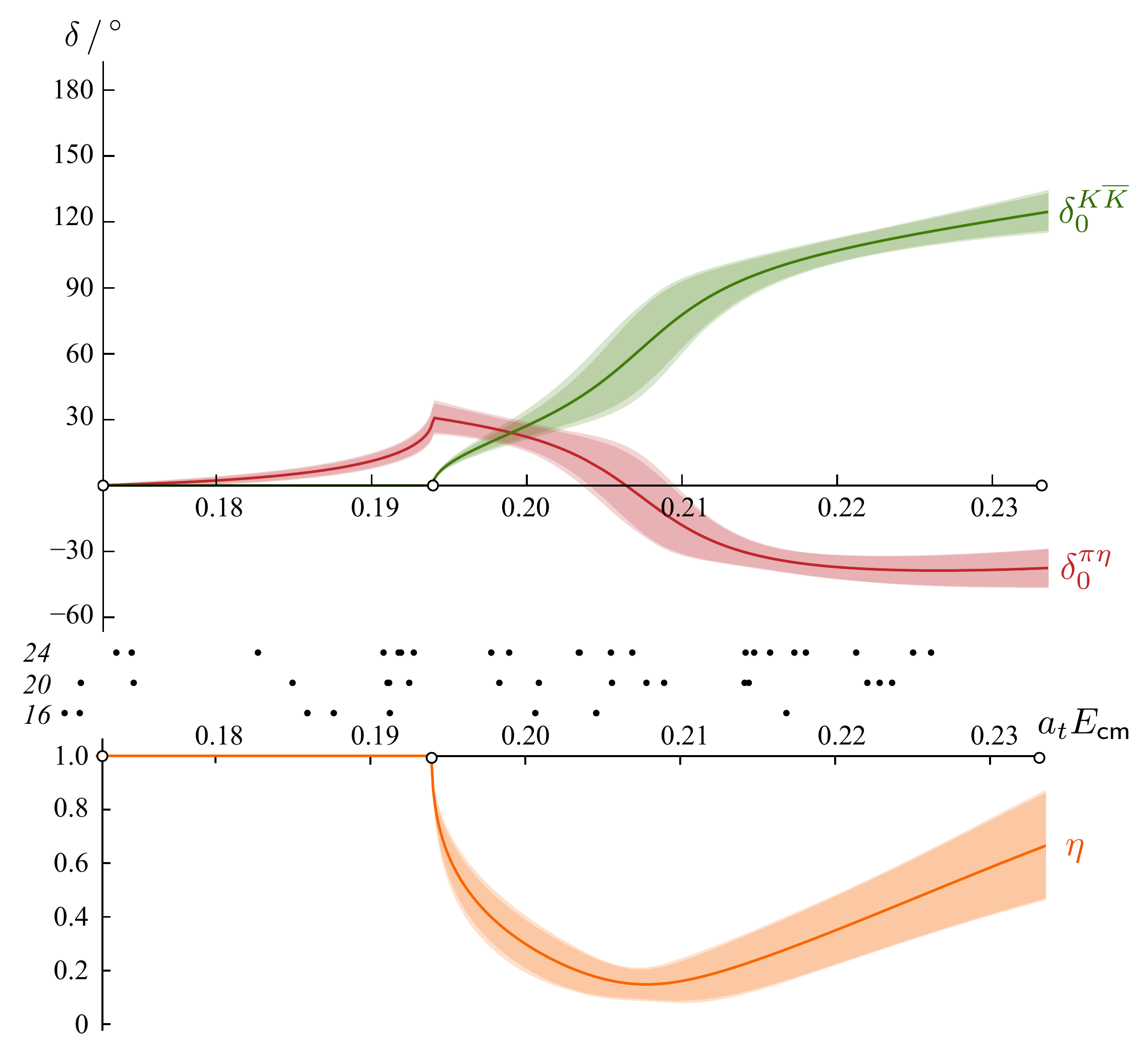}
\caption{The phase-shifts and inelasticity of the $S$-wave scattering amplitude corresponding to the minimization presented in Eq.~\ref{fit_Kmatrix_const_2x2}.}
\label{fig_kmat2x2_combined}
\end{figure}

An alternative way to display the $t$-matrix is to use phase-shifts for each channel and an inelasticity parameter, such that the diagonal elements of the $S$-matrix\footnote{related to $\bf{t}$ by $\mathbf{S} =  \mathbf{1} + 2 i \sqrt{ \boldsymbol{\rho} }\cdot \mathbf{t}\cdot \sqrt{ \boldsymbol{\rho} } $} are $\eta\, e^{2i \delta_\pe}, \eta\,  e^{2i \delta_\kkb}$. The phase-shifts and inelasticity as a function of energy are presented in Figure~\ref{fig_kmat2x2_combined}. The inelasticity deviates sharply from unity at $\kk$ threshold indicating a large coupling between channels.\footnote{It is worth noting here that the energy dependence of the two phase-shifts in strongly-coupled cases like this can undergo a complete change in character as the relative strength of coupling of a resonance to each channel is adjusted, while the corresponding scattering cross-sections change relatively little -- see Appendix~\ref{app_phases} for an illustration.} 

\begin{figure}
\includegraphics[width=0.75\columnwidth]{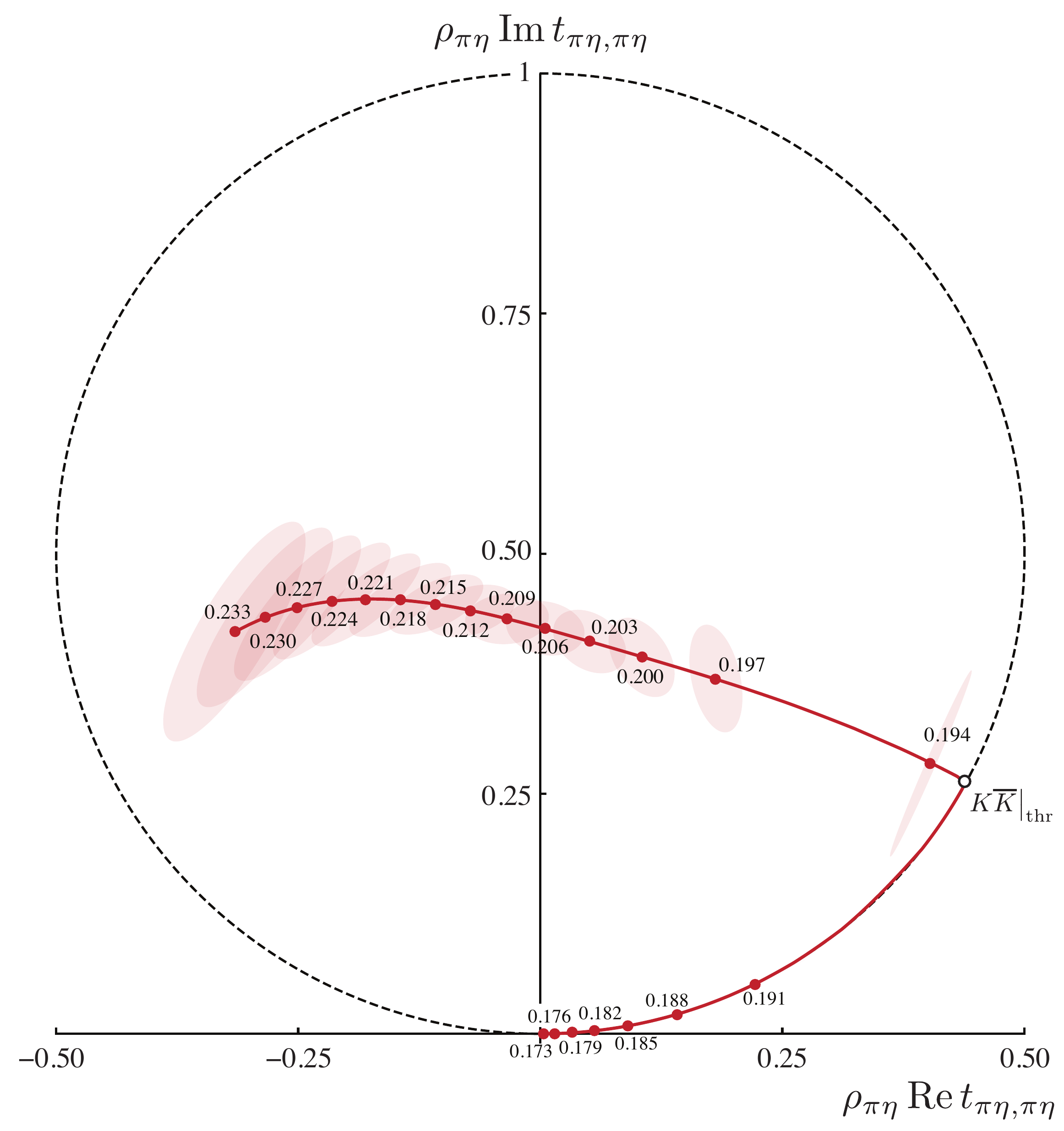}
\includegraphics[width=0.75\columnwidth]{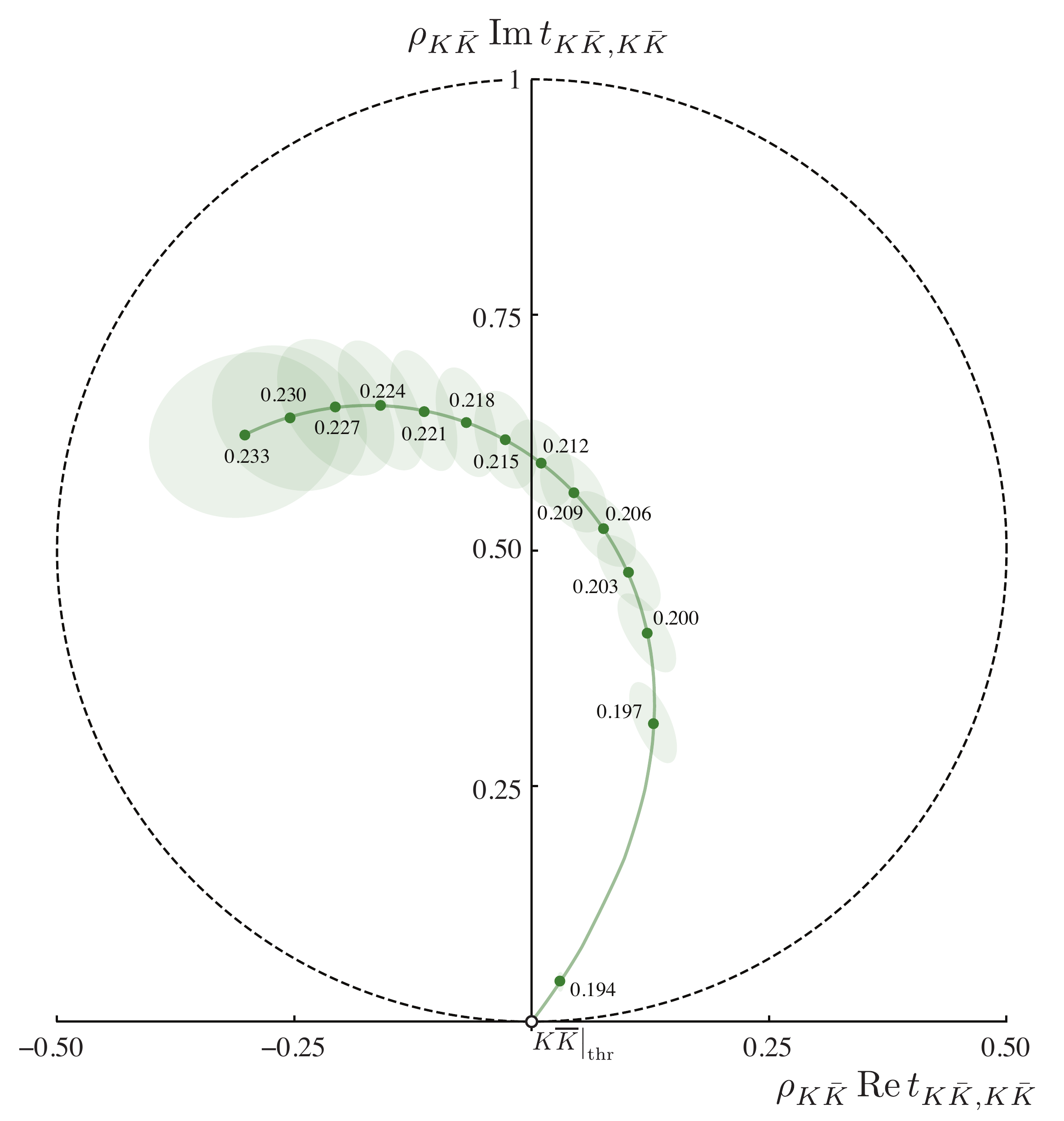}
\caption{Argand diagram representation of the amplitude in Eq.~\ref{fit_Kmatrix_const_2x2}. Dashed circle shows the unitarity bound. Points spaced equally in energy with $a_t \Delta E_\mathrm{cm} = 0.003$. Ellipses indicate the uncertainty on the amplitudes following from the statistical uncertainty on the parameters in Eq.~\ref{fit_Kmatrix_const_2x2}.}
\label{fig_argand}
\end{figure}

Finally, the amplitudes can be presented in a manner which makes explicit the constraint of unitarity, by using an `Argand' diagram which plots $\rho_i\, \mathrm{Im}\,  t_{ii}$ against $\rho_i \, \mathrm{Re}\, t_{ii}$, as shown in Figure~\ref{fig_argand}. Departure inside the dashed circle for the case $i = \pe$ corresponds to inelasticity, which we observe to turn on rapidly at the \kk~threshold.

While a cusp behavior at the opening of a new threshold is, in general, expected due to the corresponding branch-point singularity, the strength of the effect observed in Figure~\ref{fig_kmat2x2_rho_t}, and the very rapid turn-on of the \kk~amplitudes suggests that there may well be resonant behavior in this energy region. In Section \ref{poles} we will examine the continuation of our amplitude to complex values of $s$ to determine if there are additional nearby pole singularities that may provide a resonant explanation of the above observations.

A common parameterization of coupled-channel scattering that can describe an isolated resonance is the simple two-channel extension of the familiar elastic Breit-Wigner form that was proposed by Flatt\'e~\cite{Flatte:1976xv,Flatte:1976xu},
\begin{align}
t_{ij}(s)=\frac{g_i \, g_j}{m^2-s-ig^2_1 \, \rho_1(s)-ig^2_2 \, \rho_2(s)}\,,
\label{eq_flatte}
\end{align}
where the phase-space, $\rho(s)$, becomes imaginary below threshold, and where $m$, $g_1=g_\pe$, and $g_2=g_\kkb$ are free parameters. If we attempt to describe our finite-volume spectra using this form, the best fit we can obtain is,
{\small
\begin{align*}
\begin{tabular}{rll}
$m =$                         & $(0.2250 \pm 0.0024 \pm 0.0004) \cdot a_t^{-1}$   &
\multirow{3}{*}{ $\begin{bmatrix*}[r] 1 &  0.44   & 0.41    \\
                                    	&  1      &  0.84    \\
                                    	&         &  1      \end{bmatrix*}$ } \\
$g_{\pi \eta} =$                  & $(0.1194 \pm 0.012 \pm 0.003) \cdot a_t^{-1}$   & \\
$g_{\kkb}     =$                  & $(0.1362 \pm 0.013 \pm 0.004) \cdot a_t^{-1}$   & \\[1.3ex]
&\multicolumn{2}{l}{ $\chi^2/ N_\mathrm{dof} = \frac{149.5}{47-3} = 3.40 $\,,}
\end{tabular}
\end{align*}
}%
and while the key features of the spectra are reproduced, the goodness-of-fit is significantly inferior to our previous description using a ``pole plus constant'' $K$-matrix. The problem with this form appears to be that it is too restrictive -- it describes an amplitude {\it completely} dominated by a single resonance at all energies, no other behavior is accommodated. An example is that channel factorization, $\mathbf{t} \propto \mathbf{g} \otimes \mathbf{g} $ with $\mathbf{g} = [g_\pe, g_\kk]$ is forced to hold at all energies, while more generally we would expect this to hold exactly only at the complex pole position.

We note that the $K$-matrix ``pole plus constant" form defined in Eq.~\ref{K_ppc}, when used with our choice of subtraction in the Chew-Mandelstam phase-space (which ensures that $\mathrm{Re}\,  I(s=m^2) = 0$), approximately agrees with the Flatt\'e form for $s \approx m^2$ if $\rho \gamma \ll 1$.  The presence of the $\gamma$ matrix provides more freedom, allowing the amplitude to deviate from a pure resonance contribution at energies away from $s = m^2$ and breaking the exact channel factorization for real energies.

Of course, there is nothing unique about the ``pole plus constant'' form for the $K$-matrix that we found could successfully describe the finite-volume spectra, and we should explore to what extent other parameterizations can be used. Ultimately, as we will discuss in Section \ref{poles}, it is the singularities of $t(s)$ in the complex-$s$ plane, in particular poles relatively close to the real axis, that provide the least model-dependent description of the resonant content of scattering amplitudes. In what follows we will consider a variety of parameterizations of $t(s)$ to describe the finite-volume spectra and explore whether they share a common singularity structure.

\begin{figure}
\includegraphics[width=\columnwidth]{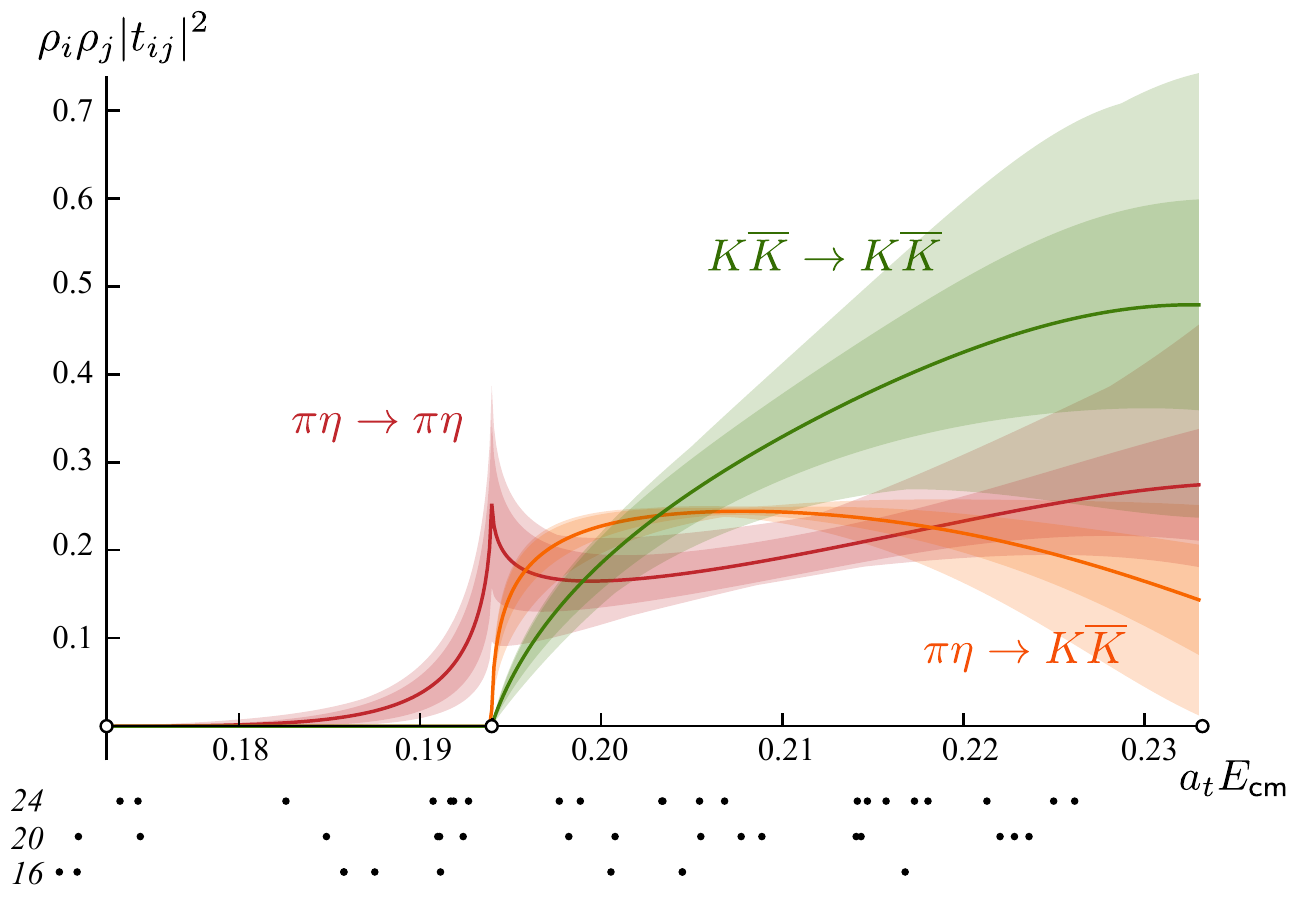}
\caption{The result of varying the parameterization used to describe the finite volume spectrum. All parameterizations with $\chi^2/N_\mathrm{dof}<1.5$ from Table~\ref{table_fit_forms} are included. The inner band corresponds to the fit in Eq.~\ref{fit_Kmatrix_const_2x2}, as plotted in Fig.~\ref{fig_kmat2x2_rho_t}. The outer band corresponds to taking the maximum and minimum values of all of the other fits, including their errors from minimization.}
\label{fig_coupled_2x2_parameterisation_variation_rho_t}
\end{figure}

\begin{figure}
\includegraphics[width=\columnwidth]{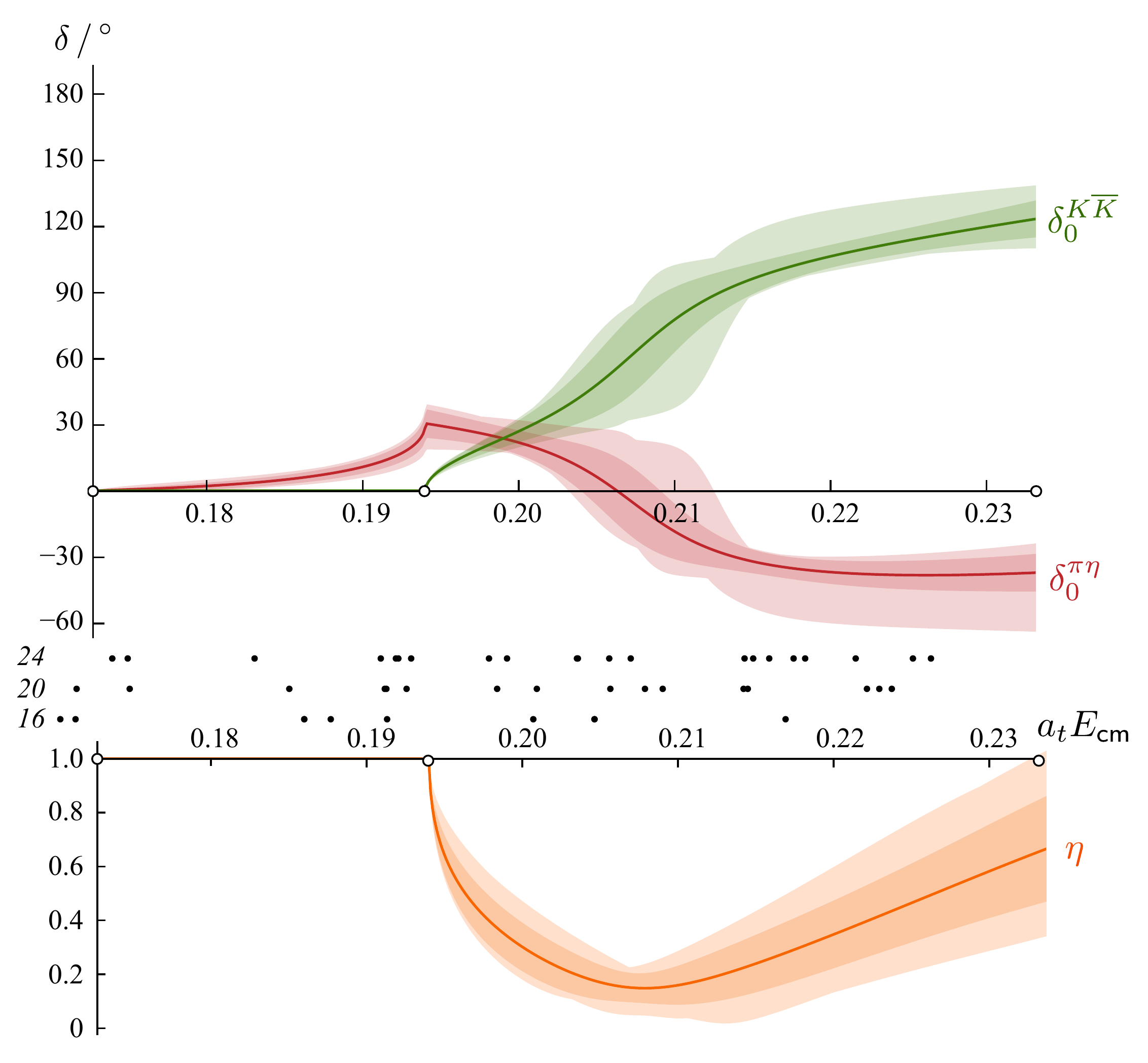}
\caption{The result of varying the parameterization used to describe the finite volume spectrum. All parameterizations with $\chi^2/N_\mathrm{dof}<1.5$ from Table~\ref{table_fit_forms} are included. The inner band corresponds to the fit in Eq.~\ref{fit_Kmatrix_const_2x2}, as plotted in Fig.~\ref{fig_kmat2x2_combined}. The outer band corresponds to taking the maximum and minimum values of all of the other fits, including their errors from minimization.}
\label{fig_coupled_2x2_parameterisation_variation}
\end{figure}

One class of parameterizations generalizes the ``pole plus constant'' $K$-matrix form, allowing multiple poles plus a polynomial in $s$,
\begin{align}
K_{ij}(s) = \sum_p\frac{g^{(p)}_{i} g^{(p)}_{j}}{m_p^2-s} + \sum_n\gamma_{ij}^{(n)}s^n.
\label{eq_kmatrix_pole_plus_poly}
\end{align}
Inclusion of (real) poles in $K$ proves to be an efficient parameterization when nearby poles are present in $t(s)$, but their use does not guarantee that there are nearby poles in $t(s)$ -- the interference with the polynomial can push them far from the region of the real energy axis constraining the amplitude. Typically we find that including a low-order polynomial is parametrically more economical than adding a pole in $K$ that appears outside the constrained region. In this study only a single pole was found to be necessary and various constant and linear-order polynomials were utilized.

Another alternative parameterizes the elements of the \emph{inverse} of $K$ by polynomials in $s$,
\begin{align}
K^{-1}_{ij}(s) = \sum_{n=0}^{N_{ij}} c_{ij}^{(n)} s^n ,
\label{eq_kmatrix_inverse_polynomial}
\end{align}
where $c_{ij}^{(n)}$ are real free parameters. When using this form we subtract the Chew-Mandelstam phase-space integral at $\pe$ threshold. 

We attempt to describe the finite-volume spectra using a variety of $K$-matrix parameterizations, varying the form and the number of parameters used. The results are summarized in Table~\ref{table_fit_forms} and the variation of the extracted amplitudes, observed to be modest, is shown in Figures~\ref{fig_coupled_2x2_parameterisation_variation_rho_t} and~\ref{fig_coupled_2x2_parameterisation_variation}. Higher order polynomials, multiple poles in Eq.~\ref{eq_kmatrix_pole_plus_poly}, and ``running'' $K$-matrix pole coupling forms that were applied in Ref.~\cite{Wilson:2015dqa} were also attempted, however they introduce more freedom than is necessary to describe the spectra and generally lead to large parameter correlations. It appears that the relatively narrow region of energy we are considering can be adequately described by just four suitably chosen parameters.

\begin{table}
\begin{tabular}{llcccc}
\hline\hline\\[-0.9ex]
Parameterization                                        & Restrictions                                               & $N_\mathrm{pars}$ &  $\chi^2/N_\mathrm{dof}$  \\[1.3ex]
\hline\hline\\[-0.9ex]
 \multirow{8}{*}{$K=\frac{g_i g_j}{m^2-s} + \gamma_{ij}^{(0)}$}  & --                                                         & 6          &   1.41  \\
                                                        & $\gamma_{\pe,\pe}^{(0)}  =0$                               & 5          &   1.38  \\
												           & $\gamma_{\pe,\kkb}^{(0)} =0$                               & 5          &   1.45  \\
                                                        & $\gamma_{\kkb,\kkb}^{(0)}=0$                               & 5          &   1.38  \\
                                                        & $\gamma_{\pe,\pe}^{(0)}  =0$, $\gamma_{\pe,\kkb}^{(0)}=0$  & 4          &   {\it 2.26}  \\
                                                        & $\gamma_{\pe,\pe}^{(0)}  =0$, $\gamma_{\kkb,\kkb}^{(0)}=0$ & 4          &   1.38  \\
                                                        & $\gamma_{\pe,\kkb}^{(0)} =0$, $\gamma_{\kkb,\kkb}^{(0)}=0$ & 4          &   {\it 1.52}  \\[1.3ex]
\hline\hline\\[-0.9ex]
 \multirow{8}{*}{$K=\frac{g_i g_j}{m^2-s} + \gamma_{ij}^{(1)} s$ }  & --                                                         & 6          &   1.40  \\
                                                        & $\gamma_{\pe,\pe}^{(1)}  =0$                               & 5          &   1.37  \\
                                                        & $\gamma_{\pe,\kkb}^{(1)} =0$                               & 5          &   1.36  \\
                                                        & $\gamma_{\kkb,\kkb}^{(1)}=0$                               & 5          &   1.36  \\
                                                        & $\gamma_{\pe,\pe}^{(1)}  =0$, $\gamma_{\pe,\kkb}^{(1)}=0$  & 4          &   {\it 1.99}  \\
                                                        & $\gamma_{\pe,\pe}^{(1)}  =0$, $\gamma_{\kkb,\kkb}^{(1)}=0$ & 4          &   1.34  \\
                                                        & $\gamma_{\pe,\kkb}^{(1)} =0$, $\gamma_{\kkb,\kkb}^{(1)}=0$ & 4          &   1.36  \\[1.3ex]
\hline\hline\\[-0.9ex]
 \multirow{9}{*}{$K^{-1}= c_{ij}^{(0)} +c_{ij}^{(1)} s$} & --                                                         & 6          &   1.40  \\
                                                        & $c_{\pe,\pe}^{(1)}  =0$                                    & 5          &   1.37  \\
                                                        & $c_{\pe,\kkb}^{(1)} =0$                                    & 5          &   1.50  \\
                                                        & $c_{\kkb,\kkb}^{(1)}=0$                                    & 5          &   1.38  \\
                                                        & $c_{\pe,\pe}^{(1)}  =0$, $c_{\pe,\kkb}^{(1)}=0$            & 4          &   {\it 1.52}  \\
                                                        & $c_{\pe,\pe}^{(1)}  =0$, $c_{\kkb,\kkb}^{(1)}=0$           & 4          &   1.36  \\
                                                        & $c_{\pe,\kkb}^{(1)} =0$, $c_{\kkb,\kkb}^{(1)}=0$           & 4          &   {\it 2.26}  \\
                                                        & $c_{ij}^{(1)}=0$                                           & 3          &   {\it 6.27}  \\[1.3ex]
\hline\hline\\[-0.9ex]
 \multirow{8}{*}{ \minitab[c]{ $K=\frac{g_i g_j}{m^2-s} + \gamma_{ij}^{(0)}$ \\[1.3ex]   $I_{ij}(s)=-i\delta_{ij}\rho_i(s)$ }   } & --                                                         & 6          &   1.48  \\
                                                        & $\gamma_{\pe,\pe}^{(0)}  =0$                               & 5          &   1.45  \\
                                                        & $\gamma_{\pe,\kkb}^{(0)} =0$                               & 5          &   1.49  \\
                                                        & $\gamma_{\kkb,\kkb}^{(0)}=0$                               & 5          &   1.47  \\
                                                        & $\gamma_{\pe,\pe}^{(0)}  =0$, $\gamma_{\pe,\kkb}^{(0)}=0$  & 4          &   {\it 1.72}  \\
                                                        & $\gamma_{\pe,\pe}^{(0)}  =0$, $\gamma_{\kkb,\kkb}^{(0)}=0$ & 4          &   1.44  \\
                                                        & $\gamma_{\pe,\kkb}^{(0)} =0$, $\gamma_{\kkb,\kkb}^{(0)}=0$ & 4          &   {\it 1.52}  \\[1.3ex]
\hline\hline\\[-0.9ex]
\multicolumn{2}{l}{$t_{ij}= g_i g_j/(m^2-s-ig_1^2\rho_1-ig_2^2\rho_2)$ }                                             & 3          &   {\it 3.40} \\[1.3ex]
\hline\hline
\end{tabular}
\caption{Variation of $S$-wave amplitude parameterization.}
\label{table_fit_forms}
\end{table}

We observe that many different forms give amplitudes having essentially the same structure, indicating that the details of the parameterization form are not important provided it contains sufficient freedom. In Section~\ref{poles} we will consider to what extent the singularity structures of each of these amplitudes are common.

\subsection{Coupled-channel $S$-wave \pe, \kk, \pep ~scattering}\label{3chan}

We may extend our $S$-wave scattering analysis up to the $\pi\pi\omega$ threshold at $a_t\, E_\mathsf{cm} = 0.2949$ if we also allow for scattering in the \pep ~channel. We consider only the extra constraint from describing the 10 additional energy levels in the rest-frame $A_1^+$ irrep in this region, as no higher partial-waves with $J<4$ contribute to this irrep, nor do any three-pseudoscalar channels. The relatively large statistical uncertainties on energy levels having large overlap with \pep--like operators (see Figure \ref{fig_spectrum_with_histograms}) limits the precision with which we will determine scattering amplitudes above \pep ~threshold.

We perform a minimization to describe the 57 energy levels using a $3\times 3$ version of Eq.~\ref{eq_kmatrix_pole_plus_poly} with a ``pole plus constant'' form. We find that a reasonable description of the spectra can be obtained allowing, in addition to the free parameters in Eq.~\ref{fit_Kmatrix_const_2x2}, also a non-zero $g_\pep$ and a non-zero $\gamma_{\pep,\pep}$. The fit yields,

\begin{widetext}
\small
\begin{center}
\begin{tabular}{rll}
$m =$                         & $(0.2275 \pm 0.0029 \pm 0.0013) \cdot a_t^{-1}$   &
\multirow{8}{*}{ $\begin{bmatrix*}[r] 1 &  0.69 & -0.09 &  0.04 & -0.33 &  0.17 &  0.22 &  0.02 &  \\
                                      	&  1    & -0.38 &  0.22 & -0.59 &  0.48 &  0.08 &  0.14  \\
                                      	&       &  1    & -0.49 &  0.06 & -0.09 &  0.24 &  0.05  \\
                                      	&       &       & 1     &  0.22 & -0.21 &  0.12 &  0.10  \\
                                      	&       &       &       &  1    & -0.87 &  0.28 & -0.12  \\
                                      	&       &       &       &       &  1    & -0.61 &  0.13  \\
                                      	&       &       &       &       &       &  1    &  0.01  \\
                                      	&       &       &       &       &       &       &  1     \\
                                       \end{bmatrix*}$ } \\
$g_{\pe}                = $ & $(\,\,\,\; 0.129 \pm 0.011 \pm 0.017) \cdot a_t^{-1}$   & \\
$g_{\kkb}               = $ & $(        -0.145 \pm 0.010 \pm 0.002) \cdot a_t^{-1}$   & \\
$g_{\pep}               = $ & $(\,\,\,\; 0.104 \pm 0.032 \pm 0.060) \cdot a_t^{-1}$   & \\
$\gamma_{\pe,  \,\pe }  = $ & $        -0.59 \pm 0.14 \pm 0.36 $   & \\
$\gamma_{\pe,  \,\kkb}  = $ & $        -0.31 \pm 0.13 \pm 0.37 $   & \\
$\gamma_{\kkb, \,\kkb}  = $ & $\,\,\,\; 0.14 \pm 0.16 \pm 0.34 $   & \\
$\gamma_{\pep, \,\pep } = $ & $\,\,\,\; 0.57 \pm 0.39 \pm 0.23 $   & \\[1.3ex]
&\multicolumn{2}{l}{ $\chi^2/ N_\mathrm{dof} = \frac{65.8}{57-8} = 1.34 $\,.}
\end{tabular}
\end{center}
\vspace{-1cm}
\begin{equation} \label{fit_minimum_3x3}\end{equation}
\end{widetext}

\begin{figure}
  \includegraphics[width=1.0\columnwidth]{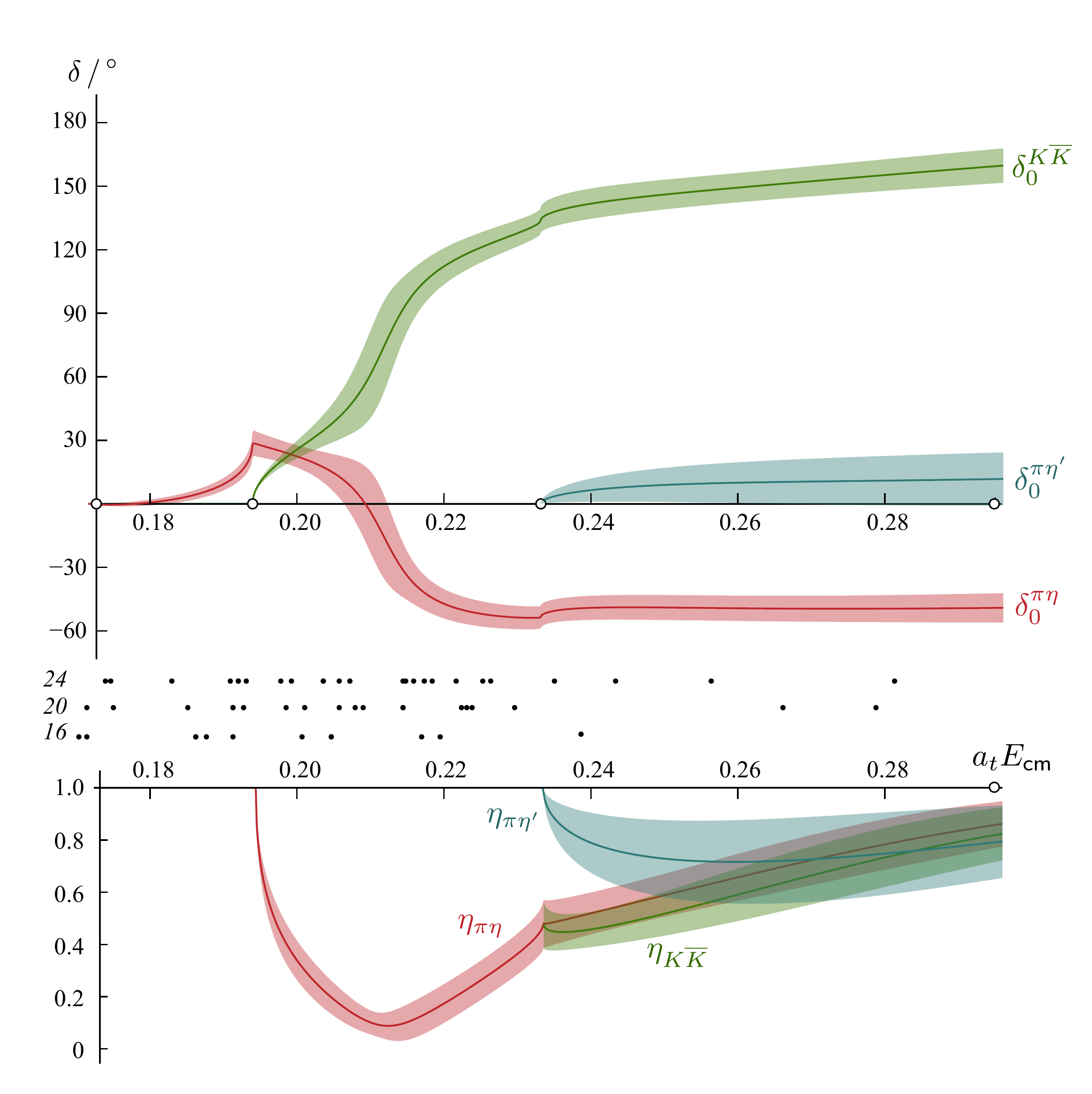}
  \caption{The phase-shifts and ``inelasticities'' of Eq.~\ref{S3} corresponding to Eq.~\ref{fit_minimum_3x3}. Below \pep~threshold, the inelasticities $\eta_\pe$ and $\eta_\kkb$ are equal.}
      \label{fig_coupled_3x3}
\end{figure}

There is not a settled convention for describing three-channel scattering amplitudes in terms of a minimal set of energy-dependent functions. We will use a scheme where the scattering is described by three phase-shifts, $\delta_i$, and three ``inelasticities", $\eta_i$, where the diagonal entries of the $S$-matrix are,
\begin{align}
\mathrm{diag}({\mathbf{S}})=(\eta_1 e^{2i\delta_1},\eta_2 e^{2i\delta_2},\eta_3 e^{2i\delta_3})\,,
\label{S3}
\end{align}
and the off-diagonal entries are determined by unitarity and invariance under time-reversal, the phase of each element of $S_{ij}$ being $\delta_i+\delta_j$. In the limit where the third channel decouples from the other two, then $\eta_3\to 1$, and $\eta_2\to\eta_1=\eta$ where $\eta$ is the two-channel inelasticity used previously.  We plot the phase-shifts and inelasticities corresponding to the amplitude described in Eq.~\ref{fit_minimum_3x3} in Fig.~\ref{fig_coupled_3x3}. The region below \pep ~threshold is, as we would expect, largely unchanged by the inclusion of the additional channel. Apart from mild cusps at the opening of \pep, there are no notable new structures below $\pi\pi \omega$ threshold. Exploring alternative parameterizations, we find some variation in the amplitudes in the region above \pep ~threshold, due to there being relatively few energy levels to constrain the additional parameters, but in all cases the region below \pep ~threshold is largely unchanged and there is no significant structure above \pep ~threshold.

\subsection{Higher partial waves}

Thus far we have assumed that in $A_1$ irreps, only \mbox{$S$-wave} scattering plays a significant role in determining the spectrum in the energy region considered. In a finite cubic volume the reduced rotational symmetry leads to a subduction of multiple partial waves of definite $\ell$ into the limited set of irreps in which we work. In the case of scattering of unequal mass particles in moving frames the subduction leads to a rather dense set of partial waves in each irrep, as summarized in Table~\ref{tab_pwa_irrep}.

The role of higher partial waves at low energies is limited by kinematic suppression proportional to $k_i^{2\ell}$, but this can be circumvented if resonances happen to appear at low energy. In this case, $\pi \eta$ in $P$-wave has exotic $J^{PC}=1^{-+}$ quantum numbers, so we do not expect any low-lying resonances\footnote{In ref.~\cite{Dudek:2010wm}, the indications are that there may be hybrid mesons~\cite{Dudek:2011bn} above $a_t \, E_\mathsf{cm}=0.35$ on these lattices} and \kk ~cannot have the required $G=-$ in $P$-wave. \pe ~and \kk ~in $D$-wave have $J^{PC}=2^{++}$, and we might expect there to be an $a_2$ resonance, a supposition that is supported by the presence in Figure~\ref{fig_spectra_Dwave} of levels near $a_t\, E_\mathsf{cm} = 0.26$, that do not lie near to non-interacting meson-meson energies.

\subsubsection{$P$-wave}

In the right pane of Fig~\ref{fig_spectra_A1_T1} we plot the spectrum in the rest-frame $T_1^-$ irrep, which receives contributions from \pe ~$P$-wave, $F$-wave and higher partial-waves. There are only very small shifts seen in the lattice QCD spectra compared to the non-interacting \pe ~levels, suggesting a very weak interaction. In the $\pi\eta$ elastic region, below \pep ~threshold, there is one energy level at ${a_tE_\cm = 0.2316(8)}$ from which we can determine a \mbox{$P$-wave} elastic scattering phase-shift, ${\delta_1= (-1.2 \pm 3.0)^\circ}$. The levels above $\pi\eta^\prime$, but below $\pi\pi \omega$, that are dominated by overlap with $\pi\eta$ operators are similarly consistent with small or no interactions. We extract one $\pi\eta^\prime$-like level on the $24^3$ lattice and this is also consistent with negligibly small interactions. Using the four levels are that are dominated by $\pi\eta$ operators we obtain a scattering length ${a_1 = (9 \pm 17) \,\cdot\,a_t^3}$ with a $\chi^2/N_\mathrm{dof}=0.61$.

The $P$-wave interactions are clearly very weak in this low-energy region, but we may include their effect when extracting the $S$-wave amplitudes from moving-frame $A_1$ irreps -- doing so we observe negligible changes in the $S$-wave amplitudes.

\subsubsection{$D$-wave} \label{D_amps}

In Fig.~\ref{fig_spectra_Dwave} we show the energy levels in irreps that have $D$-wave as their lowest subduced partial wave. The extracted levels largely lie close to non-interacting meson-meson energies, with the addition of a level that appears systematically at about $a_t\, E_\mathsf{cm} = 0.26$. No levels appear in the \pe ~elastic scattering region, but if we assume that there is negligible coupling between \pe ~and \kk ~at low energy, the lowest level in $24^3$ $[000]E^+$ at ${a_tE_\cm=0.2316(11)}$ would correspond to a $D$-wave \pe ~phase-shift of ${\delta_2 = (-0.5 \pm 1.6)^\circ}$.

With our current formalism, we are limited in how much we can determine about scattering amplitudes in this channel -- $\pi\pi\pi$ scattering can contribute here, and we have not included ``$\pi\pi\pi$-like'' operators in our basis, so we do not expect to have extracted the complete spectrum of finite-volume eigenstates. Neither do we have a complete formalism with which to determine three-body scattering amplitudes\footnote{but see~\cite{Hansen:2014eka,Hansen:2015zga,Hansen:2015azg} for promising progress in this direction}. Nevertheless, we may take a somewhat cavalier approach and proceed under the assumption that the $\pi\pi\pi$ channel plays a negligible role here -- in our previous analysis of the coupled $\pi K,\eta K$ system~\cite{Dudek:2014qha,Wilson:2014cna} we found that we were able to adequately describe the calculated $D$-wave spectrum well above the $\pi\pi K$ threshold using only meson-meson amplitudes, and extracted what appeared to be a signal for a narrow $K_2^\star$ resonance that was dominantly coupled to the $\pi K$ channel.

We will attempt something similar here, treating the description of the spectra in Fig.~\ref{fig_spectra_Dwave} as a coupled \pe,~\kk ~problem alone. Utilizing a $K$-matrix of ``pole plus constant'' form, with the necessary threshold factors from Eq.~\ref{eq_t_matrix_k} we obtain,

\begin{widetext}
\begin{center}
\begin{tabular}{rll}
$m =$                         & $(0.2658 \pm 0.0008 \pm 0.0001) \cdot a_t^{-1}$   &
\multirow{5}{*}{ $\begin{bmatrix*}[r] 1 & 0.03 &  0.03 & 0.01 & -0.05 &  0.02 \\
                                    	&  1    &  0.07 &  0.73 & -0.22 &  0.17 \\
                                    	&       & 1     &  0.07 &  0.73 &  0.27 \\
                                    	&       &       & 1     & -0.05 &  0.55 \\
                                    	&       &       &       & 1     &  0.31 \\
                                    	&       &       &       &       &  1    \end{bmatrix*}$ } \\
$g_{\pi \eta} =$                  & $(0.712 \pm 0.056 \pm 0.004) \cdot a_t$   & \\
$g_{\kkb}     =$                  & $(0.665 \pm 0.072 \pm 0.010) \cdot a_t$   & \\
$\gamma_{\pi \eta,\,\pi \eta} = $ & $(-0.1 \pm 7.7 \pm 2.7 )   \cdot a_t^4$   & \\
$\gamma_{\pi \eta,\,\kkb}     = $ & $(30. \pm 18. \pm 2. ) \cdot a_t^4$   & \\
$\gamma_{\kkb,     \,\kkb}    = $ & $(1.2 \pm 11.9 \pm 6.3 )  \cdot a_t^4$   & \\[1.3ex]
&\multicolumn{2}{l}{ $\chi^2/ N_\mathrm{dof} = \frac{45.0}{28-6} = 2.05 $\,.}
\end{tabular}
\end{center}
\vspace{-1cm}
\begin{equation} \label{D_fit_par_values}\end{equation}
\end{widetext}

\begin{figure}
\includegraphics[width=\columnwidth]{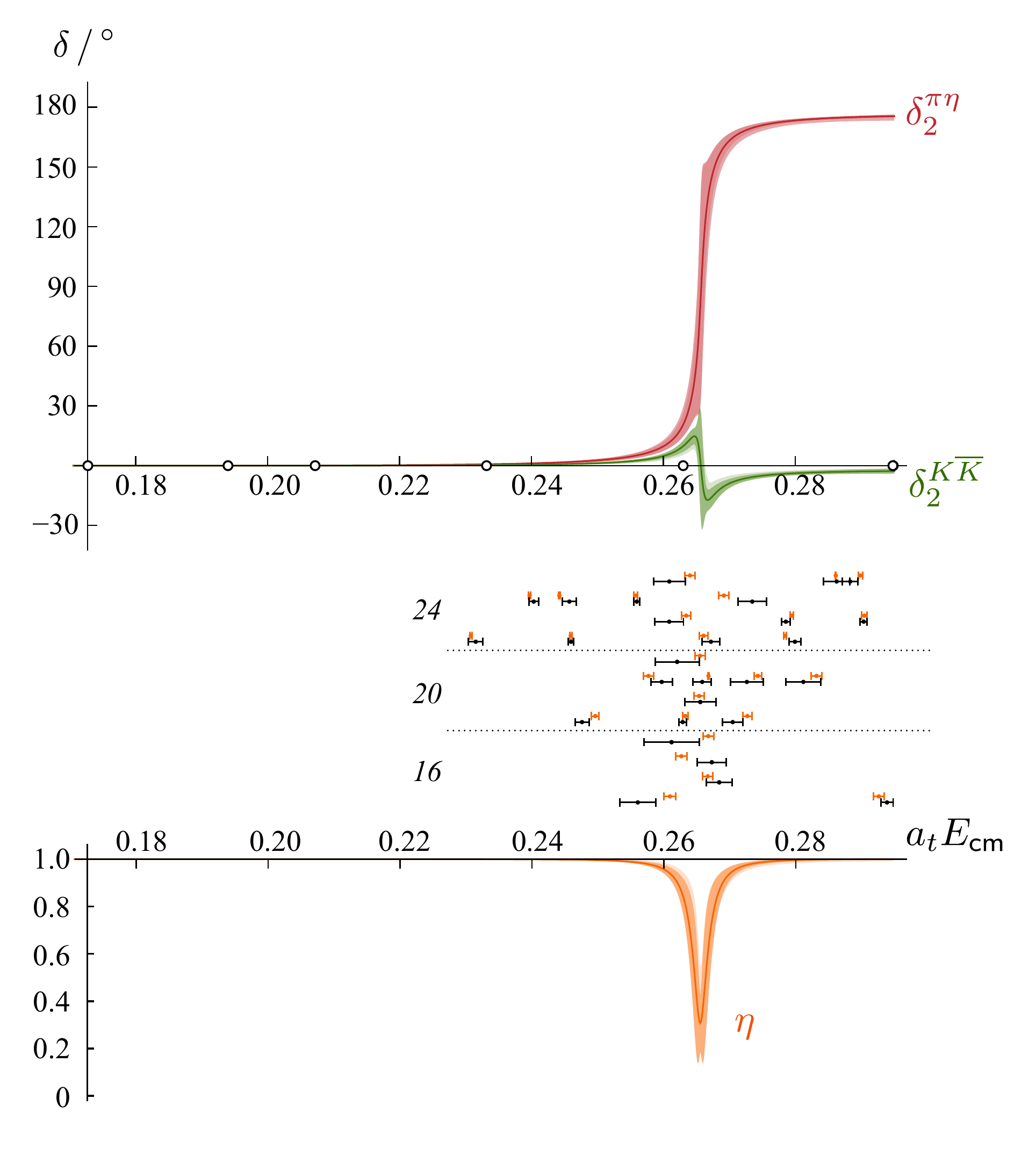}
\caption{The $D$-wave scattering amplitude corresponding to the six parameter $K$-matrix minimization in Eq.~\ref{D_fit_par_values} shown via the phase-shifts and inelasticity. In the center, we plot the energy spectra used to constrain the amplitude (in black, as in Figure~\ref{fig_spectra_Dwave}) along with finite-volume spectra corresponding to the parameterization (in orange). The empty circles on the energy axis denote threshold energies from Table~\ref{tab_masses}.}
\label{fig_coupled_D_2x2}
\end{figure}

The corresponding phase-shifts and inelasticity are plotted in Figure~\ref{fig_coupled_D_2x2}, along with a comparison of the finite-volume spectrum given by the amplitude in Eq.~\ref{D_fit_par_values} and the lattice QCD spectrum used to the constrain the amplitude. The description is reasonable, but not perfect\footnote{variation of parameterization form did not yield any descriptions of the spectrum with $\chi^2/ N_\mathrm{dof}$ significantly below this value}, and may indicate the limitations due to the assumed correctness of the spectrum without $\pi\pi\pi$ operators, and the assumed absence of coupling of the \pe, \kk ~system to $\pi\pi\pi$. The solution obtained clearly corresponds to dominance of the scattering by a narrow resonance, coupled to both \pe ~and \kk, and if this result is correct, it is quite interesting, since the experimental $a_2(1320)$ meson couples dominantly to $\pi\pi\pi$, but perhaps this coupling decreases rapidly with increasing pion mass? Further calculations at other pion masses will be required to explore this.

The presence of a narrow resonance, coupled to both \pe~and \kk, suggests that the Flatt\'e amplitude may provide an efficient description. Generalizing Eq.~\ref{eq_flatte} to non-zero angular momentum $\ell$, we have,
\begin{align*}
t_{ij}=\frac{(2 k_i)^{\ell} \, g_i \, g_j \, (2 k_j)^{\ell}}{m^2 - s - i g_1^2 \, \rho_1\,  (2 k_1)^{2\ell} - i  g_2^2\,  \rho_2 \, (2 k_2)^{2\ell}},
\end{align*}
and we find that the spectrum can be described well by,
{\small
\begin{align*}
\begin{tabular}{rll}
$m =$                         & $(0.2658 \pm 0.0008 \pm 0.0001) \cdot a_t^{-1}$   &
\multirow{3}{*}{ $\begin{bmatrix*}[r] 1 &  0.06   & 0.08    \\
                                    	&  1      &  0.30    \\
                                    	&         &  1      \end{bmatrix*}$ } \\
$g_{\pi \eta} =$                  & $(0.766 \pm 0.043 \pm 0.017) \cdot a_t$   & \\
$g_{\kkb}      =$                 & $(0.581 \pm 0.050 \pm 0.006) \cdot a_t$   & \\[1.3ex]
&\multicolumn{2}{l}{ $\chi^2/ N_\mathrm{dof} = \frac{49.0}{28-3} = 1.96 $\,.}
\end{tabular}
\end{align*}
}
This appears to be an appropriate description of the scattering amplitude having well determined parameters with low correlations corresponding to a narrow resonance coupled to both \pe~and \kk.

\section{Resonance poles}\label{poles}

\begin{figure}
\includegraphics[width=0.9\columnwidth]{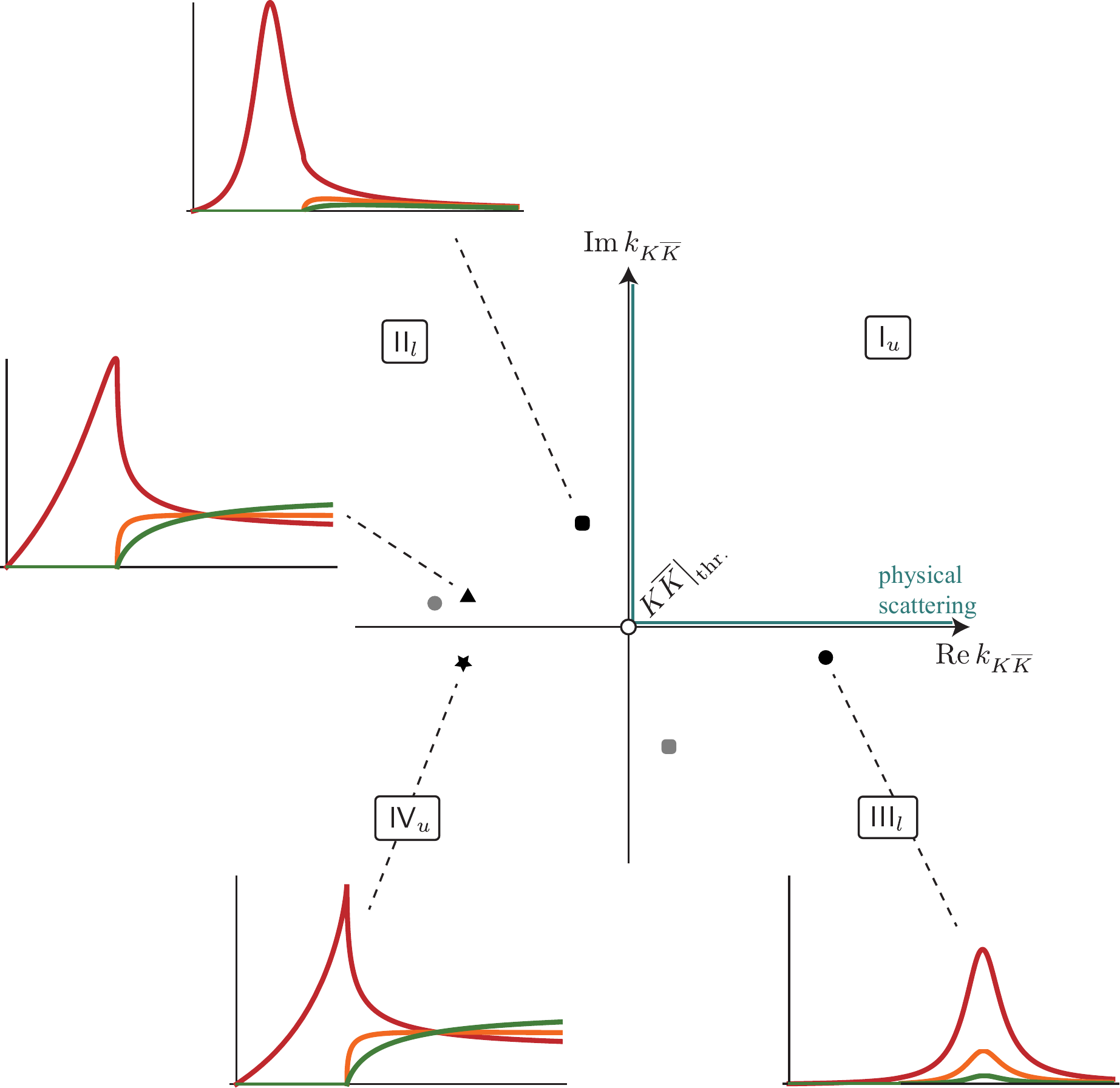}
\caption{Sheet structure around the \kk~threshold expressed in terms of the \kk~$\mathsf{cm}$ momentum. Physical scattering with increasing energy follows the blue line down the imaginary axis (subthreshold) and then along the real axis (above threshold). The lower half-plane of sheet {\sf II} is seen to be closest well below \kk~threshold and the lower half-plane of sheet {\sf III} is closest well above threshold. For energies close to the threshold, the upper half-plane of sheet {\sf IV} is also nearby. Four different possible resonant pole structures are also shown:\\
(squares) ``Breit-Wigner''-like (Flatt\'e) pair of poles for subthreshold narrow resonance -- $\pi \eta$ elastic amplitude shows a bump below threshold and a weak threshold cusp.\\
(circles) ``Breit-Wigner''-like (Flatt\'e) pair of poles for narrow resonance above threshold -- amplitudes show the canonical bump in each coupled-channel.\\
(triangle) single nearby pole on sheet {\sf II} -- leads to an asymmetric bump / strong cusp at \kk~threshold and a rapid turn-on of amplitudes leading to the \kk~final state.\\
(star) single nearby pole on sheet {\sf IV} -- leads to an asymmetric bump / strong cusp at \kk~threshold and a rapid turn-on of amplitudes leading to the \kk~final state.\\
The upper half plane of sheet {\sf IV} is continuously connected to the lower half-plane of sheet {\sf II} above threshold, so the triangle and star should be considered to be close to each other, and it should be no surprise that their amplitudes are similar. 
}
\label{k2sheet}
\end{figure}

The scattering amplitudes that we have extracted are directly constrained for real values of $s=E_\cm^2$ by the finite-volume energy levels computed in our lattice QCD calculations. In a similar way, scattering amplitudes may be determined experimentally using scattering data collected at real energies above kinematic thresholds. It has proven useful to consider the continuation of these amplitudes to values of $s$ in the complex plane since singularities, in particular pole singularities associated with resonances, are the dominating features. 

The structure of the complex $s$-plane away from the real axis becomes more complicated as new scattering channels open. Each new square-root branch-cut associated with an opening threshold splits the plane into two Riemann sheets, so that for \mbox{$n$-channel} scattering, there are $2^n$ sheets. In the two channel region there are four sheets to consider, and they may be labelled by the sign of the imaginary part of the {\sf cm}-frame momenta for the two channels:
\vspace{5pt}
\begin{center}
\begin{tabular}{c|cc}
Sheet & $\mathrm{Im}\, k_\pe $  & $\mathrm{Im}\, k_\kk $ \\[0.7ex]
\hline\\[-1.6ex]
$\mathsf{I}$         & $+$ & $+$ \\
$\mathsf{II}$        & $-$ & $+$ \\
$\mathsf{III}$       & $-$ & $-$ \\
$\mathsf{IV}$        & $+$ & $-$
\end{tabular}
\end{center}
\vspace{5pt}

Physical scattering occurs on sheet {\sf I}, just above the real $s$ axis where the branch cuts that lead to the other, ``unphysical'', sheets lie. Causality ensures that singularities cannot appear off the real axis on the physical sheet, but they can appear on any of the unphysical sheets. In particular, we can have poles at complex values of $s$, and in a region near a pole we have for an element of the $t$-matrix, $t_{ij}\sim \frac{c_i c_j}{s-s_0}$. $s_0$ is the pole position, which is often associated with a resonance mass and width as $s_0 = \left( m_R \pm \frac{i}{2} \Gamma_R \right)^2$, and the residue can be factorized into $c_i$, $c_j$, couplings that indicate the relative strength with which the pole couples to each channel\footnote{Poles always appear in complex-conjugate pairs in $\sqrt{s_0}$ and $\sqrt{s_0}^\star$ and their residues are also related by complex conjugation. Usually only one pole of the pair is in close proximity to physical scattering.}.

For physical scattering above \pe ~threshold, but well below \kk ~threshold, the lower half-plane of sheet {\sf II} is nearby, while well above \kk ~threshold it is the lower half-plane of sheet {\sf III} that is nearby. In the energy region close to \kk ~threshold all of the following are nearby: the lower half-plane of sheet {\sf II}, the lower half-plane of sheet {\sf III} and the upper half-plane of sheet {\sf IV} (or using a convenient shorthand, $\mathsf{II}_l, \mathsf{III}_l, \mathsf{IV}_u$). This can be more easily visualized by using $k_\kkb$ as a ``uniformizing variable''~\cite{Newton:1961,Morgan:1993td} -- this unfolds the multi-sheeted $s$-plane into a single sheet, with physical scattering running first down the imaginary axis towards the origin, corresponding to the \kk ~threshold, and then out to positive real values. Figure~\ref{k2sheet} illustrates the sheet proximities described above in terms of the complex-$k_\kkb$ plane.

An illustration of the kind of unphysical sheet pole structures that can arise for a single isolated resonance is provided by the two-channel Breit-Wigner extension of Flatt\'e, Eq.~\ref{eq_flatte}. If we consider the case of a narrow \mbox{$S$-wave} resonance, above the kinematic thresholds for each channel, we find poles in $\mathsf{II}_l$ and $\mathsf{III}_l$ (or in $\mathsf{IV}_u$ and $\mathsf{III}_l$ for certain relative channel couplings, $g_2/g_1$), which in the case of dominance of coupling to one channel over the other, lie at approximately mirror positions in the complex-$k_\kkb$ plane: $k_\kkb^{\mathsf{II}} \approx - k_\kkb^{\mathsf{III}}$, as shown in Figure~\ref{k2sheet} (circles). The $\mathsf{III}_l$ pole is very close to physical scattering when $E_\mathsf{cm} \approx m_R$ and this leads to the narrow bumps and rapid phase motion of the amplitudes that we typically associate with resonances. Also shown in Figure~\ref{k2sheet} is the case of a sub \kk-threshold resonance (squares) described by a pair of poles -- this case can be described by the Flatt\'e form.

While this ``pair of poles'' situation corresponds to our canonical view of a coupled-channel hadron resonance, other distributions of poles are not forbidden by any general principle. Another possibility is to have a \emph{nearby} pole on only \emph{one} unphysical sheet~\cite{Morgan:1992ge,Morgan:1993td, Baru:2004xg} -- cases of this type are illustrated in Figure~\ref{k2sheet} by the star and the triangle, where we see that they can lead to a structure at threshold that is strongly asymmetric. We will discuss the physical interpretation of such a pole distribution in Section~\ref{interpret}.

\subsection{$S$-wave poles from $K$-matrix analysis}

\begin{figure}
\includegraphics[width=0.95\columnwidth]{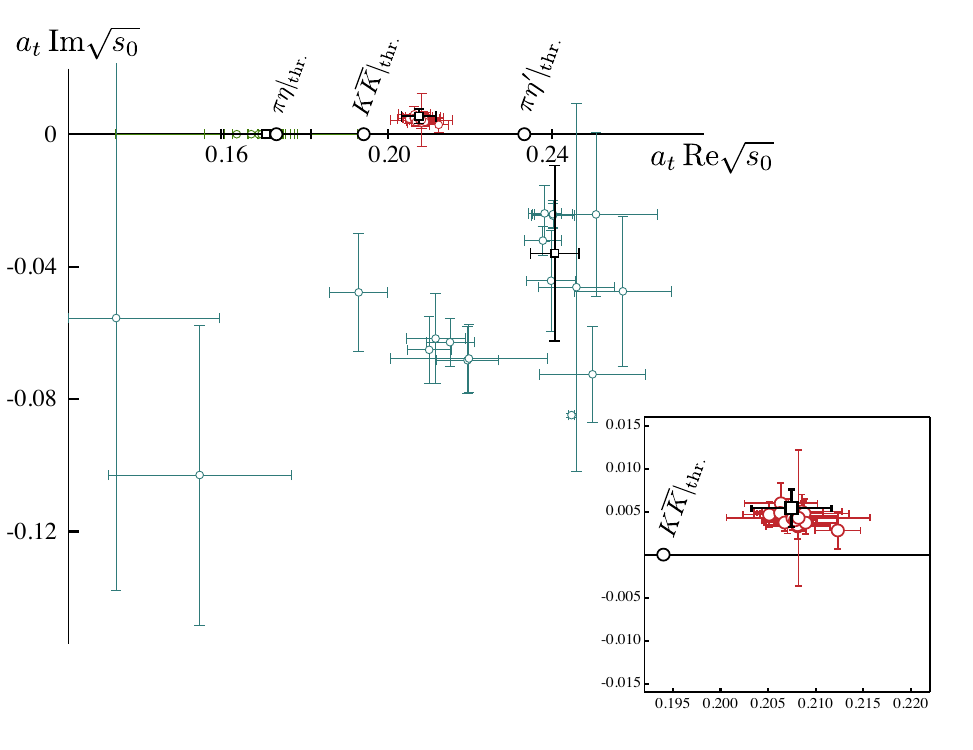}
\includegraphics[width=0.85\columnwidth]{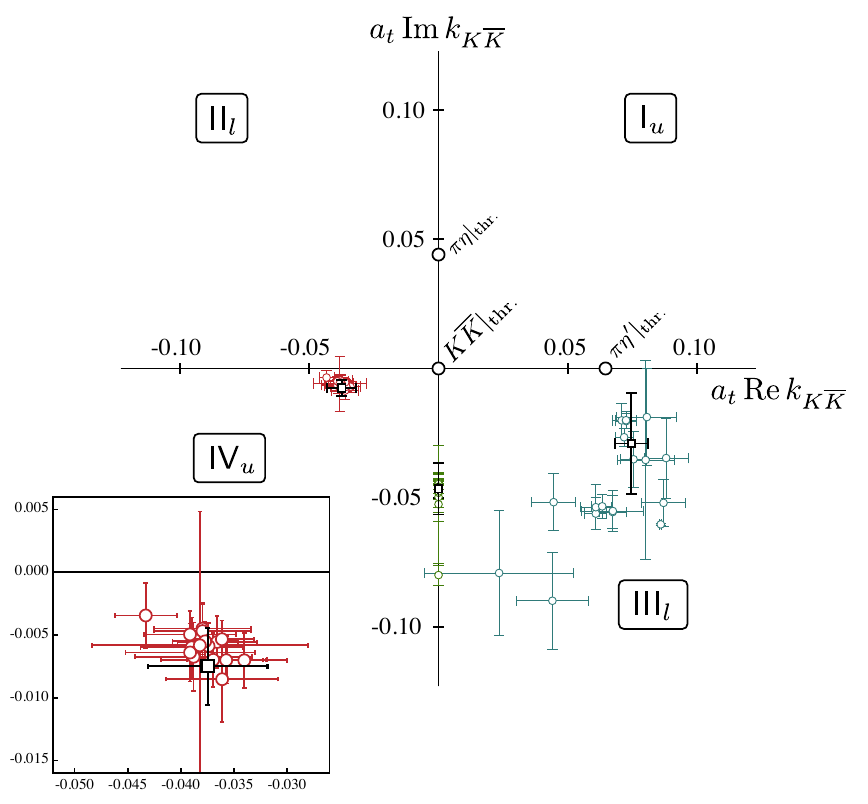}
\caption{Poles found for each $K$-matrix parameterization presented in Table~\ref{table_fit_forms} having $\chi^2/N_\mathrm{dof} < 1.5$. Red points are poles found on sheet {\sf IV}, blue points are poles found on sheet {\sf III}, and green points are poles found on the real energy axis, but on either sheet {\sf III} or sheet {\sf IV} depending upon the parameterization. The thick black points indicate the parameterization described by Eq.~\ref{fit_Kmatrix_const_2x2}. }
\label{fig_poles_Ecm}
\end{figure}

\begin{figure}
\includegraphics[width=0.75\columnwidth]{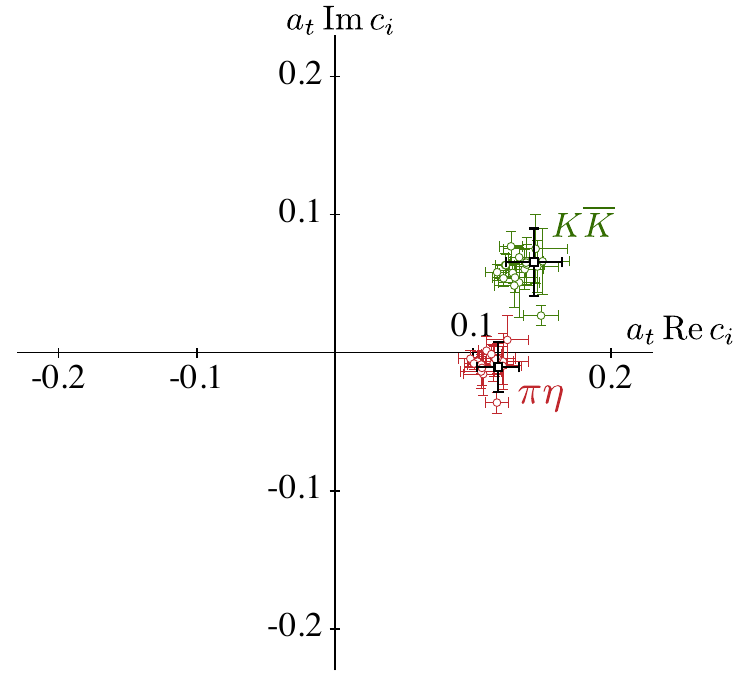}
\caption{The channel couplings extracted from the residues of the sheet~$\mathsf{IV}_u$ pole. Shown in red is $c_\pe$ and green is $c_\kkb$. We highlight those from the case of Eq.~\ref{fit_Kmatrix_const_2x2} in bold.}
\label{fig_res_sheet4_2x2}
\end{figure}

\begin{figure}
\includegraphics[width=0.75\columnwidth]{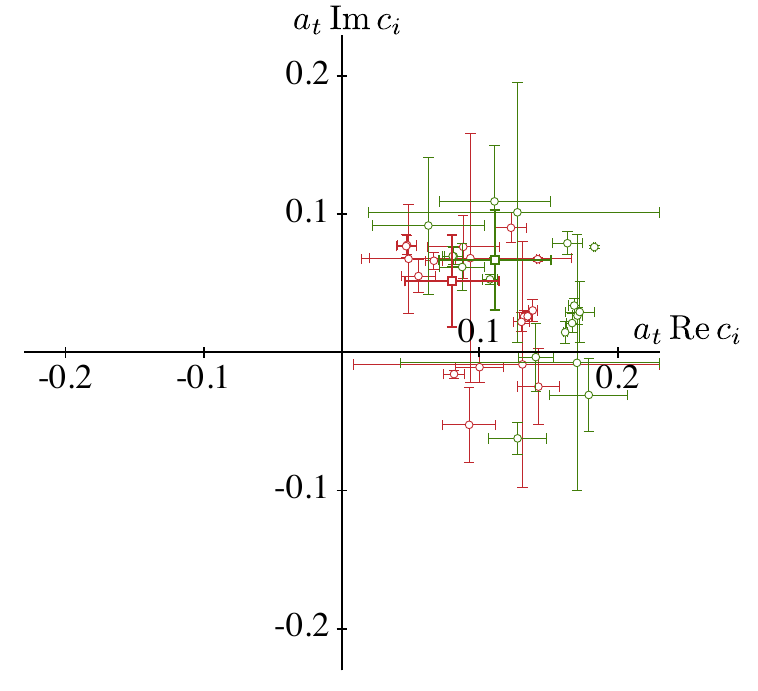}
\caption{As Figure~\ref{fig_res_sheet4_2x2} for the sheet~$\mathsf{III}_l$ pole.}
\label{fig_res_sheet3_2x2}
\end{figure}

We examined the singularity structure of the amplitudes presented in Table \ref{table_fit_forms}, and the pole positions for all descriptions with $\chi^2/N_\mathrm{dof} < 1.5$ are presented in Figure \ref{fig_poles_Ecm}. In every case we find a pole on $\mathsf{IV}_u$ located rather close to the real axis slightly above the \kk ~threshold (red points). The distribution of other nearby poles depends strongly upon the parameterization -- typically we find a $\mathsf{III}_l$ pole located significantly further into the complex plane (blue points), and we see that its position varies significantly with parameterization choice. There is also usually a pole on the real energy axis (green points), that while it may appear to be close to \pe~threshold, is actually on sheet {\sf III} or {\sf IV} and, as can be observed in the lower panel of Figure~\ref{fig_poles_Ecm}, is not close to \pe~threshold.

The lower panel of Figure~\ref{fig_poles_Ecm} makes it clear that the well-determined $\mathsf{IV}_u$ pole is actually very close to $\mathsf{II}_l$, these two sheets being continuously connected. It is also clear that a pole at this position is likely to strongly influence the behavior of amplitudes close to the \kk ~threshold, while the poorly determined pole on sheet {\sf III} is most likely influencing the higher energy behavior of the amplitudes. 

We also determine the residues of each pole, and factorize these into couplings $c_i$ -- these are shown in Figures~\ref{fig_res_sheet4_2x2}(for the $\mathsf{IV}_u$ pole) and \ref{fig_res_sheet3_2x2}(for the $\mathsf{III}_l$ pole). For the well-determined $\mathsf{IV}_u$ pole we find a slightly larger coupling to $\kk$ channel than to $\pe$, with the $\pe$ coupling being close to purely real in all of the fits, while the $\kk$ coupling has a phase of roughly $-30^\circ$. The residues of the sheet~$\mathsf{III}$ poles are statistically poorly determined, and show significant scatter over choice of parameterization. The residues of the real-axis poles show significant scatter.

The pole content of the six parameter ``pole plus constant'' $K$-matrix parameterization that we presented in Eq.~\ref{fit_Kmatrix_const_2x2} is highlighted in bold in Figs.~\ref{fig_poles_Ecm}, \ref{fig_res_sheet4_2x2} and \ref{fig_res_sheet3_2x2}. In this case the nearby poles are located at,
\begin{align}
a_t\sqrt{s_0}|_{\mathsf{IV}}  &= (0.2075\pm 0.0042)+\tfrac{i}{2}(0.0108\pm 0.0043)\nonumber\\
a_t\sqrt{s_0}|_{\mathsf{III}} &= (0.2406\pm 0.0059)-\tfrac{i}{2}(0.072\,\,\,\pm 0.053)\nonumber\\
a_t\sqrt{s_0}|_{\mathsf{IV}}  &= (0.1702\pm 0.0110), \nonumber
\end{align}
and the corresponding factorized residues of these poles in each channel are found to be
\begin{center}
\begin{tabular}{c|ll}
sheet          & $a_t c_{\pe}$      & $a_t c_{\kkb}$ \\[0.3ex]
\hline\\[-2ex]
$\mathsf{IV}$  & $0.119(15)\, e^{-i\pi \, 0.028(48)}$ & $0.158(19)\, e^{+i\pi \, 0.136(51)}$\\[0.3ex]
$\mathsf{III}$ & $0.095(35)\, e^{+i\pi \, 0.18(10)}$  & $0.130(43)\, e^{+i\pi \, 0.173(83)}$\\[0.3ex]
$\mathsf{IV}$  & $0.13(25)$ & $0.13(20)$
\end{tabular}
\end{center}

\noindent and there are also poles on $\mathsf{IV}_l$ and $\mathsf{III}_u$ located at the complex conjugate positions, with factorized residues that are the complex conjugates of those above.

Taking a conservative average over many parameterizations we find for the sheet $\mathsf{IV}_u$ pole,
\begin{center}
\begin{tabular}{rl}
$a_t\sqrt{s_0} =$ & $(0.2077\pm 0.0047) + \tfrac{i}{2}(0.0086\pm 0.0059)$ \\[0.5ex]
 $a_t c_{\pe} = $                  & $(0.115\pm 0.023)\,e^{-i\pi \, (0.032\pm 0.098)}$ \\[0.5ex]
 $a_t c_{\kkb} = $                 & $(0.149\pm 0.030)\,e^{+i\pi \, (0.122\pm 0.073)}$ \,.
\end{tabular}
\end{center}

In the three-channel (\pe,~\kk,~\pep) analysis discussed in Section~\ref{3chan} we find a very similar pole position and residues in good agreement, for both signs of $k_\pep$. We are also able to determine $c_\pep$ even though this channel is kinematically closed at the position of the real part of the pole, finding $|a_t\, c_\pep|\simeq 0.1$ and phases consistent with zero but with significant uncertainty.

\subsection{Jost functions}

The $K$-matrix formalism we used to describe the finite-volume spectra in Section~\ref{amps} has the distinct advantage of ensuring that unitarity is automatically satisfied at all the real energies we consider. On the other hand, the pole content of the resulting $t$-matrix is obscure -- we do not know precisely how many nearby poles the amplitude will feature until we've determined the particular parameters needed to describe the spectra. It would be convenient to have a parameterization of the $t$-matrix in which we can manually specify the number of poles and on which unphysical sheets they appear. We describe here an approach which attempts, in a limited way, to do that.

What we will refer to as the Jost function parameterisation~\cite{Jost:1947,LeCouteur:1960,Newton:1961,Kato:1965,Morgan:1993td} leverages relations for the two-channel $S$-matrix, when written as a function of the $\cm$ momenta for each channel, that allow the four sheets of the complex energy plane to be unfolded into a single sheet. The elements of the $S$-matrix can be written in terms of an auxilliary function of the two complex momenta, $\mathfrak{J}(k_1, k_2)$, the Jost function,
\begin{align*}
S_{11}&=\frac{\mathfrak{J}(-k_1,k_2)}{\mathfrak{J}(k_1,k_2)} \\
S_{22}&=\frac{\mathfrak{J}(k_1,-k_2)}{\mathfrak{J}(k_1,k_2)}\\
\mathrm{det}\,{\textbf S}&=\frac{\mathfrak{J}(-k_1,-k_2)}{\mathfrak{J}(k_1,k_2)}.
\end{align*}
Through a simple mapping \cite{Kato:1965} of $(k_1, k_2)$, we can write these as functions of a variable $\omega$,
\begin{equation*}
\omega=\frac{k_1+k_2}{\sqrt{k_1^2-k_2^2}},\;\;
\omega^{-1}=\frac{k_1-k_2}{\sqrt{k_1^2-k_2^2}}\,,
\end{equation*}
and by making the identifications
\begin{align*}
\mathfrak{J}(k_1,k_2)  \,\to\,  & \mathfrak{D}(\omega)\\
\mathfrak{J}(-k_1,k_2) \,\to\,  & \mathfrak{D}(-\omega^{-1})\\
\mathfrak{J}(k_1,-k_2) \,\to\,  & \mathfrak{D}(\omega^{-1})\\
\mathfrak{J}(-k_1,-k_2) \,\to\, & \mathfrak{D}(-\omega),
\end{align*}
we find that the $S$-matrix may be written
\begin{align}
S_{11}&=\frac{\mathfrak{D}(-\omega^{-1})}{\mathfrak{D}(\omega)}\,, \nonumber \\
S_{22}&=\frac{\mathfrak{D}(\omega^{-1})}{\mathfrak{D}(\omega)}\,, \nonumber \\
\mathrm{det}\,{\textbf S }&=\frac{\mathfrak{D}(-\omega)}{\mathfrak{D}(\omega)}\,.
\label{eq_S_jost}
\end{align}
There are restrictions on the function $\mathfrak{D}(\omega)$, notably $\mathfrak{D}(\omega)=\mathfrak{D}^\star(-\omega^\star)$ which follows from the Hermitian analyticity of the scattering amplitude. Following ref.~\cite{Kato:1965} we might write a relatively simple parameterization of $\mathfrak{D}(\omega)$ as a product of zeroes:
\begin{align}
\mathfrak{D}(\omega)=\frac{1}{\omega^2}
\left(1-\frac{\omega}{\omega_{p_1}}\right)
\left(1+\frac{\omega}{\omega_{p_1}^\star}\right)
\left(1-\frac{\omega}{\omega_{p_2}}\right)
\left(1+\frac{\omega}{\omega_{p_2}^\star}\right).
\label{eq_dkato}
\end{align}
The zeroes at $\omega=\omega_{p_i}$ and $\omega=-\omega_{p_i}^\star$ become poles of the $S$-matrix when used in Eq.~\ref{eq_S_jost}. The utility of this form is that these poles are input parameters whose real and imaginary parts can be manipulated as desired. A complication is that not all aspects of unitarity are certain to be obeyed by this amplitude -- while $|S_{11}| = |S_{22}|$ automatically, $|S_{11}| \le 1$ is not guaranteed, and for certain parameter choices may be violated. In practice we must always verify that this constraint is satisfied before we can accept an amplitude of this type.

\begin{figure}
\includegraphics[width=0.95\columnwidth]{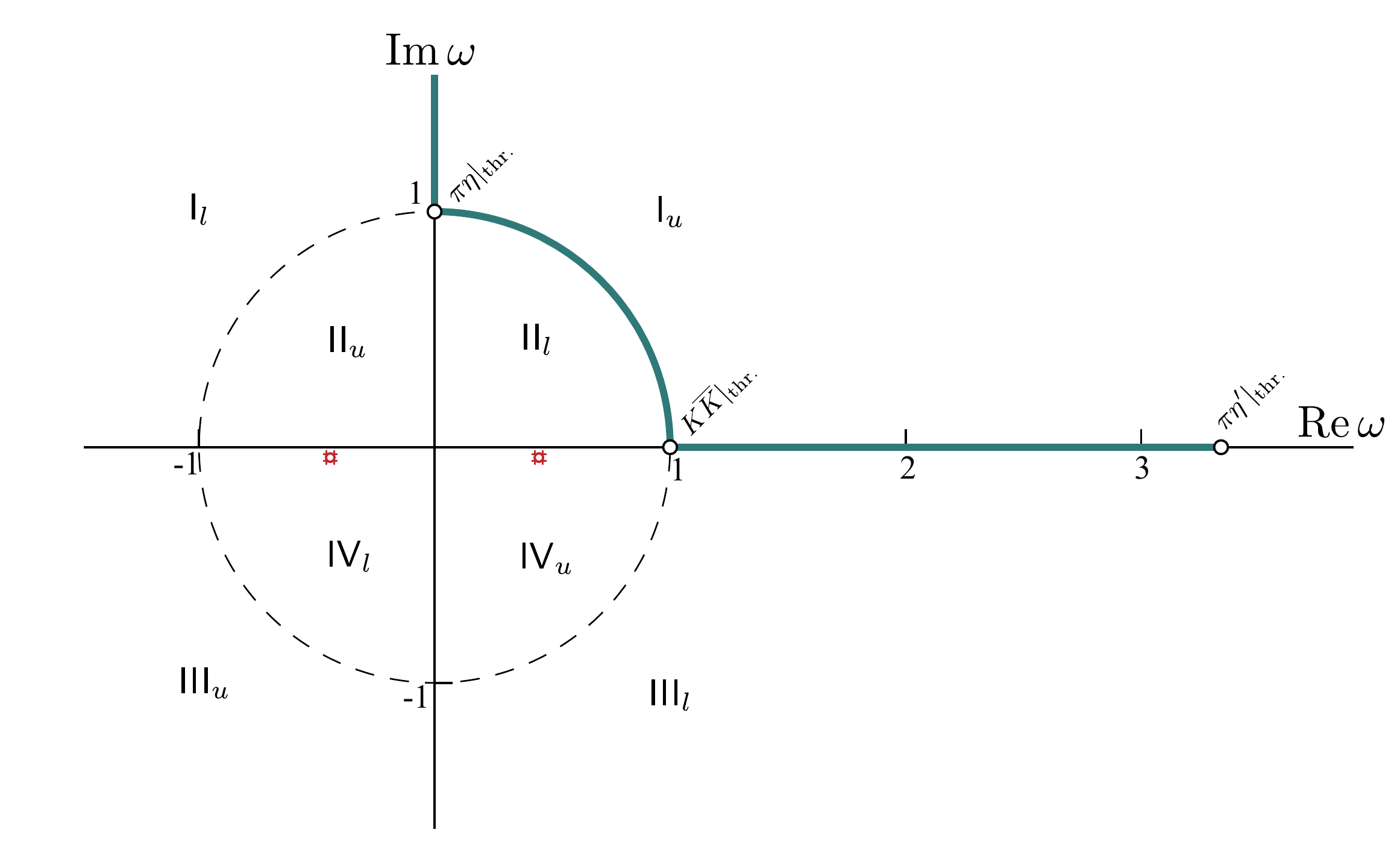}
\caption{ Complex $\omega$ plane. Physical scattering occurs along the blue line. The Riemann sheet structure in $s$ is shown.
Shown in red are the poles corresponding to $\omega_{p1}$ in Eq.~\ref{jost_min}.}
\label{fig_polesomegaplane}
\end{figure}

For orientation we plot the complex-$\omega$ plane in Fig.~\ref{fig_polesomegaplane}. Real energies below \pe~threshold appear on the imaginary $\omega$ axis above $\omega = i$. As energy increases above the \pe~threshold physical scattering follows the unit circle clockwise to $\omega=1$ where $\kk$ threshold opens, and from there it moves along the positive $\omega$ axis with increasing energy. In this application we do not consider energies above $a_t E_\cm=0.233$ where the $\pep$ channel opens. The sheets are labelled using the usual numbering scheme with a $u$ suffix denoting the upper half of the complex-$s$ plane and $l$ denoting the lower half-plane.

\begin{figure}
\includegraphics[width=\columnwidth]{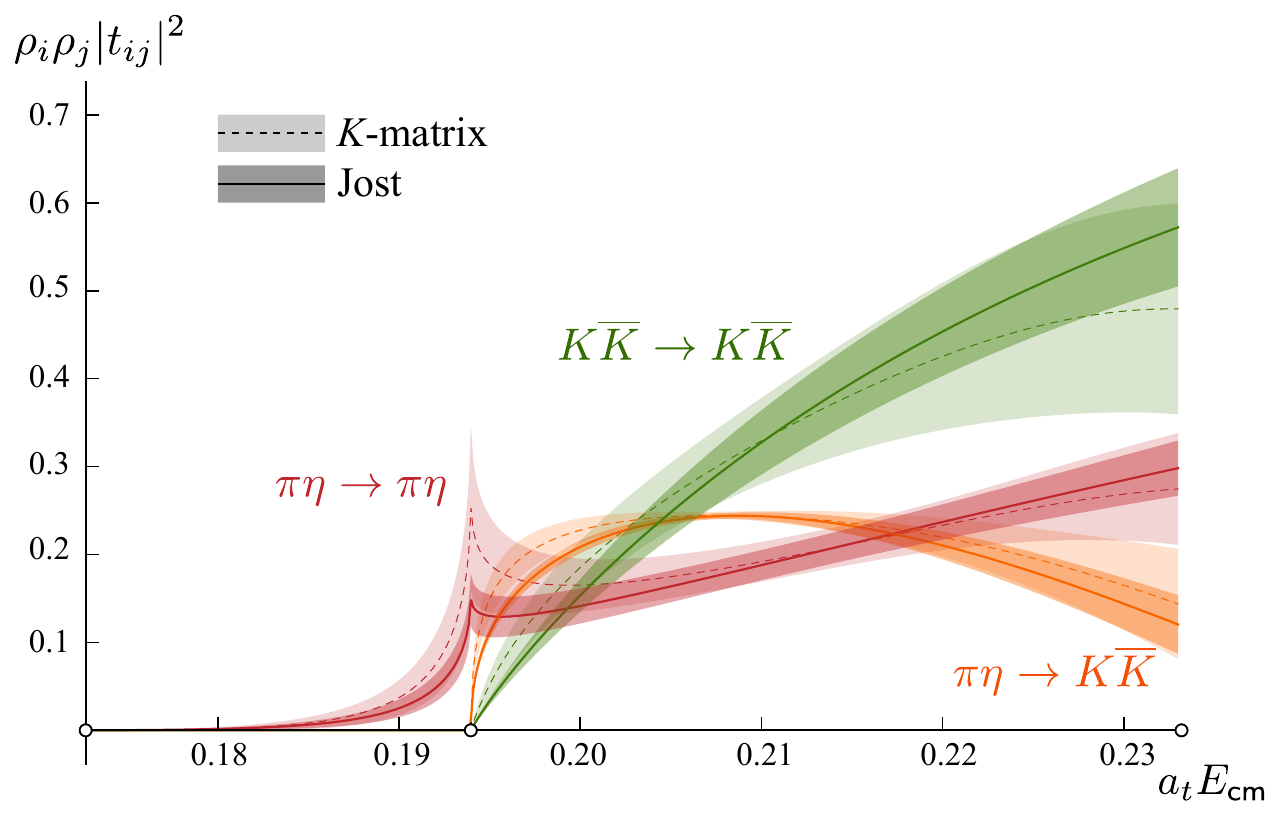}
\caption{The Jost amplitude, Eq.~\ref{jost_min} (solid line), compared with the $K$-matrix ``pole plus constant'' form of Eq.~\ref{fit_Kmatrix_const_2x2} (dashed line).}
\label{fig_jost_vs_KC}
\end{figure}

\begin{figure}
\includegraphics[width=0.85\columnwidth]{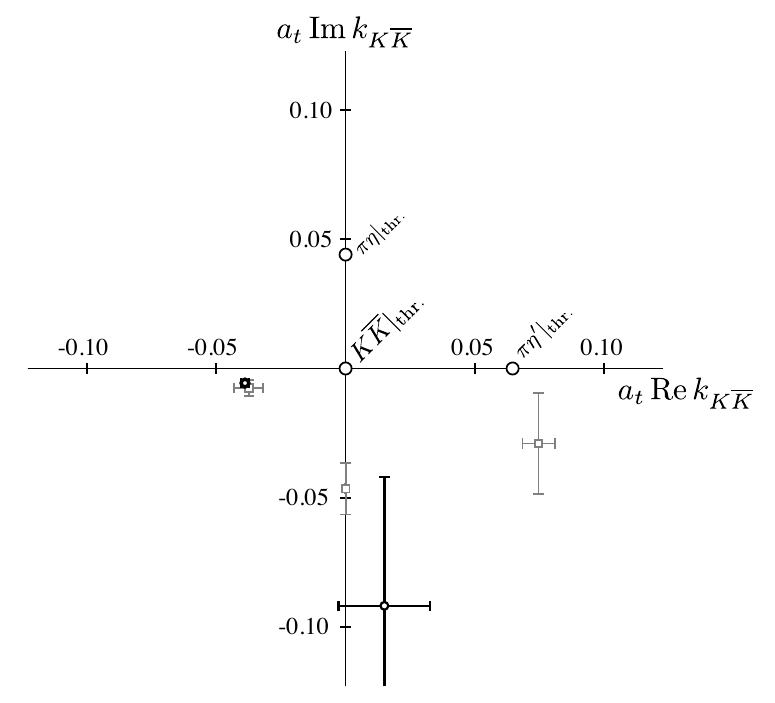}
\caption{Complex $k_\kkb$ plane: Jost amplitude poles, Eq.~\ref{jost_min} (black points), compared with those of the $K$-matrix ``pole plus constant'' form of Eq.~\ref{fit_Kmatrix_const_2x2} (grey points).}
\label{fig_jostkcmecm}
\end{figure}

Attempting to describe our standard set of 47 energy levels using the parameterization $\mathfrak{D}(\omega)$ in Eq.~\ref{eq_dkato} with the real and imaginary parts of the poles as free parameters we find a best fit given by,
{\small
\begin{align}
\begin{tabular}{rll}
$\mathrm{Re}\,\omega_{p_1} =$                  & $( \,\,\,\; 0.443 \pm 0.016 \pm 0.006)$   &
\multirow{4}{*}{ $\begin{bmatrix*}[r] 1 &  -0.35    & -0.01  &   0.47    \\
                                    	&  1        &  0.01  &  -0.02    \\
                                    	&           &  1     &   0.10    \\
                                        &           &        &   1      \end{bmatrix*}$ } \\
$\mathrm{Im}\,\omega_{p_1} =$                  & $(-0.044 \pm 0.014 \pm 0.003) $   & \\
$\mathrm{Re}\,\omega_{p_2} =$                  & $(\,\,\,\; 0.00 \,\,\, \pm 2.22\,\,\, \pm 0.02)$   & \\
$\mathrm{Im}\,\omega_{p_2} =$                  & $(-3.83\,\,\, \pm 0.20 \,\,\, \pm 0.08) $   & \\
&\multicolumn{2}{l}{ $\chi^2/ N_\mathrm{dof} = \frac{58.9}{47-4} = 1.37 $\,,}
\end{tabular}\label{jost_min}
\end{align}
}
{\noindent}where the nearby poles in $\omega$ are shown in Fig.~\ref{fig_polesomegaplane}. The resulting amplitude for real energies is qualitatively similar to our $K$-matrix parameterizations, as can be seen in Fig.~\ref{fig_jost_vs_KC} where it is compared to the six parameter $K$-matrix ``pole plus constant'' fit given by Eq.~\ref{fit_Kmatrix_const_2x2}. $\omega_{p1}$ corresponds to a pole on sheet {\sf IV} in the same location found for all successful $K$-matrix parameterizations -- Figure~\ref{fig_jostkcmecm} shows these poles in the complex-$k_\kkb$ plane.

\begin{figure*}
\includegraphics[width=0.75\textwidth]{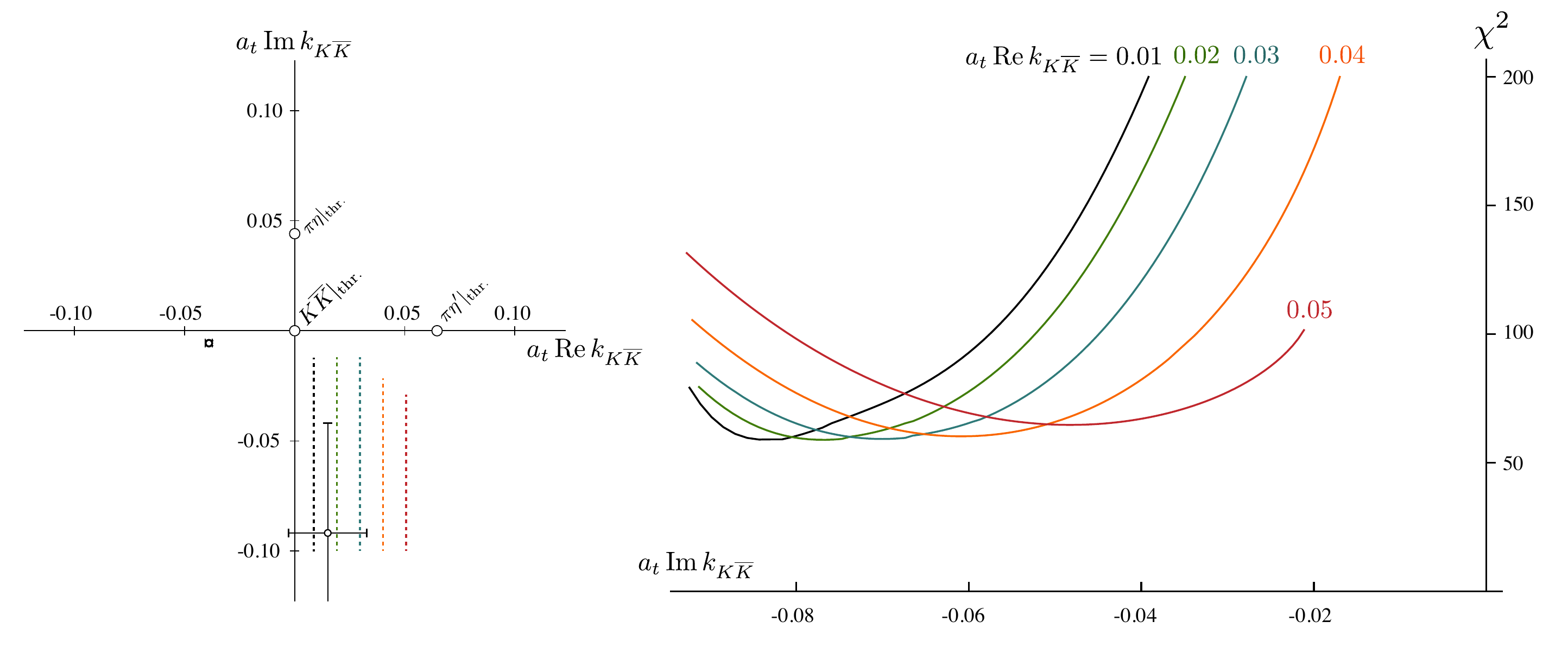}
\caption{Left: Jost amplitude poles, Eq.~\ref{jost_min} (black points) and the lines in $k_{\kkb}$ scanned over with a fixed sheet $\mathsf{IV}$ pole. Right: The $\chi^2$ along the lines of constant $\mathrm{Re}\, k_{\mathsf{cm}}^{\kkb}$ as $\mathrm{Im}\, k_{\mathsf{cm}}^{\kkb}$ is varied.}
\label{fig_jostchisq}
\end{figure*}

We can explore the sensitivity of the description of the finite-volume spectrum to the precise location of the ``second'' pole by scanning the $\chi^2$ as a function of that pole's position. If we fix the first pole parameter at the position found in the two pole fit above, ${\omega_{p_1}= 0.443 -0.044i}$ and scan the second pole over positions in sheet~$\mathsf{III}$, we find the result presented in Figure~\ref{fig_jostchisq}. We see clearly that a second pole close to the $\mathrm{Re} \, k_\kkb$ axis is not preferred\footnote{values very close to the real axis lead to amplitudes which violate the $|S_{11}| \le 1$ unitarity condition}.

Our well-determined pole is on the upper half-plane of sheet {\sf IV}, but is very close to the lower half-plane of sheet {\sf II}. Within the Jost parameterization, we find qualitatively similar results, but with a larger total $\chi^2$, if we fix $\omega_{p1}$ so that the pole is shifted just onto sheet~$\mathsf{II}$.

These simple applications of the Jost parameterization appear to confirm our observation from $K$-matrix studies, that the finite-volume spectrum requires the presence of a nearby pole on sheet {\sf IV} (but very close to sheet {\sf II}), while the presence of other nearby poles is not strongly suggested.

\subsection{$D$-wave scattering}

\begin{figure}[h]
\includegraphics[width=\columnwidth]{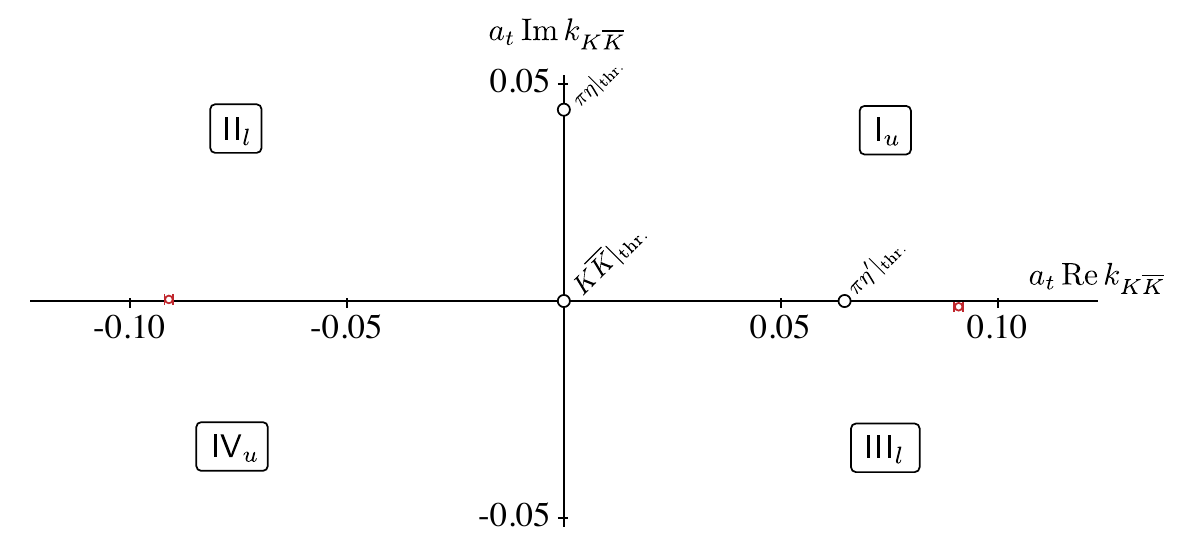}
\caption{The $D$-wave poles plotted in the complex $k_\kkb$ plane. We see a pair of poles with $k_\mathsf{II}\simeq-k_\mathsf{III}$ as is expected for a narrow resonance far above threshold (see e.g. Fig.~\ref{k2sheet}).}
\label{fig_D_poles}
\end{figure}

The $D$-wave scattering amplitudes discussed in Section~\ref{D_amps} all contain a narrow resonance with strong couplings to both $\pe$ and $\kk$. We reiterate that these amplitudes were obtained from spectra without $\pi\pi\pi$-like operator constructions and no attempts were made to incorporate three-body effects in the finite-volume formalism (efforts in this direction~\cite{Beane:2007qr, Polejaeva:2012ut, Briceno:2012rv, Hansen:2014eka, Meissner:2014dea, Hansen:2015zga, Hansen:2015azg, Hansen:2015zta} are not yet applicable in our situation). Well determined poles were obtained on sheets~$\mathsf{III}$ and $\mathsf{II}$, and these are illustrated in Fig.~\ref{fig_D_poles} where we observe that this resonance appears to be of the ```canonical'' coupled-channel type having a mirror-pole pair with the sheet {\sf III} pole dominating the amplitude at real values of energy close to the pole position (although recall that we are ignoring any complications from the presence of other two-body and three-body channels in this first analysis).

Analyzing the $K$-matrix with six parameters shown in Eq.~\ref{D_fit_par_values} we find poles at,
\begin{align*}
a_t\sqrt{s_0}|_{\mathsf{II}_l}  &= 0.26576(77)-\tfrac{i}{2}0.00108(45)\\
a_t\sqrt{s_0}|_{\mathsf{III}_l} &= 0.26577(77)-\tfrac{i}{2}0.00359(48),
\end{align*}
and the corresponding factorized residues of these poles in each channel are found to be,
\begin{center}
\begin{tabular}{c|cc}
sheet          & $a_t c_{\pe}$      & $a_t c_{\kkb}$ \\[0.3ex]
\hline\\[-2ex]
$\mathsf{II}_l$  & $0.0286(23)e^{-i\pi \, 0.0108(43)}$ & $0.0221(24)e^{-i\pi \, 0.0077(75)}$\\[0.3ex]
$\mathsf{III}_l$ & $0.0287(23)e^{-i\pi \, 0.0098(46)}$ & $0.0221(24)e^{-i\pi \, 0.0053(76)}$
\end{tabular}
.
\end{center}

\section{Interpretation and Summary}\label{interpret}

\begin{figure*}
\includegraphics[width=0.44\textwidth]{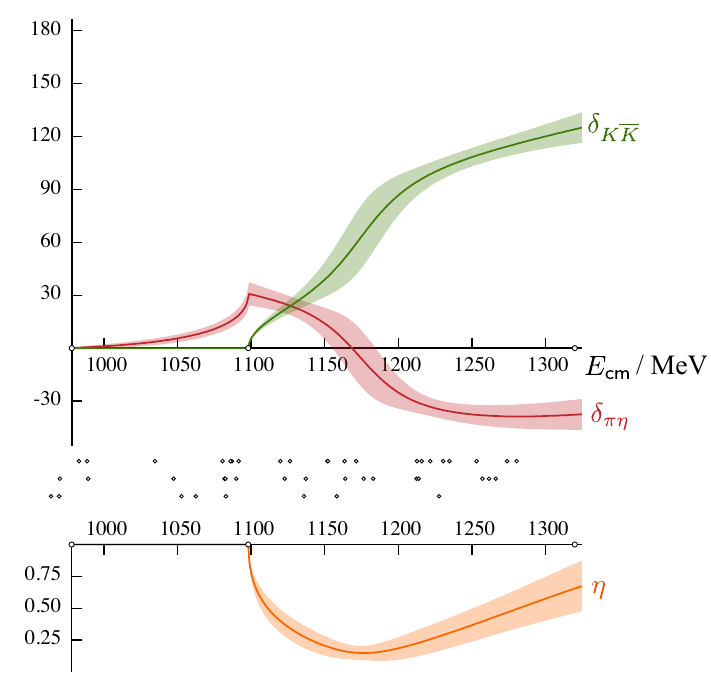}
\includegraphics[width=0.54\textwidth]{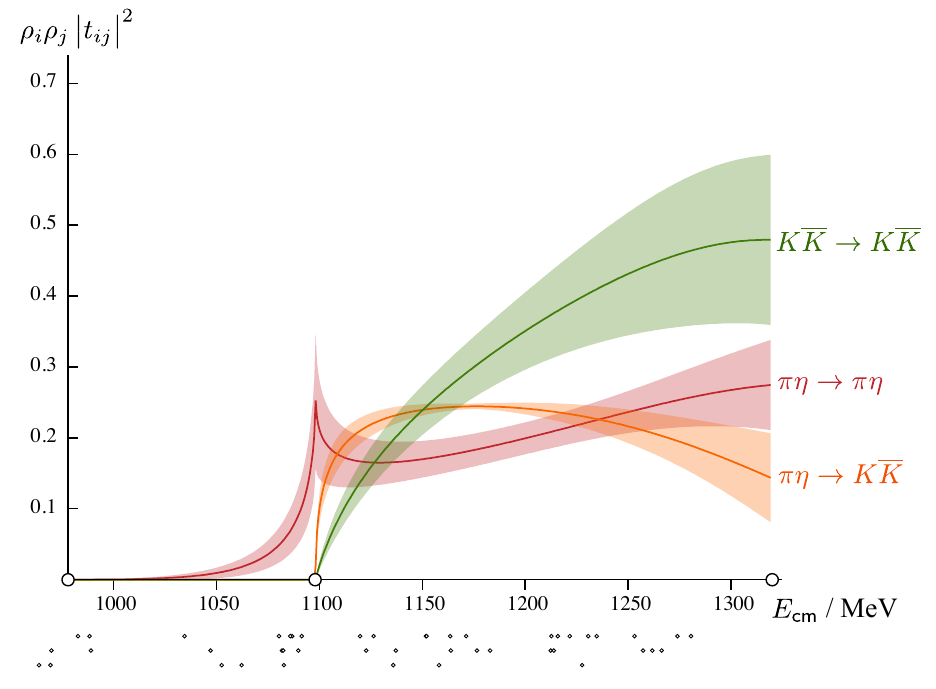}
\caption{$S$-wave coupled-channel \pe,~\kk~scattering amplitudes using ``pole plus constant'' $K$-matrix parameterization. Left: phase-shifts and inelasticity. Right: amplitude magnitudes. Open circles on axis indicate the positions of the \pe,~\kk~and \pep~thresholds. Lattice QCD energy levels constraining the amplitude are shown as dots below the figures (increasing lattice volume from bottom to top). In this calculation the stable meson masses are $m_\pi=391$ MeV, $m_K=549$ MeV, $m_\eta=587$ MeV and $m_\eta^\prime=929$ MeV.}
\label{fig_phys}
\end{figure*}

All parameterizations of the $S$-wave $t$-matrix that we found capable of describing the finite-volume spectrum share the same essential features: a strong cusp-like enhancement in $\pi \eta \to \pi \eta$ at the \kk~threshold coupled with a rapid turn-on of amplitudes leading to the \kk~final-state. This is illustrated in Figure~\ref{fig_phys} for a typical successful parameterization. The strength of these effects likely indicates a resonance close to the threshold that is strongly coupled to both channels, but clearly one that does not manifest itself in the canonical way as a simple symmetrical bump in the amplitudes. 

Upon examining the singularity structure of our parameterized amplitudes we found in all cases a statistically well-determined pole near to the \kk~threshold, located on sheet {\sf IV}, but very close to the boundary with sheet {\sf II}. Expressed in physical units (using the scale setting procedure described in Section~\ref{levels}) this pole lies at
\begin{align*}
{\sqrt{s_0}}=\left((1177\pm 27) + \frac{i}{2}(49\pm 33)\right)\,\, \mathrm{MeV},
\end{align*}
and has a residue which factorizes into couplings
\begin{align*}
\big| c_\pe \big|  &= 652(130)\,\mathrm{MeV} \\
\big| c_\kkb \big|  &= 844(170)\,\mathrm{MeV},
\end{align*}
indicating comparable coupling to each channel, $\big| c_\kkb / c_\pe \big| = 1.30(37)$.

Many parameterizations indicated another pole, on the real energy axis at energy values close to $m_\pi + m_\eta$, but actually lying on sheet {\sf III} or {\sf IV} and hence far from the physical \pe~threshold. The position of this pole changed with parameterization choice and was typically not statistically well determined -- its influence on the amplitude behavior near \kk~threshold is not completely clear. A further pole on sheet {\sf III} lying far from the \kk~threshold may be present, but it showed significant sensitivity to parameterization choice and it is unlikely that it is relevant to the threshold activity. The sheet {\sf IV} pole and the corresponding behavior around the \kk~threshold is the robust result of this analysis.

A canonical two channel resonance, as described for example by the Flatt\'e form, features a \emph{pair} of poles, located either on sheets {\sf II} and {\sf III}, or sheets {\sf IV} and {\sf III}, depending upon the relative size of the couplings to the two channels. The alternative case observed in this analysis, featuring dominance of a \emph{single} pole near a kinematic threshold has previously been discussed~\cite{Morgan:1992ge,Morgan:1993td} as a possible signal for a state which is not one bound tightly by short-range (``confining'') forces, but rather one which binds through the long-range interaction between a pair of mesons, i.e. a ``hadron molecule''. The resonance pole we have determined is only $79(27)$ MeV above the \kk~threshold, and we find it to have a large coupling to the \kk~channel, and as such we expect \kk~components to play a significant role in the wavefunction of the state.

Since this is a calculation with artificially heavy $u,d$ quarks, direct comparison to experiment is not justified, but we note that most analyses of experimental data (see Ref.~\cite{Baru:2004xg} for a summary in the context of the Flatt\'e amplitude) suggest that the $a_0(980)$ appears with a phase-shift $\delta_\pe$ which rises with increasing energy, and a corresponding pole lying slightly above \kk~threshold on sheet {\sf II}. This superficially differs from our result, but as discussed in Appendix A, relatively small changes in the coupling to the \kk~channel or to the resonance `bare-mass' parameter would lead to the sheet {\sf IV} pole migrating to sheet {\sf II}\footnote{and see Ref~\cite{Baru:2004xg} who find the same evolution using a Flatt\'e form when the resonance `bare-mass' is above threshold and the coupling to \kk~is large} and a sudden `flip' to a rising phase-shift $\delta_\pe$. This evolution may indicate a possible destiny for this state as the quark mass is reduced toward the physical value, ending in agreement with the experimental data. Explicit lattice QCD calculations at lower quark mass are warranted to explore this.

If we restrict our attention to the behavior of the ${\pe \to \pe}$ $S$-wave amplitude below \kk~threshold, we observe that an effective range parameterization with a scattering length $a_0 = (0.09 \pm 0.06 )\,\mathrm{fm}$ and zero effective range can well describe the finite volume spectra. The scattering length extracted from the low-energy behavior of the coupled-channel $K$-matrix amplitudes, ${a_0 = (0.02 \pm 0.04)\, \mathrm{fm}}$, is compatible with this. These values prove to be much smaller than those we found for $\pi K$ scattering in the $I=1/2$ channel at the same value of the $u,d$ quark mass in Refs.~\cite{Dudek:2014qha,Wilson:2014cna}, $a_0^{\pi K} = (0.65 \pm 0.07)\, \mathrm{fm}$. Although our quarks are far from having exact chiral symmetry, this observation is in line with expectations of chiral effective theory~\cite{Bernard:1990kx,Bernard:1991xb, Albaladejo:2015aca}, and may reflect the absence of a ($\kappa$, $\sigma$)-like state in the $\pi \eta$ channel.

Beyond $S$-wave scattering, we were able to infer from the spectrum in the $T_1^-$ irrep that there is negligible $P$-wave $\pi \eta$ scattering at low energy. This was expected as this is an exotic $J^{PC} = 1^{-+}$ channel in which $q\bar{q}$ mesons cannot appear -- hybrid mesons are indicated at much higher energy~\cite{Dudek:2010wm}.

In $D$-wave we extracted coupled $\pe, \kk$ amplitudes under the assumption that $\pi\pi\pi$ and other three-meson channels remain irrelevant at the quark masses considered here. We found our spectrum could be well described by an amplitude featuring a very narrow resonance coupled to both \pe~and \kk~whose nearby sheet {\sf III} pole is located at ${ \sqrt{s_0} =\left(1506(4)-\frac{i}{2}20(3)\right) \,\mathrm{MeV} }$ with couplings $\big| c_\pe \big | = 162(14)\,\mathrm{MeV}$, $\big| c_\kkb \big | = 125(14)\,\mathrm{MeV}$.

\subsection*{Summary}
We have presented the first extraction of a strongly coupled-channel meson-meson scattering system in lattice QCD finding an $S$-wave resonance which may be associated with the experimental $a_0(980)$ state. The resonance lies in a two coupled-channel (\pe, \kk) region, and we also extended the analysis to higher energy with a limited first consideration of three-channel scattering (\pe, \kk, \pep). In order to proceed to still higher energies a formalism is required to extract scattering amplitudes featuring three-meson channels, and significant progress is being made in this direction~\cite{Beane:2007qr, Polejaeva:2012ut, Briceno:2012rv, Hansen:2014eka, Meissner:2014dea, Hansen:2015zga, Hansen:2015azg, Hansen:2015zta}.

In the near future we will apply similar methods to those considered here to the $I=0$ $\pi\pi, K\overline{K} \ldots$ coupled system in which we expect to see physics corresponding to the low-lying scalar mesons, $\sigma$ and $f_0(980)$. Once this is done we will have a first survey of the $I=0,\tfrac{1}{2},1$ scalar meson sector from first-principles QCD computation. Exploring the structure of these states, including the possible role of hadronic molecule components, will then be a priority, and possible tools at our disposal include coupling to external currents in order to determine ``form-factors'' of the resonances. It has recently been shown that finite-volume lattice QCD calculations can give access to such quantities~\cite{Briceno:2015dca}. The behavior of the states with varying quark mass will also inform our descriptions, in particular their relative proximity to the \kk~threshold.

The extraction of the first strongly coupled-channel meson resonance is a major milestone in the path towards our goal of studying highly excited hadron resonances in first-principles QCD calculations.

\begin{acknowledgments}

We thank our colleagues within the Hadron Spectrum Collaboration and M. R. Pennington for fruitful discussions. {\tt Chroma}~\cite{Edwards:2004sx} and {\tt QUDA}~\cite{Clark:2009wm,Babich:2010mu} were used to perform this work on clusters at Jefferson Laboratory under the USQCD Initiative and the LQCD ARRA project. Gauge configurations were generated using resources awarded from the U.S. Department of Energy INCITE program at Oak Ridge National Lab, the NSF Teragrid at the Texas Advanced Computer Center and the Pittsburgh Supercomputer Center, as well as at Jefferson Lab. RGE and JJD acknowledge support from U.S. Department of Energy contract DE-AC05-06OR23177, under which Jefferson Science Associates, LLC, manages and operates Jefferson Laboratory. JJD acknowledges support from the U.S. Department of Energy Early Career award contract DE-SC0006765. DJW received support from a grant from the Isaac Newton Trust/University of Cambridge Early Career Support Scheme [RG74916].

\end{acknowledgments}

\bibliographystyle{apsrev4-1}
\bibliography{bib}


\appendix
\section{Amplitudes, phase-shifts and poles in a ``pole plus constant'' $K$-matrix}
\label{app_phases}

In this appendix we'll explore parameter variations in a ``pole plus constant'' $K$-matrix and the corresponding changes to the amplitudes. Notably we'll find that while the magnitudes of the amplitudes evolve with changing parameter values rather smoothly, the phase-shifts can undergo discontinuous change, and these changes are correlated with poles moving between Riemann sheets. Similar evolutions of pole position have been reported previously in Refs.~\cite{Janssen:1994wn,Baru:2004xg} where a Flatt\'e form is used.

Let's consider a relatively simple parameterization of the two-channel $t$-matrix,
\begin{align}
t^{-1}_{ij}=K^{-1}_{ij}-i\rho_i\delta_{ij},
\label{eq_Kmat_noCM}
\end{align}
with
\begin{align}
\mathbf{K} = \frac{1}{m^2-s}\begin{bmatrix} g_\pe^2 & g_\pe \, g_\kkb \\ g_\pe \, g_\kkb & g_\kkb^2 \end{bmatrix}      + \begin{bmatrix} 0 & \gamma \\ \gamma & 0   \end{bmatrix},
\label{eq_Kmat_onlyCC}
\end{align}
where we are choosing to use the ordinary phase-space (rather than the Chew-Mandelstam form) for simplicity. We note that this parameterization is able to successfully describe our lattice QCD spectra with ${\chi^2/N_\mathrm{dof}=\frac{61.9}{47-4}=1.44}$, with parameter central values: $g_\pe=0.127$, $g_\kkb=0.178$, $m=0.2221$ and $\gamma=0.570$. We will explore the behavior of the amplitude as we vary the value of $g_\kkb$ or $m$.

\begin{figure*}
\includegraphics[width=0.9\textwidth]{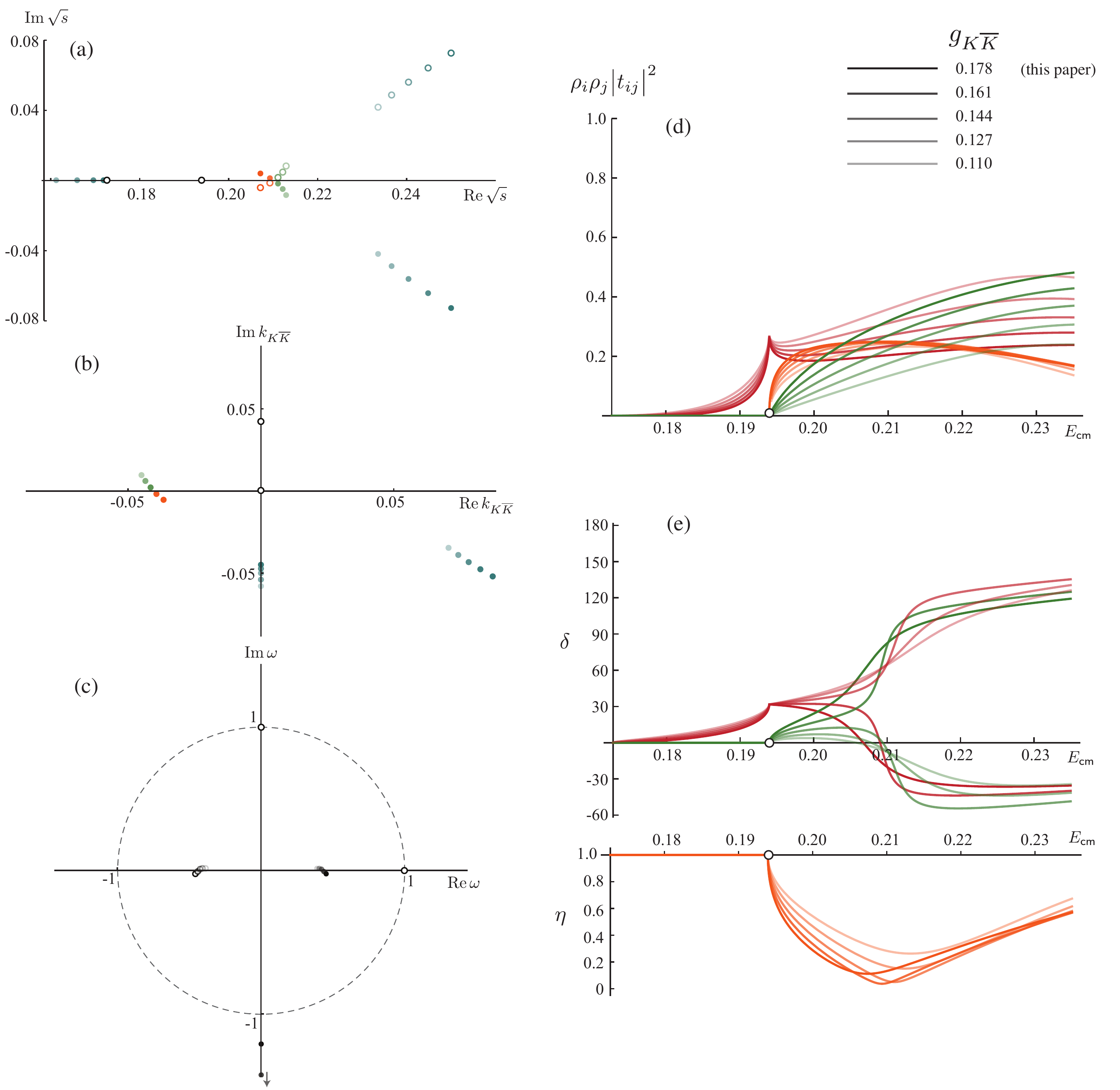}\\
\caption{Amplitude of Eq.~\ref{eq_Kmat_onlyCC} for varying $g_\kkb$. \\
Pole positions in (a) complex $\sqrt{s}$ plane, (b) complex $k_\kkb$ plane and (c) complex $\omega$ plane. Poles in green on sheet {\sf II}, poles in blue on sheet {\sf III} and poles in orange on sheet {\sf IV}.\\
(d) Amplitude magnitudes, $\rho^2_\pe \, |t_{\pe,\pe}|^2$ (red), $\rho_\pe \rho_\kkb \, |t_{\pe,\kkb}|^2$ (orange), $\rho^2_\kkb \, |t_{\kkb,\kkb}|^2$ (green).\\
(e) Phase-shifts ($\delta_\pe$ in red and $\delta_\kkb$ in green) and inelasticity (orange). }
\label{varying_gKK}
\end{figure*}

Figure~\ref{varying_gKK} shows the amplitude evolution with varying $g_\kkb$ considering five values in equal steps from $0.178$ down to $0.110$. Note that between $g_\kkb = 0.161$ and $0.144$, the phase-shift graph, (e), `flips' from one in which $\delta_\kkb$ (green) has the rising behavior to one where it is $\delta_\pe$ (red) which rises. However this change in character is not associated with any discontinuous change in the magnitude of the amplitudes, $\rho_i \rho_j |t_{ij}|^2$ (d). We observe in (a), (b), and (c), that this transition in phase-shift form corresponds to the nearby pole moving smoothly from sheet {\sf IV} to sheet {\sf II}.

Other behavior visible in Figure~\ref{varying_gKK} is that the real axis pole on sheet {\sf III} moves further away as $g_\kkb$ is reduced while the off-axis pole on sheet {\sf III} moves closer to physical scattering.

\begin{figure*}
\includegraphics[width=0.9\textwidth]{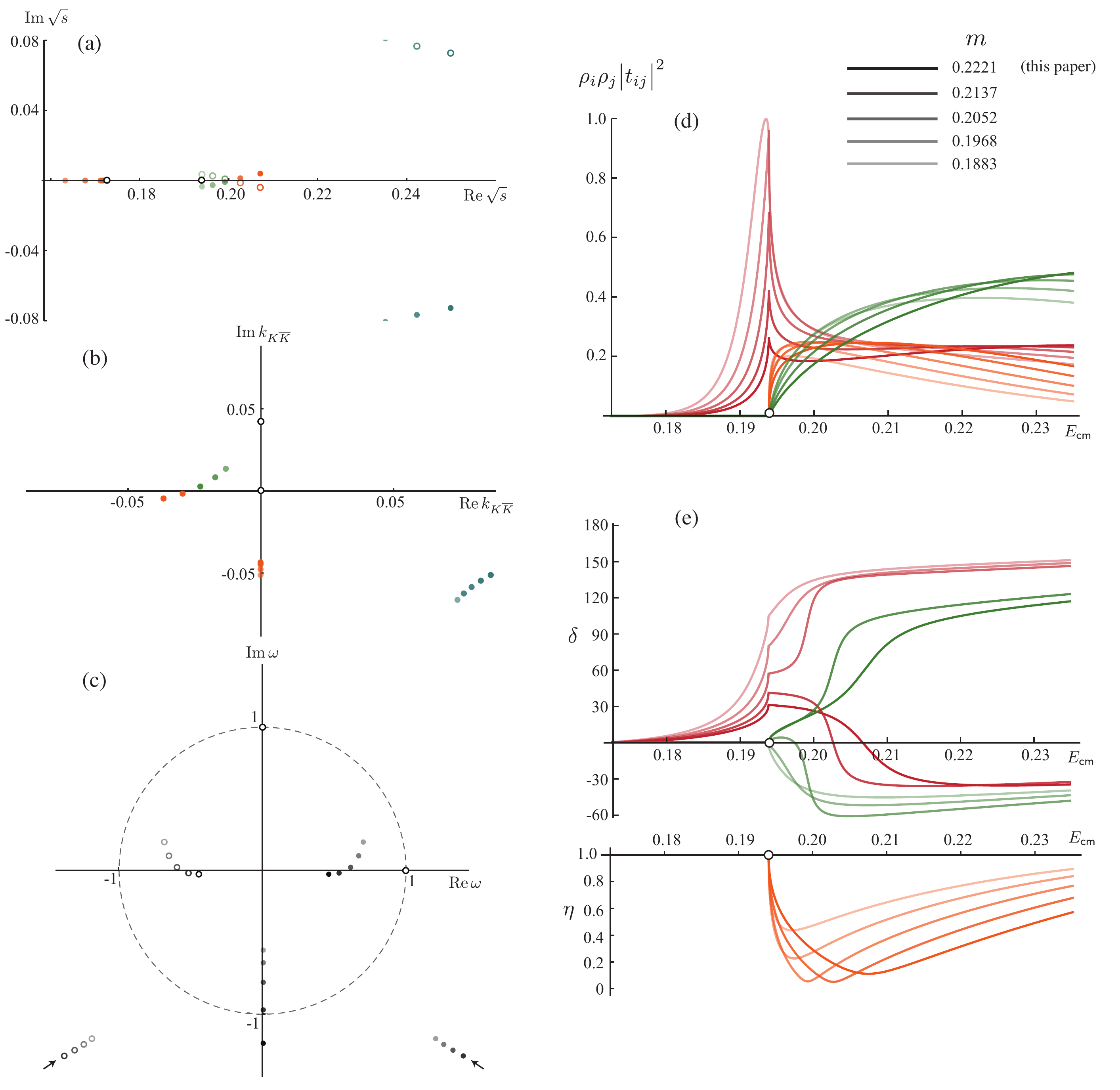}\\
\caption{As Figure~\ref{varying_gKK} but for varying parameter, $m$.}
\label{varying_m}
\end{figure*}

Figure~\ref{varying_m} shows the amplitude evolution with varying $m$ considering five values in equal steps from $0.2221$ down to $0.1883$. Again we observe a `flip' in the character of the phase-shifts, while the magnitudes of the amplitudes evolve smoothly, and we can again trace this behavior to the transition of a nearby pole from sheet {\sf IV} to sheet {\sf II}. Note also that there is more pole motion between sheets in this case with the pole on the real energy axis moving from sheet {\sf III} to sheet {\sf IV} as $m$ is reduced -- the off-axis sheet {\sf III} pole evolves rather slowly

The right top panel shows gradual adjustment from a `cusp'-like $\pe \to \pe$ amplitude, which we have found describes the lattice QCD spectra obtained in this document, to one which is more of an asymmetric bump near \kk threshold as $m$ is reduced. This observation is interesting, since one possible evolution of this system with decreasing quark mass would be for the resonance mass to decrease faster than the energy of the $\kk$ threshold, which this in some ways resembles.

\section{Operator Tables}
\label{app_op_tables}

In this appendix we summarize the operator bases that were used to construct matrices of correlation functions, from which the finite volume spectra used throughout this study were obtained. Table~\ref{tab_opsused} contains the operators used in irreps that contain subductions of $S$-wave and higher partial-waves, while Table~\ref{tab_opsused_PD} shows the operators used in irreps whose lowest partial wave is either $P$ or $D$-wave.

\begin{table*}
\begin{ruledtabular}
\begin{tabular}{ccc|ccc|ccc}
\multicolumn{3}{c|}{$[000]\, A_1^+$}   &
\multicolumn{3}{c|}{$[100]\, A_1$}  	&
\multicolumn{3}{c}{$[110] \, A_1$} 	\\[0.3ex]
$16^3$ & $20^3$ & $24^3$ & $16^3$ & $20^3$ & $24^3$ & $16^3$ & $20^3$ & $24^3$\\
\hline \hline
%
%
  $\pi_{000} \,\eta_{000}$  & $\pi_{000} \,\eta_{000}$  & $\pi_{000} \,\eta_{000}$                      
& $\pi_{000} \,\eta_{100}$  & $\pi_{000} \,\eta_{100}$  & $\pi_{000} \,\eta_{100}$                      
& $\pi_{000} \,\eta_{110}$  & $\pi_{000} \,\eta_{110}$  & $\pi_{000} \,\eta_{110}$ 	\\              
  $\pi_{100} \,\eta_{\sm 100}$  & $\pi_{100} \,\eta_{\sm 100}$  & $\pi_{100} \,\eta_{\sm 100}$          
& $\pi_{100} \,\eta_{000}$  & $\pi_{100} \,\eta_{000}$  & $\pi_{100} \,\eta_{000}$                      
& $\pi_{100} \,\eta_{010}$  & $\pi_{100} \,\eta_{010}$  & $\pi_{100} \,\eta_{010}$ 	\\              
                            & $\pi_{110} \,\eta_{\sm 1 \sm 1 0}$  & $\pi_{110} \,\eta_{\sm 1 \sm 1 0}$  
&                           & $\pi_{001} \,\eta_{10\sm 1}$  & $\pi_{001} \,\eta_{10 \sm 1}$             
& $\pi_{110} \,\eta_{000}$  & $\pi_{110} \,\eta_{000}$  & $\pi_{110} \,\eta_{000}$ 	\\              
                            &                           &                                               %
&                           & $\pi_{101} \,\eta_{00\sm 1}$  & $\pi_{101} \,\eta_{00 \sm 1}$             
& $\pi_{001} \,\eta_{11\sm 1}$  & $\pi_{001} \,\eta_{11\sm 1}$  &$\pi_{001} \,\eta_{11\sm 1}$	\\      
  & &    & & &  $\pi_{\sm 100} \,\eta_{200}$
&                           &                           &  $\pi_{101} \,\eta_{01\sm 1}$	  \\
  & &    & & &
&                           &                           &  $\pi_{111} \,\eta_{00\sm 1}$	  \\[1.5ex]     
%
%
  $  K_{000} \, \kb_{000}$  & $  K_{000} \,\kb_{000}$   & $  K_{000} \,\kb_{000}$
& $  K_{100} \, \kb_{000}$  & $  K_{100} \,\kb_{000}$   & $  K_{100} \,\kb_{000}$
& $  K_{000} \, \kb_{110}$  & $  K_{000} \,\kb_{110}$   & $  K_{000} \,\kb_{110}$ 	    \\
  $  K_{100} \, \kb_{\sm 100}$  & $K_{100} \,\kb_{100}$ & $  K_{100} \,\kb_{100}$
&                           & $K_{101} \,\kb_{00\sm 1}$ & $  K_{101} \,\kb_{00\sm 1}$
& $K_{100} \,\kb_{010}$     & $K_{100} \,\kb_{010}$     & $  K_{100} \,\kb_{010}$           \\
                            &                           & $  K_{110} \,\kb_{\sm 1 \sm 10}$
&                           &                           &
&                           &                           & $  K_{101} \,\kb_{0 1 \sm 1}$  \\[1.5ex]
%
%
  $\pi_{000} \, \ep_{000}$  & $\pi_{000} \,\ep_{000}$   & $\pi_{000} \,\ep_{000}$
&                           & $\pi_{000} \,\ep_{100}$   & $\pi_{000} \,\ep_{100}$
& $\pi_{000} \, \ep_{110}$  & $\pi_{000} \,\ep_{110}$   & $\pi_{000} \,\ep_{110}$ 	\\
                            & $\pi_{100} \,\ep_{\sm 100}$   & $\pi_{100} \,\ep_{\sm 100}$
&                           & $\pi_{100} \,\ep_{000}$   & $\pi_{100} \,\ep_{000}$
& $\pi_{100} \,\ep_{010}$   & $\pi_{100} \,\ep_{010}$   & $\pi_{100} \,\ep_{010}$ 	\\[1.5ex]
$\bar{\psi}\mathbf{\Gamma} \psi  \times  13$ & $\bar{\psi}\mathbf{\Gamma} \psi  \times  4$ & $\bar{\psi}\mathbf{\Gamma} \psi  \times  13$ &
$\bar{\psi}\mathbf{\Gamma} \psi  \times  9$ & $\bar{\psi}\mathbf{\Gamma} \psi  \times  8$ & $\bar{\psi}\mathbf{\Gamma} \psi  \times  10$ &
$\bar{\psi}\mathbf{\Gamma} \psi  \times  4$ & $\bar{\psi}\mathbf{\Gamma} \psi  \times  18$ & $\bar{\psi}\mathbf{\Gamma} \psi  \times  6$ \\
\end{tabular}
\end{ruledtabular}

\vspace{.2cm}

\begin{ruledtabular}
\begin{tabular}{ccc|ccc}
\multicolumn{3}{c|}{$[111]\, A_1$}   &
\multicolumn{3}{c}{$[200]\, A_1$}  	\\[0.3ex]
$16^3$ & $20^3$ & $24^3$ $\qquad \qquad$& $16^3$ & $20^3$ & $24^3$ \\
\hline \hline
%
%
  $\pi_{000} \,\eta_{111}$  & $\pi_{000} \,\eta_{111}$  & $\pi_{000} \,\eta_{111}$      $\qquad \qquad$           
& $\pi_{000} \,\eta_{200}$  & $\pi_{000} \,\eta_{200}$  & $\pi_{000} \,\eta_{200}$ 	\\              
  $\pi_{100} \,\eta_{011}$  & $\pi_{100} \,\eta_{011}$  & $\pi_{100} \,\eta_{011}$       $\qquad \qquad$      
& $\pi_{100} \,\eta_{100}$  & $\pi_{100} \,\eta_{100}$  & $\pi_{100} \,\eta_{100}$ 	\\              
  $\pi_{110} \,\eta_{001}$  & $\pi_{110} \,\eta_{001}$  & $\pi_{110} \,\eta_{001}$       $\qquad \qquad$ 
& $\pi_{200} \,\eta_{000}$  & $\pi_{200} \,\eta_{000}$  & $\pi_{200} \,\eta_{000}$ 	\\              
  $\pi_{111} \,\eta_{000}$  & $\pi_{111} \,\eta_{000}$  & $\pi_{111} \,\eta_{000}$       $\qquad \qquad$                 
&                           & $\pi_{101} \,\eta_{10\sm 1}$  & $\pi_{101} \,\eta_{10\sm 1}$   \\
                            &                           &
&                           &                           &  $\pi_{111} \,\eta_{1\sm 1 \sm 1}$     \\[1.5ex]
%
%
  $  K_{111} \, \kb_{000}$  & $  K_{111} \,\kb_{000}$   & $  K_{111} \,\kb_{000}$         $\qquad \qquad$
& $  K_{000} \, \kb_{200}$  & $  K_{200} \,\kb_{200}$   & $  K_{000} \,\kb_{200}$ 	    \\
  $  K_{110} \, \kb_{001}$  & $  K_{110} \,\kb_{001}$   & $  K_{110} \,\kb_{001}$         $\qquad \qquad$
& $  K_{100} \,\kb_{100}$   & $  K_{100} \,\kb_{100}$   & $  K_{100} \,\kb_{100}$         \\
                            &                           &
&                           & $  K_{101} \,\kb_{10\sm 1}$   & $  K_{101} \,\kb_{10\sm 1}$ 	   \\[1.5ex]
%
%
                            &                           & $\pi_{000} \,\ep_{111}$       $\qquad \qquad$
& $\pi_{000} \, \ep_{200}$  & $\pi_{000} \,\ep_{200}$   & $\pi_{000} \,\ep_{200}$ 	\\
                            &                           & $\pi_{100} \,\ep_{011}$       $\qquad \qquad$
&                           & $\pi_{100} \,\ep_{100}$   & $\pi_{100} \,\ep_{100}$ 	\\
                            &                           &
&                           &                           & $\pi_{200} \,\ep_{000}$      \\[1.5ex]
$\bar{\psi}\mathbf{\Gamma} \psi  \times  5$ & $\bar{\psi}\mathbf{\Gamma} \psi  \times  6$ & $\bar{\psi}\mathbf{\Gamma} \psi  \times  5$ $\qquad \qquad$ &
$\bar{\psi}\mathbf{\Gamma} \psi  \times  7$ & $\bar{\psi}\mathbf{\Gamma} \psi  \times  5$ & $\bar{\psi}\mathbf{\Gamma} \psi  \times  9$ \\
\end{tabular}
\end{ruledtabular}

\caption{The operator bases used in each lattice irrep in this calculation. For each irrep we list the ``$\pi\eta$-like'', ``$\kk$-like'' and ``$\pi\eta^\prime$-like'' operators that were used as well as the number of ``single-meson-like'' operators. We use a notation which indicates the momentum (in units of $2\pi/L$) of the pseudoscalar meson operators, recalling that the directions of momentum are summed over with generalized Clebsch-Gordan weights to ensure the operator lies in the stated irrep~\cite{Dudek:2012gj,Dudek:2012xn}.}
\label{tab_opsused}
\end{table*}

\begin{table*}
\begin{ruledtabular}
\begin{tabular}{ccc|ccc|ccc}
\multicolumn{3}{c|}{$[000]\, T_1^-$}   &
\multicolumn{3}{c|}{$[000]\, E^+$}  	&
\multicolumn{3}{c}{$[000]\, T_2^+$} 	\\[0.3ex]
$16^3$ & $20^3$ & $24^3$ & $16^3$ & $20^3$ & $24^3$ & $16^3$ & $20^3$ & $24^3$\\
\hline \hline
%
%
  $\pi_{100} \,\eta_{\sm 100}$  & $\pi_{100} \,\eta_{\sm 100}$  & $\pi_{100} \,\eta_{\sm 100}$                      
& $\pi_{100} \,\eta_{\sm 100}$  & $\pi_{100} \,\eta_{\sm 100}$  & $\pi_{100} \,\eta_{\sm 100}$                      
&                               &                               &                               	\\          
                                   & $\pi_{110} \,\eta_{\sm 1 \sm 10}$  & $\pi_{110} \,\eta_{\sm 1 \sm 10}$             
& $\pi_{110} \,\eta_{\sm 1\sm 10}$ & $\pi_{110} \,\eta_{\sm 1 \sm 10}$  & $\pi_{110} \,\eta_{\sm 1 \sm 10}$            
& $\pi_{110} \,\eta_{\sm 1\sm 10}$ & $\pi_{110} \,\eta_{\sm 1 \sm 10}$  & $\pi_{110} \,\eta_{\sm 1 \sm 1 0}$ 	\\[1.5ex]     
%
%
                              &                            &
& $  K_{100} \,\kb_{\sm 100}$ &$  K_{100} \,\kb_{\sm 100}$ & $  K_{100} \,\kb_{\sm 100}$
&                             &                            & $  K_{110} \,\kb_{\sm 1 \sm 10}$ 	   \\[1.5ex]
%
%
                                & $\pi_{100} \, \ep_{\sm 100}$   & $\pi_{100} \, \ep_{\sm 100}$
& $\pi_{100} \, \ep_{\sm 100}$  & $\pi_{100} \, \ep_{\sm 100}$   &
&                               &                                &                        	\\[1.5ex]
$\bar{\psi}\mathbf{\Gamma} \psi  \times  4$ & $\bar{\psi}\mathbf{\Gamma} \psi  \times  4$ & $\bar{\psi}\mathbf{\Gamma} \psi  \times  13$ &
$\bar{\psi}\mathbf{\Gamma} \psi  \times 17$ & $\bar{\psi}\mathbf{\Gamma} \psi  \times 17$ & $\bar{\psi}\mathbf{\Gamma} \psi  \times  12$ &
$\bar{\psi}\mathbf{\Gamma} \psi  \times 22$ & $\bar{\psi}\mathbf{\Gamma} \psi  \times  6$ & $\bar{\psi}\mathbf{\Gamma} \psi  \times  5$ \\
\end{tabular}
\end{ruledtabular}

\vspace{.2cm}

\begin{ruledtabular}
\begin{tabular}{ccc|ccc}
\multicolumn{3}{c|}{$[100]\, B_1$}   &
\multicolumn{3}{c}{$[100]\, B_2$}  	\\[0.3ex]
$16^3$ & $20^3$ & $24^3$ $\qquad \qquad$ & $16^3$ & $20^3$ & $24^3$ \\
\hline \hline
%
%
  $\pi_{010} \,\eta_{1\sm 10}$     & $\pi_{010} \,\eta_{1\sm 10}$      & $\pi_{010} \,\eta_{1\sm 10}$           $\qquad \qquad$
&                                  &                                   & $\pi_{011} \,\eta_{1\sm 1\sm 1}$    	\\
  $\pi_{110} \,\eta_{0\sm 10}$     & $\pi_{110} \,\eta_{0\sm 10}$      & $\pi_{110} \,\eta_{0\sm 10}$           $\qquad \qquad$
&                                  &                                   & $\pi_{111} \,\eta_{0\sm 1\sm 1}$	\\[1.5ex]
%
%
  $  K_{010} \, \kb_{1\sm 10}$    &  $  K_{010} \, \kb_{1\sm 10}$      & $  K_{010} \, \kb_{1\sm 10}$            $\qquad \qquad$
&                                 &                                    & $  K_{111} \, \kb_{0\sm 1\sm 1}$         \\[1.5ex]
$\bar{\psi}\mathbf{\Gamma} \psi  \times 11$  & $\bar{\psi}\mathbf{\Gamma} \psi  \times 11 $ & $\bar{\psi}\mathbf{\Gamma} \psi  \times 7$ $\qquad \qquad$ &
$\bar{\psi}\mathbf{\Gamma} \psi  \times 11$  & $\bar{\psi}\mathbf{\Gamma} \psi  \times 11 $ & $\bar{\psi}\mathbf{\Gamma} \psi  \times 10$ \\
\end{tabular}
\end{ruledtabular}

\caption{As above but for irreps not featuring $S$-wave subductions.}
\label{tab_opsused_PD}
\end{table*}

\end{document}